\documentclass[%
nofootinbib,
 amsmath,amssymb,
 aps,
prd,showpacs,twocolumn,superscriptaddress
]{revtex4-2} 

\usepackage{graphicx} % Include figure files
\usepackage{subfigure} 
\usepackage{dcolumn} % Align table columns on decimal point

\usepackage{amsmath} 
\usepackage{amssymb} 
\usepackage{bm} 
\usepackage{color}

\def\red#1{\textcolor{red}{#1}}

\usepackage[normalem]{ulem} 
\usepackage[dvipsnames]{xcolor} 
\usepackage{hyperref} 
\usepackage{bm} % bold math
\usepackage{subfiles} % Best loaded last in the preamble
\usepackage{amsthm,amsmath,amssymb} 
\usepackage{mathrsfs} 
\usepackage{url} 
\usepackage{orcidlink}%插入orcid

\newcommand\be{\begin{equation}}
\newcommand\ee{\end{equation}}
\newcommand\ba{\begin{eqnarray}}
\newcommand\ea{\end{eqnarray}}
\def\ba#1\ea{\begin{align}#1\end{align}}
\newcommand{\delay}[1]{\mathbf{D}_{#1}}

\begin{document}

\title{Assessing the systematic errors of extreme-mass-ratio inspirals waveforms  for testing general relativity} 
\author{Ping Shen~\orcidlink{0009-0002-8366-9473}} 
 \affiliation{Shanghai Astronomical Observatory, Shanghai, 200030, China} 
\affiliation{School of Astronomy and Space Science, University of Chinese Academy of Sciences,
Beijing, 100049, China} 
\author{Qiuxin Cui~\orcidlink{0009-0008-7882-8867}}
 \affiliation{Shanghai Astronomical Observatory, Shanghai, 200030, China} 
\affiliation{School of Astronomy and Space Science, University of Chinese Academy of Sciences,
Beijing, 100049, China} 
\author{Wen-Biao Han~\orcidlink{0000-0002-2039-0726}} 
 \email{wbhan@shao.ac.cn} 
 \affiliation{Shanghai Astronomical Observatory, Shanghai, 200030, China} 
\affiliation{School of Astronomy and Space Science, University of Chinese Academy of Sciences,
Beijing, 100049, China} 
\affiliation{School of Fundamental Physics and Mathematical Sciences, Hangzhou Institute for Advanced Study, UCAS, Hangzhou 310024, China} 
\affiliation{Shanghai Frontiers Science Center for  Gravitational Wave Detection, 800 Dongchuan Road, Shanghai 200240, China} 
\date{\today} % It is always \today, today,

\begin{abstract} 

Gravitational wave (GW) observations from extreme-mass-ratio inspirals (EMRIs) are powerful tools for testing general relativity (GR). However, systematic errors arising from waveform models could potentially lead to incorrect scientific conclusions. These errors can be divided into two main categories: fundamental bias (due to limitations in the validity of the Einstein field equations) and modeling error (due to inaccuracies in waveform templates). Using Bayesian inference, we investigate the impact of these systematic errors on tests of GR. Regarding fundamental bias, we find that at low signal-to-noise ratios (SNR), there is a risk of misidentifying a non-GR EMRI signal as a GR-EMRI one, and vice versa. However, this risk diminishes as the SNR increases to around 40 or higher. Additionally, modeling errors might reduce the SNR of detected EMRI signals and could be misinterpreted as deviations from GR, leading Bayesian inference to favor non-GR scenarios, especially at high SNR. We emphasize the importance of developing sufficiently accurate waveform templates based on alternative gravity theories for testing GR.

\end{abstract} 

\maketitle
\section{\label{intro} Introduction} 

Since the discovery of gravitational waves (GWs) from binary black holes (BBH) coalescence in 2015 \cite{PhysRevLett.116.061102}, the ground-based GW detectors (LIGO \cite{Aasi_2015}, Virgo \cite{Acernese_2015}, etc.)  have opened a new window on the universe. GW observations provide opportunities to test  General Relativity (GR) at the strong-field regime and enhance our understanding of gravity and fundamental physics. Several tests \cite{PhysRevLett.116.221101,PhysRevD.94.084002,PhysRevLett.120.031104,Li_2024,PhysRevD.100.104036} have been carried out on the data from the ground-based GW detectors, with no  conclusive evidence for the deviation from GR to date.

The future space-borne  GW detectors, such as the Laser Interferometer Space Antenna (LISA)  \cite{amaroseoane2017laserinterferometerspaceantenna}, Taiji   \cite{Hu2017TheTP,zhong2023exploring,Ren_2023}  and TianQin   \cite{Luo_2016}, will detect the GWs radiating from stellar-mass compact objects [SCOs: neutron stars, white dwarfs or black holes (BHs)] inspiralling into supermassive black holes (SMBHs). These events are known as extreme-mass-ratio inspirals (EMRIs) \cite{2007Intermediate,PhysRevD.95.103012}, which have the potential to facilitate stringent tests of GR \cite{gair2013testing,barack2007using,PhysRevD.56.1845,barack2007using,CH04,Fransen_2022,PhysRevD.100.084055,PhysRevD.104.064008,emri_GR_REVIEW}. 
Due to their extreme mass-ratios ($10^{-4}\sim 10^{-7}$), SCO acts like a test particle moving in the background of the central supermassive BHs. The radiated GWs reflect rich information about the geometry and environment around the central object.
If extracting this information from the GW signals, we will be able to accurately distinguish whether the central
object is indeed a Kerr BH or another object.

To achieve this scientific goal, we must accurately model EMRI waveforms. The accuracy requirement of the models depends on our purpose to which we put the model. For detection purposes (determining whether an EMRI signal is in your data or not), it is required that the dephasing $\Delta \phi \lesssim 1$ radian throughout the signal's duration \cite{hughes2016adiabatic}. For measurement purposes (extracting source parameters), our model must be accurate enough that systematic errors (due to inadequate modeling) are smaller than statistic errors (due to noise). A crude rule of thumb is that the template’s phase must match the signal to within $\Delta \phi \lesssim 1/{\rm SNR}$ \cite{hughes2016adiabatic,PhysRevD.104.064047}. As in Ref. \cite{Barack_2018}, the GW phase $\phi$ can be expanded as 
\begin{equation}
    \phi=\frac{1}{\nu}\{\nu^0 \phi_0 + \nu^{1}\phi_1 +O(\nu^2)\}\label{eq:phiPA},
\end{equation}
where $\nu$ is the mass ratio. The leading term in Eq.~(\ref{eq:phiPA}) is refered to as ``adiabatic'' order [$0$th post-adiabatic (0PA)], and the $n$th subleading term as $n$th post-adiabatic ($n$PA). Models that get  $\phi_0$ might be enough to detect most signals, but models that get  both $\phi_0$ and $\phi_1$ should be enough for precise parameter extraction \cite{lisaconsortiumwaveformworkinggroup2023waveform}.
 In order to search across the large parameter space, we need to generate waveforms in less than $1 \rm{s}$ on a single central processing unit (CPU) core \cite{PhysRevD.104.064047,lisaconsortiumwaveformworkinggroup2023waveform}. The accuracy and speed requirements have led to
the development of two classes of EMRI models: gravitational self-force (GSF) models for accuracy and “kludge” models for speed.

The ongoing GSF program \cite{Barack_2018} is a specific expansion within the black hole perturbation theory (BHPT) that aims to generate EMRI waveforms satisfying the accuracy requirements of the EMRI science. In this approach, EMRI is treated as a point mass orbiting a black hole and the dynamics can be described by the equation of motion of the mass, including the influence of the interaction with the self-field, i.e. the GSF \cite{10.1093/ptep/ptv092}.
Among the GSF formalisms, the recent PA waveforms \cite{PhysRevLett.130.241402,PhysRevLett.124.021101} for nonspinning compact binaries under a quasicircular inspiral  are the most accurate waveforms to date. 
In contrast to the slow GSF waveform model, the fast EMRI ``kludge" models [(analytical kludge (AK) \cite{barack2004lisa}, numerical kludge (NK)     \cite{babak2007kludge}, augmented analytical kludge (AAK) \cite{chua2017augmented})] are designed for rapidly generating waveforms, and have been widely used for LISA data analysis.

Tests of GR cannot rely on waveform families that assume GR is correct. Instead, such tests should employ more generic waveforms that allow for GR deviations \cite{VYS11}. A framework already exists to parametrically deform the metric tensor through the construction of so-called bumpy spacetimes \cite{CH04,Glampedakis_2006,VH10,VYS11,GY11,Bnbumps,PhysRevD.82.104041,Moore_2017}. Collins and Hughes \cite{CH04} (hereafter CH04) introduced the bumpy BHs for the Schwarzschild case, which are almost, but not quite, GR's BHs.  
Vigeland and Hughes \cite{VH10} extended the bumpy BHs concept in CH04 to Kerr BHs and dealed with the non-smooth nature of bumps presented in CH04.
Vigeland, Yunes, and Stein \cite{VYS11}  (hereafter VYS11) generalized the bumpy BH framework to allow for alternative gravity (AG) theory deformations. They map the parametrically deformed metrics  to known specific non-GR BH metrics, such as those in the dynamical Chern-Simons (CS) gravity one \cite{cs2009} and the dynamical quadratic one \cite{modifyGR_nospin}. Gair and Yunes \cite{GY11} construct approximate,  ``analytic-kludge’’ waveforms for EMRIs with parametrized post-Einsteinian (ppE) corrections that allow for generic, model-independent deformations of the supermassive BH background away from the Kerr metric. The deformations represent modified gravity effects and have been analytically mapped to several modified gravity black hole solutions in four dimensions \cite{cs2009,modifyGR_nospin}.

In our work, we focus on evaluating the systematic errors arising from waveform templates, which could potentially result in incorrect scientific conclusions. These errors can be divided into two main categories: modeling error \cite{PhysRevD.76.104018} and fundamental bias \cite{PhysRevD.80.122003}. The systematic error generated by the use of inaccurate template families can be broadly thought of as a \textit{modeling error}. It arises from simplified physical assumptions or unverified assumptions about the accuracy of the solutions used to model the given event. In addition, concerns about the validity of the Einstein field equations themselves represent a \textit{fundamental bias}.

Firstly, we study the systematic errors arising from fundamental biases. 
Following the approach of \cite{GY11}, we obtain the bumpy-kludge waveform templates by adding corrections to the corresponding evolution equations in the kludge waveform models. Our analysis also includes the detector response function, generating the second-generation Time-Delay Interferometry (TDI) variables with realistic orbits produced by Taiji. We compare the parameter estimation results obtained using AG-EMRI waveform templates (e.g., bumpy-AAK) with those from GR-EMRI waveform templates (e.g., AAK). Our findings indicate that in low SNR scenarios, there is a risk of mistaking an AG-EMRI signal for a GR-EMRI one, and vice versa. However, as the SNR increases to around $40$ or higher, the potential for confusion between AG and GR signals diminishes. 

Secondly, we discuss the systematic errors stemming from modeling errors. We explore the effectiveness of using less accurate AAK models to infer the parameters of EMRI signals generated by the more accurate NK models. We find that modeling errors may reduce the SNR of detected EMRI signals due to the mismatch between the NK signal and the AAK model. Additionally, the calculated Bayesian factors tend to favor the bumpy signal hypothesis, especially at high SNR. In such high SNR scenarios, even slight mismatches in waveform templates become noticeable and may be misconstrued as differences in the deformation parameter between Kerr and bumpy EMRI signals. This misinterpretation could lead to an overestimation of support for the bumpy signal hypothesis, potentially skewing the analysis and resulting in incorrect conclusions about the underlying EMRI signal.

This paper is organized as follows. Section \ref{AAK-bumpy} introduces the bumpy-kludge waveforms in the context of modified gravity theories. Sec. \ref{bayes} represents the method of parameter estimation and model selection. Sec. \ref{sec:fundamental bais} and Sec. \ref{sec:modeling error} evaluate the systematic errors induced by fundamental bais and modeling error, respectively. In Sec. \ref{conclusion} we draw conclusions, discuss our results, and suggest directions for future work. Throughout this paper, we use geometric units with $G = c = 1$.

\section{waveform models} \label{AAK-bumpy} 
The generalized model considered in this work is based on the kludge family waveforms, summarized in Sec. \ref{sec:kludge}. These GR-based EMRI kludge models are the most computationally efficient ones available and have been widely used in data analysis for space-borne GW detectors. In Sec. \ref{sec:bumpy}, we describe the construction of kludge waveforms on a family of generic modified-gravity BH spacetimes. These spacetimes are parameterized by metric deformations (or ``bumps") of different sizes \cite{GY11}, which show up in the resultant ``bumpy-kludge" model as perturbations to the phase evolution at different orders. In Sec.~\ref{sec:TDI}, we consider the detector response function, producing second-generation TDI variables featuring realistic orbits generated by Taiji.

\subsection{Kludge waveform models}\label{sec:kludge}

We begin with the kludge family waveforms which include the AK \cite{barack2004lisa}, NK \cite{babak2007kludge}, and AAK \cite{chua2017augmented,aak_2015} models. The AK model is very fast to compute, but is less accurate than the NK model, which combines Kerr geodesics with post-Newtonian (PN) orbital evolution to improve accuracy. The AAK model possesses both the speed of the AK model and the accuracy of the NK model, and has been widely used for many space-borne GW detectors.

In general, the kludge family describes the inspiral of a CO, treated as a point mass $\mu$, around a supermassive Kerr BH with mass $M$ and spin $a$. 
Assuming the spin of the compact object is negligible, an EMRI can be described by 14 parameters \cite{chua2017augmented,barack2004lisa}: 

\begin{equation}
	\begin{aligned}
     \Theta^i_{\rm kludge} &\equiv (\Theta^1,...,\Theta^{14}) \\
		&=[\mu,M,a,p_0,e_0,\iota_0,\Tilde{\gamma}_0,\Phi_0,\alpha_0,\theta_S,\phi_S,\theta_K,\phi_K,D]
	\end{aligned}
\end{equation}
where $\Vec{S}$ is the central black hole's spin vector, $\Vec{L}$ is the CO's orbital angular momentum vector, and $\hat{S}$ and $\hat{L}$ are the corresponding unit vectors, respectively. The directions of $\hat{S}$ are described by the angles ($\theta_K, \phi_K$), and the angle between $\hat{L}$ and $\hat{S}$ is labeled as $\iota$ (inclination). The azimuthal angle of $\hat{L}$  in the spin-equatorial plane is $\alpha$ , and the angle in orbital plane between $\hat{L} \times \hat{S}$ and the pericenter is $\tilde{\gamma}$. The parameters $e$, $p$, and $\Phi$ denote the orbital eccentricity, the semi-latus rectum, and the mean anomaly, respectively. The subscript $0$ in $\{p_0,e_0,\iota_0,\Tilde{\gamma}_0,\Phi_0,\alpha_0\}$ indicates the values of these parameters at the initial time $t_0$. The location of the system is determined by the angles ($\theta_S, \phi_S$) in ecliptic-based coordinates and the luminosity distance $D$.

The main components of a kludge waveform model can be schematically described as follows \cite{chua2017augmented}:

(i) the construction of the inspiral trajectory in ``phase-space'', using PN or fitted fluxes $\Vec{F}$:
\begin{equation}
\dot{\Vec{C}}=\Vec{F}(\mu,M,\Vec{S},\Vec{C});
\end{equation}

(ii) the construction of the CO's worldline (the ``configuration-space'' trajectory), using geodesic or flux-derived expressions $\Vec{G}$:
\begin{equation}
\dot{\Vec{X}}=\Vec{G}(\mu,M,\Vec{S},\Vec{C});
\end{equation}

(iii) the generation of the waveform strain $h$ at the detector, using some weak-field multipole formula $H$:
\begin{equation}
h(t)=H(\Vec{X},\Vec{R}).
\end{equation}
where $\Vec{C}$ contains the orbital constants describing the CO's (instantaneous) orbit, $\Vec{X}$ represents the CO's position vector with respect to the BH, and $\Vec{R}$ denotes the system's position vector with respect to the Solar System (i.e. $\theta_S,\phi_S$, and $D$).

\subsubsection{Analytic kludge}\label{subsec:AK}

In the AK model, the trajectory is constructed from rotating Keplerian ellipses. Radiation reaction is introduced in phase space, evolving the orbital constants of a Keplerian ellipse using PN equations. In configuration space, the ellipse's orientation is also evolved with PN equations to simulate relativistic precession. 
The waveform is then obtained with $n$-harmonics by using the Peter-Matthews method \cite{PhysRev.131.435} in the quadrupole approximation, which is described by Eqs. (7)--(10) in \cite{barack2004lisa}.

\begin{equation}\label{eq:mode_sum}
h_+=\sum_{n=1}^\infty{h^+_n},\quad h_\times=\sum_{n=1}^\infty{h^\times_n}
\end{equation}
with
\begin{equation}
h^+_n=(1+(\hat{\vec{R}}\cdot\hat{\vec{L}})^2)(b_n\sin{2\tilde{\gamma}}-a_n\cos{2\tilde{\gamma}})+(1-(\hat{\vec{R}}\cdot\hat{\vec{L}})^2)c_n,
\end{equation}
\begin{equation}
h^\times_n=2(\hat{\vec{R}}\cdot\hat{\vec{L}})(b_n\cos{2\tilde{\gamma}}+a_n\sin{2\tilde{\gamma}}),
\end{equation}
where the functions $(a_n,b_n,c_n)$ (Eq. (7) in \cite{barack2004lisa}) describe the changing mass quadrupole moment of a Keplerian orbit with mean anomaly $\Phi(t)$, eccentricity $e$ and orbital angular frequency $\nu$.

\subsubsection{Numerical kludge}\label{subsec:NK}

The NK model improves upon the AK model by providing a more accurate description of the orbital motion and radiation reaction in a Kerr spacetime. In the NK model, the inspiral trajectory is constructed from the Kerr geodesic, fully characterized by three constants of motion: the orbital energy $E$, the projection $L_z$ of the orbital angular momentum $\vec{L}$ onto $\vec{S}$, and the quadratic Carter constant $Q$. These constants are evolved using Teukolsky-fitted PN equations to account for radiation reaction. The Kerr geodesic equations are then numerically integrated along this inspiral trajectory to obtain the Boyer-Lindquist coordinates [$r(t),\theta(t),\phi(t)$]  of the inspiraling object as a function of time.

\begin{equation}\label{eq:NK_geodesic}
    \begin{aligned}
\Sigma\frac{dr}{d\tau}&=\pm\sqrt{V_r},\\
\Sigma\frac{d\theta}{d\tau}&=\pm\sqrt{V_\theta},\\
\Sigma\frac{d\phi}{d\tau}&=V_\phi,\\
\Sigma\frac{dt}{d\tau}&=V_t,
\end{aligned}
\end{equation}
where $\tau$ is the proper time and $\Sigma=r^2+a^2\cos^2\theta$. The potentials $V_{r,\theta,\phi,t}$ are functions of the constants $(E,L_z,Q)$ and the coordinates $(r,\theta)$ (Eqs.~(2a)-(2d) in \cite{babak2007kludge}).
Finally, the waveform is derived from the inspiral trajectory using the quadrupole (or quadrupole-octupole) approximation via

\begin{equation}\label{eq:modes}
h_+=\frac{1}{2}h_{ij}H^+_{ij},\quad h_\times=\frac{1}{2}h_{ij}H^\times_{ij}
\end{equation}
with
\begin{equation}\label{eq:quadrupole}
h_{ij}=\frac{2}{|\vec{R}|}\left(P_{ik}P_{jl}-\frac{1}{2}P_{ij}P_{kl}\right)\ddot{I}_{kl},
\end{equation}
where $\ddot{I}_{ij}(t)$ is the second time derivative of the source's mass quadrupole moment $I_{ij}(t)$, $H^{+,\times}_{ij}$ is the polarisation tensor,  and  $P_{ij}$ is transverse projection tensor.

\subsubsection{Augmented analytical kludge}\label{subsec:AAK}
The AAK model combines the speed of the AK model and the accuracy of the NK model. It first generates a short segment of the inspiral trajectory using the NK model, which is then used to adjust the parameters of the AK model, ensuring that the AK waveform closely matches the more accurate NK results.
Finally, the waveform field is generated as in the AK model. The flowchart summary of the entire AAK algorithm is shown in Fig. 2 of \cite{chua2017augmented}.

The main idea of AAK model is to extend parameters of AK model beyond their physical meaning to match the frequencies of NK waveforms. Specifically, given the orbit evolution in the NK model, we can define the dimensionless fundamental frequencies $\omega_{r,\theta,\phi}$ as a function of $(M,a,p)$. 
These fundamental frequencies are related to the orbital frequencies as follows:
\begin{equation}\label{eq:map}
\begin{split}
\dot{\Phi}(\tilde{M},\tilde{a},\tilde{p})&=\omega_r(M,a,p),\\
\dot{\gamma}(\tilde{M},\tilde{a},\tilde{p})&=\omega_\theta(M,a,p)-\omega_r(M,a,p),\\
\dot{\alpha}(\tilde{M},\tilde{a},\tilde{p})&=\omega_\phi(M,a,p)-\omega_\theta(M,a,p).
\end{split}
\end{equation}
The left hand side is given by the AK orbital equations.
By solving these equations, we obtain the unphysical set $(\tilde{M},\tilde{a},\tilde{p})$, which is defined as the root closest to the physical set $(M,a,p)$ with a Euclidean metric on parameter space \cite{chua2017augmented}. Substituting $(\tilde{M},\tilde{a},\tilde{p})$ for $(M,a,p)$ in the AK model provides a correction of its frequencies along the inspiral trajectory. 
Next, the waveform can be generated using the AK framework with the improved orbital motion. To reduce computational costs, the mapping is performed on a small section, and the correction along the local trajectory is extrapolated to the global inspiral using fitted polynomials. Further details can be found in \cite{chua2017augmented}.

\subsection{Bumpy-kludge waveforms}\label{sec:bumpy}
\begin{figure*}
    \centering
    \includegraphics[scale=0.35]{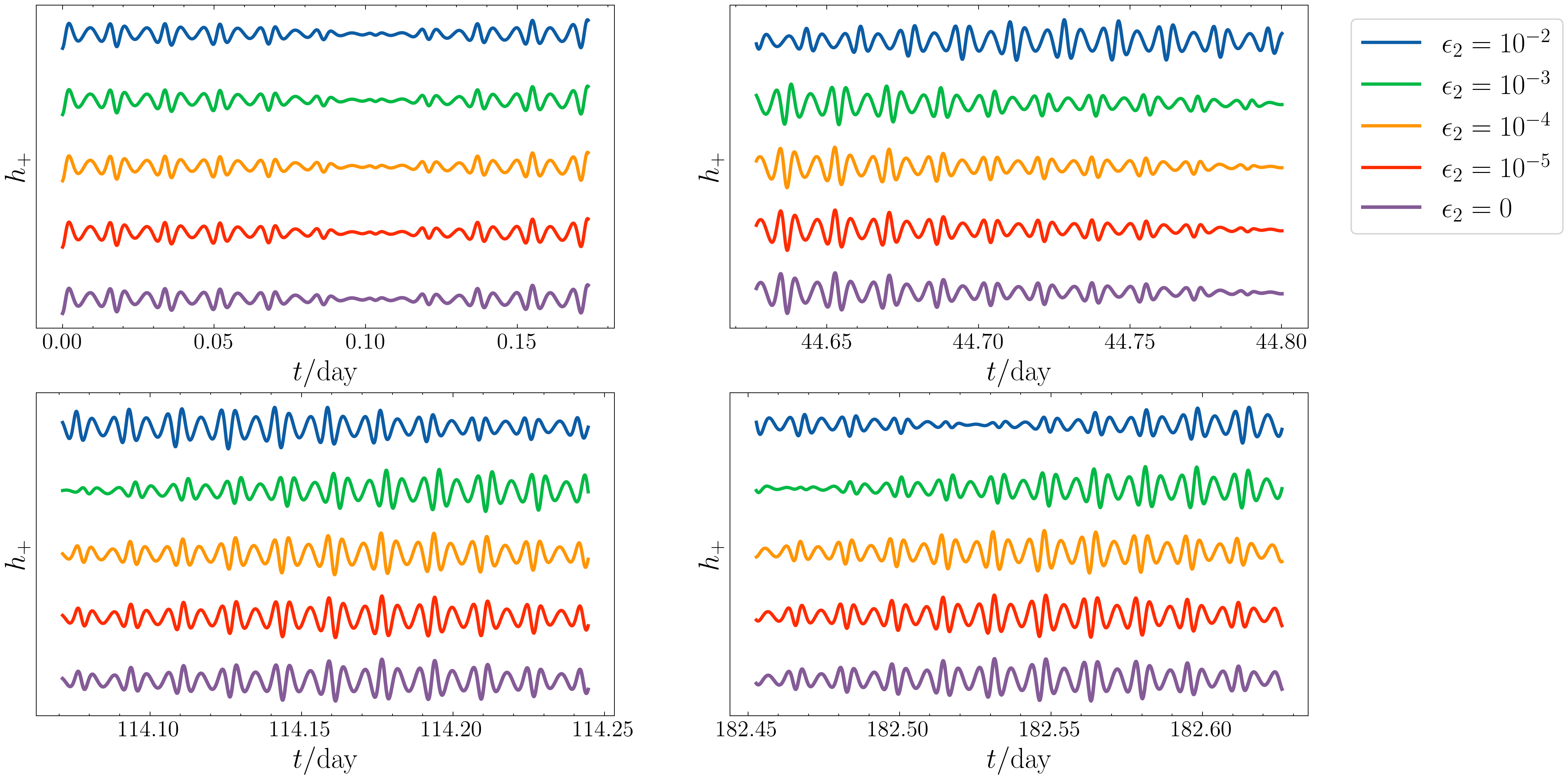}
    \caption{Comparision of bumpy-AAK waveforms with common source parameters and different deformation parameters $\epsilon_2$.
    The blue, green, orange, red, and purple lines represent $\epsilon_2=10^{-2},10^{-3},10^{-4},10^{-5}$, and $\epsilon_2=0$, respectively.
    Note that $\epsilon_2=0$ denotes the case of AAK waveforms. We fix the EMRI parameters to  $M=10^6 M_{\odot}$, $\mu=10 M_{\odot}$, $a=0.7$, $p_0=10 M$, $e_0=0.25$, $\iota_0=0.78$, $\theta_S=0.78$, $\phi_S=1.34$, $\theta_K=0.39$, $\phi_K=0$, $\Tilde{\gamma}_0=0$,$\Phi_0=0$, $\alpha_0=0$ at the distance of 1 Gpc. }
    \label{fig:hplus}
\end{figure*}
We decompose the metric tensor to describe the background spacetime of a supermassive BH as \cite{VYS11}:
\begin{equation}
    g_{\mu \nu}=g^{K}_{\mu \nu}+\epsilon h_{\mu \nu},
\end{equation}
where $g^{K}_{\mu \nu}$ is the traditional Kerr part and $\epsilon \ll 1$ is the ``deformation'' book-keeping parameter for the metric deformation $ h_{\mu \nu} $. We note that the background spacetime reduces to the ``normal'' Kerr black hole metric when  $\epsilon \rightarrow 0$.
In Boyer--Lindquist coordinates, the components of the Kerr metric for a BH with mass $M$ and dimensionless Kerr spin parameter $a$ are given by 
\begin{equation}\label{eq:g_munv}
    \begin{aligned}
g^\mathrm{K}_{tt}&=-\left(1-\frac{2Mr}{\rho^2}\right),\qquad g^\mathrm{K}_{t\phi}=-\frac{2M^2ar\sin^2{\theta}}{\rho^2},\\
g^\mathrm{K}_{\theta\theta}&=\rho^2, \qquad g^\mathrm{K}_{rr}=\frac{\rho^2}{\Delta} ,\qquad  g^\mathrm{K}_{\phi\phi}=\frac{\Sigma}{\rho^2}\sin^2{\theta}.
\end{aligned}
\end{equation}
where $\rho^2:=r^2+M^2 a^2\cos^2{\theta}$, $\Delta:=r^2-2Mr+M^2 a^2$, and $\Sigma:=(r^2+M^2a^2)^2-M^2a^2 \Delta \sin^2{\theta}$.

The only nonzero components of $ h_{\mu \nu} $ are $h_{tt}$, $h_{t\phi}$, $h_{rr}$, and $h_{\phi \phi}$, 
 which depend on the black-hole parameters $(M, a)$ and three arbitrary radial functions $\gamma_i (i=1, 3, 4)$ (Eq.~(56) in \cite{VYS11}). These deformed parametrizations $\gamma_i (i=1, 3, 4)$ can also be mapped to known alternative theory BH metrics \cite{VYS11}, such as the dynamical Chern-Simons (CS) gravity one \cite{cs2009} and the dynamical quadratic one \cite{modifyGR_nospin}.
 Gair and Yunes \cite{GY11} simplify the metric perturbations $h_{\mu \nu}$ prescription in \cite{VYS11} by considering expansions in $ M/r \ll 1$:

\begin{equation}
    h_{\mu \nu}=\sum_n h_{\mu \nu,n}(\frac{M}{r})^n,  2 \leq  n\leq 5
    \label{eq:h_munu}
\end{equation}
with
\begin{equation}
	\gamma_i=\left\{
	\begin{aligned}
		\sum^{\infty}_{n=0} \gamma_{i,n} (\frac{M}{r})^n,i=1,4\\
		\frac{1}{r} \sum^{\infty}_{n=0} \gamma_{i,n} (\frac{M}{r})^n, i=3
	\end{aligned}
	\right.
 \label{eq:gamma_i}
\end{equation}
where $\gamma_{i,n}$ are dimensionless constants.
The first few non-zero terms of Eq.~(\ref{eq:h_munu}) are given by:

\begin{widetext}
\allowdisplaybreaks[1]
\ba
h_{tt,2} &=  \gamma_{1,2} + 2 \gamma_{4,2} - 2 a \gamma_{3,1} \sin^{2}{\theta}\,,
\qquad
h_{tt,3} = \gamma_{1,3} - 8 \gamma_{4,2} - 2 \gamma_{1,2} + 2 \gamma_{4,3} + 8 a \gamma_{3,1} \sin^{2}{\theta}\,,
\nonumber \\
h_{tt,4} &= -8 \gamma_{4,3} - 2 \gamma_{1,3} + 2 \gamma_{4,4} + 8 \gamma_{4,2} + \gamma_{1,4} - 8 a  \gamma_{3,1} \sin^{2}{\theta} + a^{2} \left( \gamma_{1,2} + 2 \gamma_{4,2} \right) \sin^{2}{\theta} + 2 a^{3} \gamma_{3,1} \cos^{2}{\theta} \sin^{2}{\theta}\,,
\label{met-pert-eq1}
\ea
%\nonumber \\ 
\ba
h_{rr,2} &= - \gamma_{1,2}\,,
\qquad
h_{rr,3} = - \gamma_{1,3} - 2 \gamma_{1,2}\,,
\qquad
h_{rr,4} = - \gamma_{1,4} - 2 \gamma_{1,3} - 4 \gamma_{1,2} + (1/2) \gamma_{1,2} a^{2} \left(1 - \cos{2 \theta} \right)\,,
\ea
%\nonumber \\ 
\ba
h_{t \phi,2} &= - M \sin^{2}{\theta} \left[ \gamma_{3,3} + a \left( \gamma_{1,2} + \gamma_{4,2} \right) + a^{2} \gamma_{3,1} \right]\,,
\nonumber \\
h_{t \phi,3} &= -8 M a^{2} \gamma_{3,1} \sin^{4}{\theta} + M \sin^{2}{\theta} \left[  \left(2 \gamma_{3,3} - \gamma_{3,4} \right) + a \left( 6 \gamma_{4,2} - \gamma_{4,3} + 2 \gamma_{1,2} - \gamma_{1,3} \right) + 2 \gamma_{3,1} a^{2} \right]\,,
\nonumber \\
h_{t \phi,4} &= M \sin^{4}{\theta} \left[ a^{2} \left( 8 \gamma_{3,1} - \gamma_{3,3} \right) + a^{3}  \left(-\gamma_{1,3} - \gamma_{4,2} \right) - a^{4} \gamma_{3,1} \right] + \sin^{2}{\theta} \left[ \left(2 \gamma_{3,4} - \gamma_{3,5}\right) 
\right. 
\nonumber \\
&+ \left.
a \left( - \gamma_{4,4} - 8 \gamma_{4,2} + 6 \gamma_{4,3} - \gamma_{1,4} + 2 \gamma_{1,3} \right) - a^{2} \gamma_{3,3} \right]\,
\nonumber \\
\ea
%\nonumber \\ 
\ba
h_{\phi \phi,0} &= 2 M^{2} a \gamma_{3,1} \sin^{4}{\theta} \,,
\qquad
h_{\phi \phi,1} = 0\,,
\qquad
h_{\phi \phi,2} = M^{2} \sin^{4}{\theta} \left[ 2 a \gamma_{3,3} + a^{2} \gamma_{1,2} + a^{3} \gamma_{3,1} \left(4- 2 \cos^{2}{\theta} \right)\right]\,,
\nonumber \\
h_{\phi \phi,3} &=  8 M^{2} a^{3} \gamma_{3,1} \sin^{6}{\theta} + M^{2} \sin^{4}{\theta} \left[ a  \left( - 4 \gamma_{3,3} + 2 \gamma_{3,4} \right) + a^{2} \left(-2 \gamma_{1,2} - 4 \gamma_{4,2} + \gamma_{1,3} \right) - 4 a^{3} \gamma_{3,1} \right]\,,
\label{eq:hrr_decompose}
\ea
\end{widetext}
which implies that the full metric has two event horizons (the outer horizon $r_{+}$ and the inner one $r_{-}$) the same as a Kerr BH  determined by $\Delta=0$ \cite{GY11}. However, other quantities, such as he ergosphere and the innermost-stable circular orbit, will be different in deformed space-time relative to their Kerr values \cite{GY11}.

The perturbation  depends on how many terms in $M/r$ are kept relative to the leading-order Kerr metric. For example, up to $\mathcal{O} [(M/r)^2]$, the metric deformation is given by the 4 constants $\mathcal{B}_2= \{\gamma_{1,2},\gamma_{3,1},\gamma_{3,3},\gamma_{4,2}\}$. 
Up to $\mathcal{O} [(M/r)^5]$, it is given by the 13 constants $\mathcal{B}_2 \cup \mathcal{B}_3 \cup \mathcal{B}_4 \cup \mathcal{B}_5$, where $\mathcal{B}_n= \{\gamma_{1,n},\gamma_{3,n+1},\gamma_{4,n}\}, 3\leq n\leq 5$.
If the inclination angle of a geodesic orbit is approximated as constant, a metric deformation turns out to be fully specified by a set of three coefficients $\mathcal{B}_n := \{\gamma_{1,n},\gamma_{4,n},\gamma_{3,n+1}\}$, $2\leq n\leq 5$, which is regarded as a $\mathcal{B}_n$ bump.  
For the rest of this paper, when referring to the $\mathcal{B}_n$ limit we will mean that all the constants $\gamma_{m,n} = 0$ except for those quantities in the set $\mathcal{B}_n$.

Under $\mathcal{B}_2$ assumptions and setting $\gamma_{3,1}=0$, the expressions for the leading-order corrections to the the kludge model are ($\gamma_{3,n+1}$ is at sub-leading order): 

\begin{equation}\label{eq:delta_f_dot}
    \begin{aligned}
      M\;  \delta \dot{e}_{{\cal{B}}_{2}} =&-\frac{16}{5}\eta\frac{\left(2\pi M\nu\right)^{10/3}}{\left(1-e^{2}\right)^{7/2}}   \left( \frac{93}{4}e +\frac{67}{4}e^{3} + \frac{1}{4}e^{5} \right)   \epsilon_2,  \\
     2\pi M^{2} \;\delta \dot{\nu}_{{\cal{B}}_{2}} =& \frac{16}{5}\eta\frac{\left( 2\pi M\nu \right)^{13/3}}{\left( 1-e^{2} \right)^{9/2}}  \left( 18+78e^{2}+\frac{99}{4}e^{4} \right)  \epsilon_2,  \\
     M\;\delta \dot{\gamma}_{{\cal{B}}_{2}} =& \frac{\left( 2\pi M\nu \right)^{5/3}}{2\left(1-e^{2}\right)}\epsilon_2, \\
     M\;\delta \dot{\alpha}_{{\cal{B}}_{2}} = &-\frac{a\left( 2\pi M \nu \right)^{2}}{\left( 1-e^{2} \right)^{3/2}} \epsilon_2,
    \end{aligned}
\end{equation}
where $\eta =\mu/M$ is the mass ratio, and $\epsilon_2=\epsilon (\gamma_{1,2}+2\gamma_{4,2})$ is the deformation parameter, which determines the magnitudes of the $\mathcal{B}_2$ bumps.
These expressions are taken from Eqs.~(327)--(330) in \cite{GY11} with the restriction $\gamma_{3,1}=0$. 

Based on these corrections, we obtain the bumpy-kuldge waveforms (i.e.  bumpy-NK, bumpy-AK, and bumpy-AAK).
This involves (i) adding the corrections Eq.~(\ref{eq:delta_f_dot}) to the corresponding evolution equations in the kludge waveform model \textbf{EMRI Kludge Suite} \cite{aak_2015,chua2017augmented,barack2004lisa,babak2007kludge} \footnote{which can be found at the following URL: https://github.com/alvincjk/EMRI\_Kludge\_Suite} and (ii) extending the set of model parameters to
\begin{align}
    \Theta_{\rm bumpy-kludge}&=\Theta_{\rm kludge} \cup \Theta_{\rm bumpy},\\
     \Theta_{\rm bumpy}&=\epsilon_2.
\end{align}

Fig.~\ref{fig:hplus} shows the dephasing of bumpy-AAK (bAAK) waveforms over time relative to the AAK waveforms (where $\epsilon_2=0$).
Initially, these bAAK waveforms are in phase with AAK waveforms, but they will gradually dephase as time goes by.
Waveforms with smaller $\epsilon_2$ values correspond to smaller metric deformations, and thus dephase more slowly.

\subsection{Time Delay Interferometry (TDI)}\label{sec:TDI}

For the ground-based laser interferometers, the laser frequency noise can be cancelled very precisely by keeping the same arm length up to the picometer level \cite{Armstrong_1999,Wang_2023}. 
For the space-borne detectors, due to the coupling with unequal armlengths caused by the relative motion of the satellites, time delay interferometry (TDI) \cite{PhysRevD.62.042002,YANG2023106900} is needed to achieve targeting  sensitivity. As a data post-processing technique, TDI works by time-delaying and linearly combing the interferometric data streams and thus construct virtual equal armlength interferometers. 

\begin{figure}[h]
    \centering
    \includegraphics[scale=0.45]{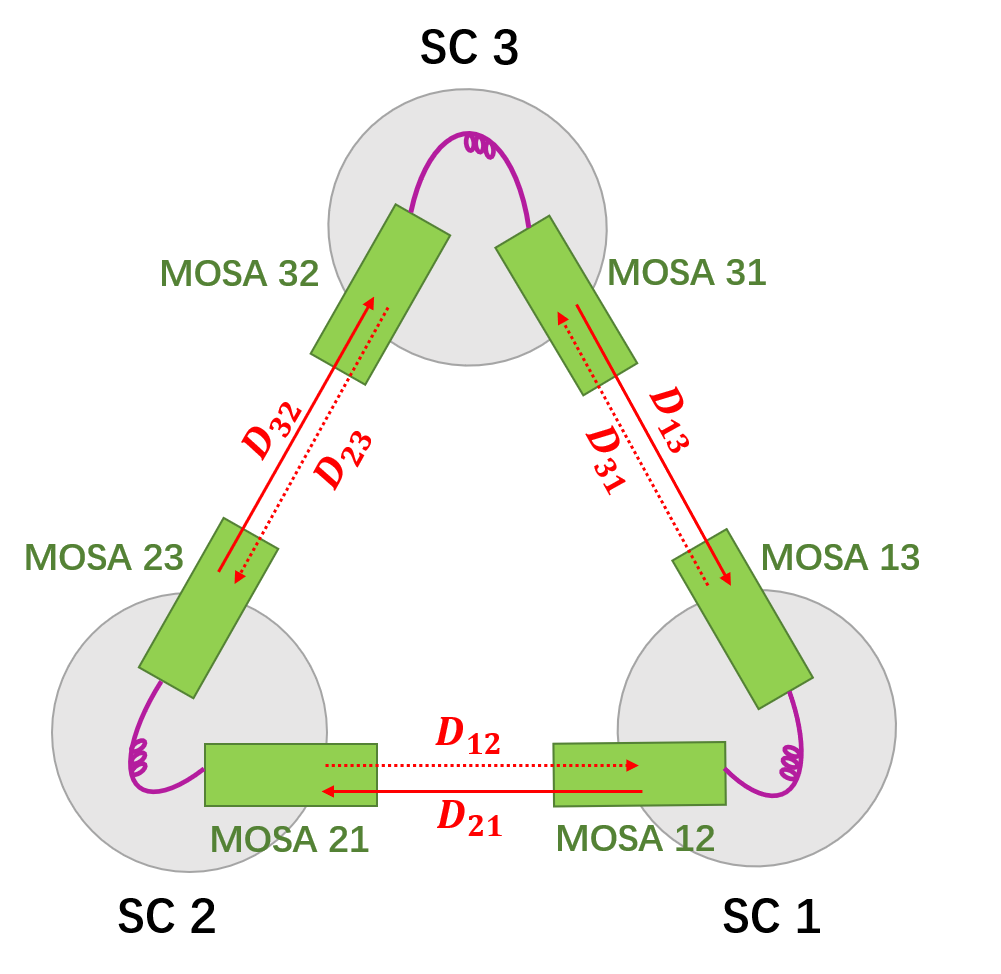}
    \caption{ Indexing conventions in TDI}
    \label{fig:tdi}
\end{figure}

\begin{figure*}
    \centering
     \includegraphics[scale=0.55]{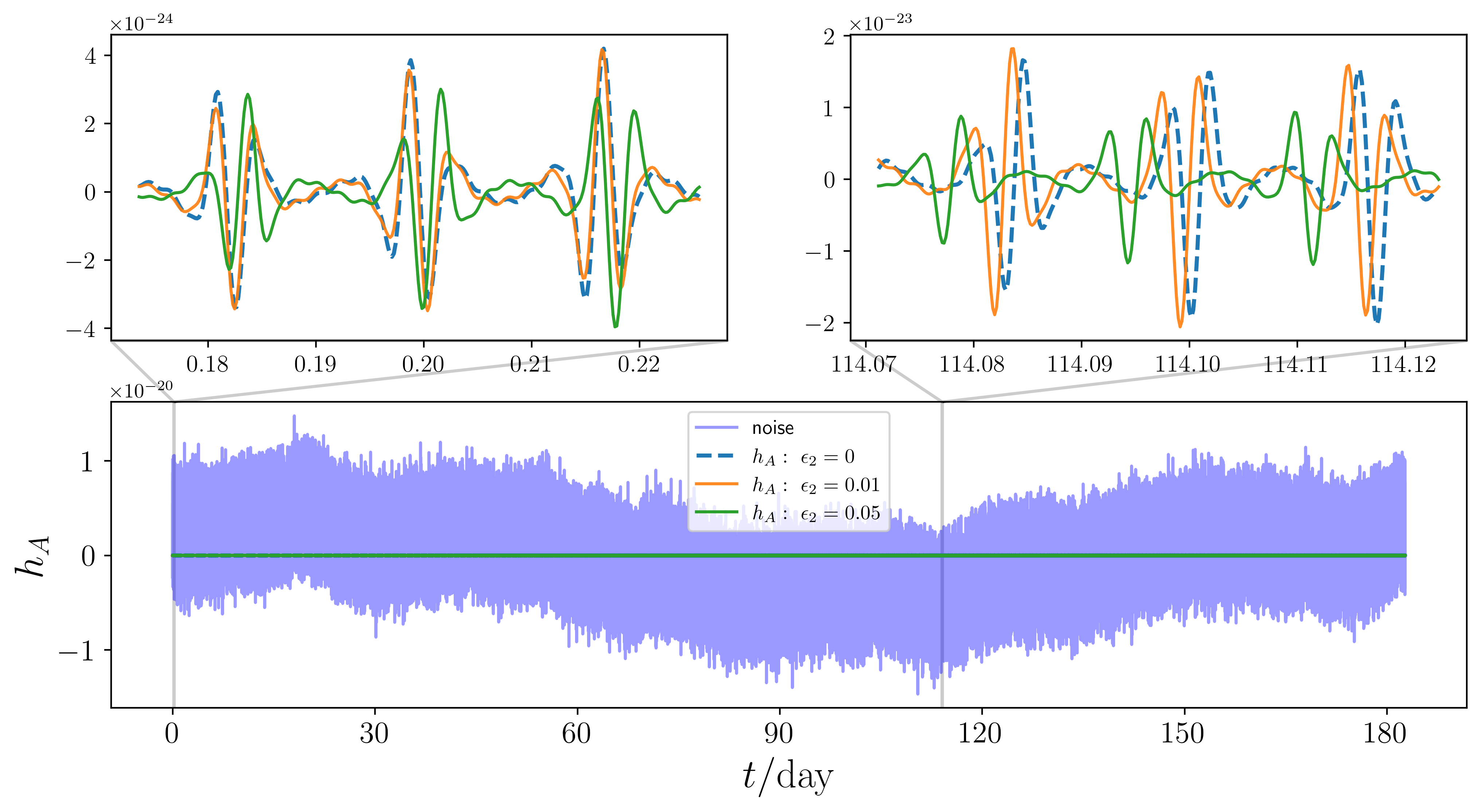} 
     \caption{The bumpy-AAK waveforms of EMRI systems in the  TDI-A channel versus the deformation parameters $\epsilon_2$. The light blue lines denote the noise of TDI-A Channel and the deep blue, orange, and green lines represent waveforms with $\epsilon_2=0$, $\epsilon_2=0.01$, and $\epsilon_2=0.05$, respectively. Note that $\epsilon_2=0$ denotes the case of Kerr black hole. %The purple lines are the detector noise. 
     We plot the half a year evolving waveforms with $M=10^6 M_{\odot}$, $\mu=10 M_{\odot}$, $a=0.7, p_0=10 M$, $e_0=0.25$, $\iota_0=0.78$, $\gamma=0$, $\psi=0$, $\alpha=0$, $\theta_S=0.78$, $\phi_S=1.34$, $\theta_K=0.39$, and $\phi_K=0$ at the distance of 1 Gpc.  
 }
     \label{fig:waveform_tdi}
\end{figure*}

Fig.~\ref{fig:tdi} shows the standard indexing conventions in TDI.
Spacecrafts (SCs) are labeled from 1 to 3 clockwise when looking down on the $z$-axis. Moveable optical sub-assemblies (MOSAs) are indexed with two numbers $ij$, where $i$ is the index of the local spacecraft, and $j$ is the index of the distant spacecraft the light is received from  \cite{PhysRevD.106.103001}.

Assuming that the constellation is static, the expression for the first generation Michelson combination $X_1$ is given by \cite{PhysRevD.69.082001}
\begin{equation}
    \begin{aligned}
      X_1 &= 
        y_{13} + \delay{13} y_{31} + \delay{131} y_{12} + \delay{1312} y_{21}
        \\
        &\quad - [y_{12} + \delay{12} y_{21} + \delay{121} y_{13} + \delay{1213} y_{31}]  
    \end{aligned}
\end{equation}
with 
\begin{equation}
    \delay{ij} x(t) = x(t - L_{ij}(t))
\end{equation}
\begin{equation}
    \delay{i_1, i_2, \dots, i_n} x(t) = x\ (t - \sum_{k=1}^{n-1}{L_{i_k i_{k+1}}(t)} ).
\end{equation}
where $y_{ji}$ is the relative frequency shift experienced by light as it travels along link $ji$, $\delay{}$ is the time-delay operator, $L_{ij}(t)$ is the propagation time along link $ij$ at reception time $t$, and $x(t)$ is the arbitrary data stream. The other two Michelson combinations $Y_1$ and $Z_1$ are obtaibed by the same method.

The first generation Michelson combinations only eliminated the laser phase noise of static constellation or rigid constellation rotating at a constant speed. In practice, the relative motion between spacecraft can not be ignored, so the second generation TDI combinations $X_2$ are developed:  

\begin{equation}
    \begin{aligned}
     X_2 &= X_1 + \delay{13121} y_{12} + \delay{131212} y_{21} + \delay{1312121} y_{13}
        \\
        &\quad+ \delay{13121213} y_{31} - [\delay{12131} y_{13} + \delay{121313} y_{31}
        \\
        &\quad+ \delay{1213131} y_{12} + \delay{12131312} y_{21}]
    \end{aligned}
\end{equation}
$Y_2$ and $Z_2$ can be obtained via cyclic permutation of the SC indices.
 These Michelson combinations have correlated noise properties.
An uncorrelated set of TDI variables, $ (A, E, T)$, can be obtained from linear combinations of $(X,Y,Z)$ given by 

\begin{equation}
\begin{aligned}
    A =& \frac{1}{\sqrt{2}}\left(Z-X\right) , \\
    E =& \frac{1}{\sqrt{6}}\left(X-2Y+Z\right) , \\
    T =&\frac{1}{\sqrt{3}}\left(X+Y+Z\right).
\end{aligned}
\end{equation}
These channels  $A,E,T$ are only exactly orthogonal (or uncorrelated in noise properties) in the equal-armlength limit. We notice that the $A$ and $E$ channels are sensitive to GWs and regarded as scientific channels, while the $T$ channel is insensitive to GW, which is used to characterize the instrument noise.

In general, the observed signal, $d(t)$, can be described as the sum of the GW signal $h(t)$ and the detector noise $n(t)$
\begin{equation}
    d_i(t)=h_i(t)+n_i(t),i=\{A,E,T\}
\end{equation}
where the value of $i$ represents different the second generation TDI channels $(A,E,T)$. In this work, we use the TDI package \textbf{fastlisaresponse} \cite{chua_2020_3981654,PhysRevD.106.103001} to generate GW waveforms in the  TDI-A channel. We assume the noise is stationary Gaussian, and the power spectral density (PSD) of noise is taken to be the sky averaged one for Taiji \cite{Ren_2023}.

As an example, Fig.~\ref{fig:waveform_tdi} shows three waveforms in the TDI-A channel for different deformation parameters $\epsilon_2$. The bottom row displays the waveforms over half a year, while the top row zooms in on two segments. 
 In the early stages (top left), the disparity between two EMRI waveforms with minor deformation parameters ($\epsilon_2 = 0, 0.01$) is minimal. However, as time progresses, differences due to deformation accumulate. In the later stages (top right), significant differences emerge between the waveforms with different deformation parameters. Notably, the waveform corresponding to $\epsilon_2 = 0.05$ (green line) deviates more substantially from the other two waveforms (blue and orange lines). This demonstrates that as both the value of 
$\epsilon_2$ and time increase, the waveform gradually deviates from the Kerr black hole waveform.  

\section{\label{bayes}parameter estimation and model selection} 

Given the ppE waveforms (bumpy-kludge waveforms) mentioned above, which include the deformed parameter, we can extract parameter information from GW signals to test GR. For stationary Gaussian noise, the likelihood is given by
%define h(\theta)
\begin{equation}
\mathcal{L}(\theta) \propto \exp[-\frac{1}{2} \langle d-h(\theta)|d-h(\theta) \rangle]\, ,
\end{equation}
The noise-weighted inner product is  defined as 
\begin{equation}
    \langle a|b\rangle \equiv 2\int_0^{\infty} df\, \frac{a^*(f)b(f)+a(f) b^*(f)}{S_n(f)}
\end{equation}
where $S_n(f)$ is the PSD of the noise $n$, which is taken to be the sky-averaged PSD for Taiji throughout the paper. The SNR of a given source is equivalent to $\sqrt{\langle d|h\rangle}$.
We utilize the package \textbf{dynesty} \cite{2020MNRAS4933132S,sergey8408702}, employing dynamic nested sampling \cite{nested_sampling}, to infer the Bayesian posterior distribution of parameters and evidence. This approach is particularly well-suited for complex, multimodal distributions \cite{Wang_2023}.

It is known that using various models to estimate parameters will get different results.
So, which model is statistically preferred by the data and by how much? 
In this work, we take advantages of the Bayesian inference to do model selection. 
In the Bayesian inference, the relative probability of two or more alternative hypotheses given observed data $d$ is described by the
odds ratio $\mathcal{O}$.  If GR is the general relativity hypothesis (model $H_\mathrm{GR}$ with parameters $\nu$) and AG is the hypothesis corresponding to some alternative theory of gravity (model $H_\mathrm{AG}$ with parameters $\theta$), the odds ratio in favor of AG is given by \cite{Thrane_Talbot_2019}:

\begin{equation}
   O^{\mathrm{AG}}_{\mathrm{GR}} \equiv \frac{ \mathcal{Z}_\mathrm{AG} } { \mathcal{Z}_\mathrm{GR} } \frac{\pi_\mathrm{AG}}{\pi_\mathrm{GR}}=\mathcal{B}^{\mathrm{AG}}_{\mathrm{GR}} \frac{\pi_\mathrm{AG}}{\pi_\mathrm{GR}}
		\label{eq:oddsratio} 
\end{equation}
with
\begin{equation}
    \mathcal{Z}_\mathrm{AG} =\int d\theta \mathcal{L}(d|\theta,H_\mathrm{AG} )\pi (\theta)
\end{equation}
\begin{equation}
    \mathcal{Z}_\mathrm{GR} =\int d\nu \mathcal{L}(d|\nu,H_\mathrm{GR} )\pi (\nu)
\end{equation}
\begin{equation}
      {\mathcal{B}^{\mathrm{AG}}_{\mathrm{GR}}}=\frac{\mathcal{Z}_\mathrm{AG}}{\mathcal{Z}_\mathrm{GR}} 
\end{equation}
\begin{equation}
     \log_e {\mathcal{B}^{\mathrm{AG}}_{\mathrm{GR}}}=\log_e {\mathcal{Z}_\mathrm{AG}}-\log_e 
 {\mathcal{Z}_\mathrm{GR}}
\end{equation}
where $\mathcal{L}$ is the likelihood, $\mathcal{Z}$ is marginalised likelihood (or evidence) of observing the data given the specific model $H$, and $\mathcal{B}^{\mathrm{AG}}_{\mathrm{GR}}=\frac{ \mathcal{Z}_\mathrm{AG} } { Z_\mathrm{GR} }$ is the Bayes factor.
We typically set the prior odds ratio to unity, and so the odds ratio is equal to the Bayes factor. If $\log_e {\mathcal{B}^{\mathrm{AG}}_{\mathrm{GR}}}> 0$, then the evidence is in favor of the first hypothesis AG. If $\log_e {\mathcal{B}^{\mathrm{AG}}_{\mathrm{GR}}}< 0$, then the evidence is in favor of the second hypothesis GR.
The threshold $|\log_e \mathcal{B} |= 8$  is often treated as the level of “strong evidence” in favor of one hypothesis over another \cite{Thrane_Talbot_2019}.

\section{fundamental bias}\label{sec:fundamental bais}
If templates are used based solely on GR models, while the corresponding events may be detected, any unexpected information the signals may contain about the nature of gravity will be filtered out. Therefore, tests of GR cannot rely on waveform families that assume GR is correct. Instead, such tests should employ more generic waveforms that allow for deviations from GR \cite{VYS11}. These deviations, arising from the assumption based on GR or AG, are considered fundamental biases that question the validity of the Einstein field equations themselves.

In this section, based on the bumpy-kludge waveforms mentioned in Sec.~\ref{sec:bumpy}, we investigate the influence of systematic error caused by fundamental bias. Note that the waveform templates used in this section are all based on the AAK case but with different theoretical assumptions (GR or AG). 
The strategies are outlined below:

(1) Given a GR signal and a GR template, how well can the latter extract the former? How well can intrinsic parameters be estimated?

(2) Given a GR signal and an AG template, how well can the latter extract the former? How well can intrinsic and deformation parameters be estimated? 

(3) Given a non-GR signal and an AG template, how well can the latter extract the former? How well can intrinsic and deformation parameters be estimated?

(4) Given a non-GR signal and a GR template, how much fundamental bias-induced systematic error is generated in the estimation of parameters? Can the signal even be extracted?

\subsection{Given a GR signal} \label{sec:fundamental_GR}

\begin{figure*}[] 
\centering
     \begin{subfigure}[]{
     \includegraphics[scale=0.48]{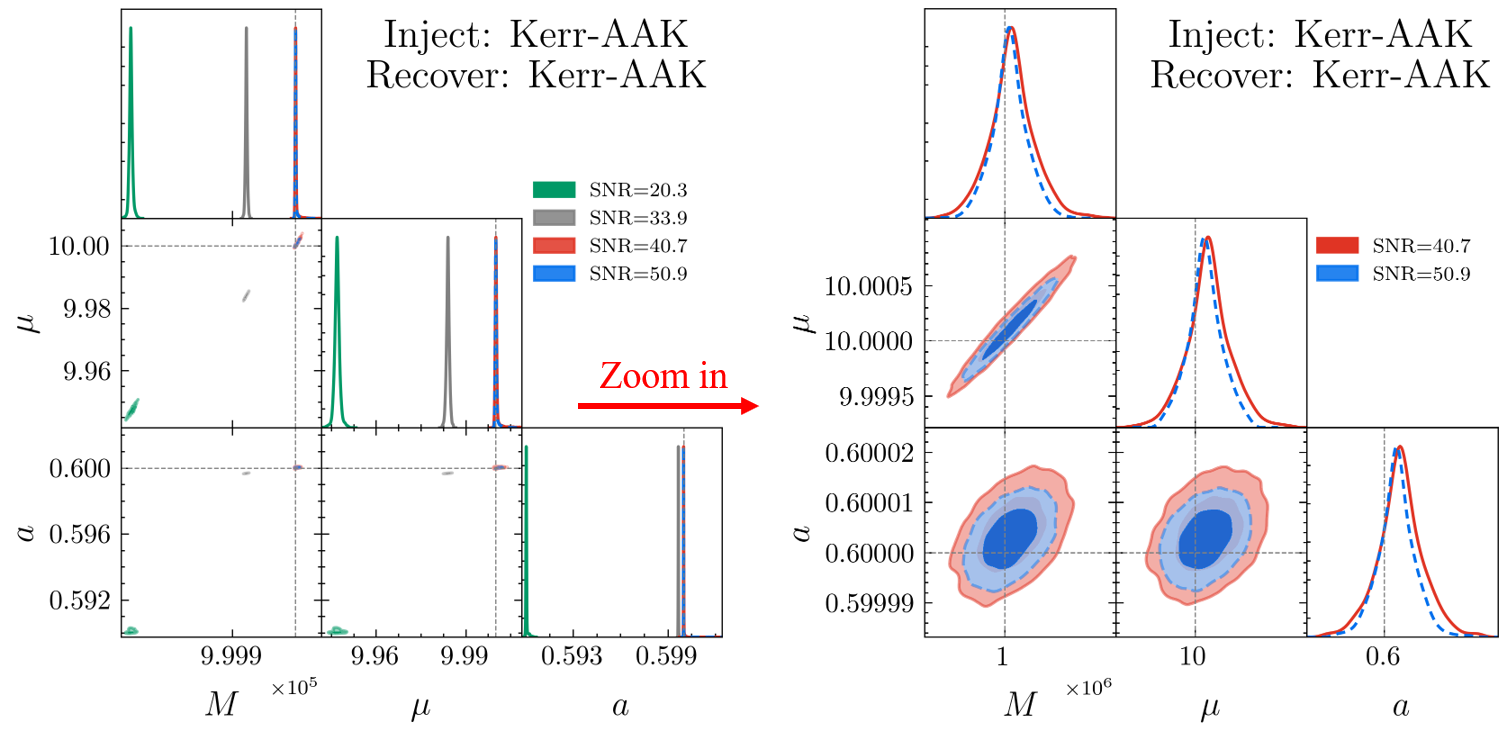}  
}
	\end{subfigure} 
    \begin{subfigure}[]{
	\includegraphics[scale=0.4]{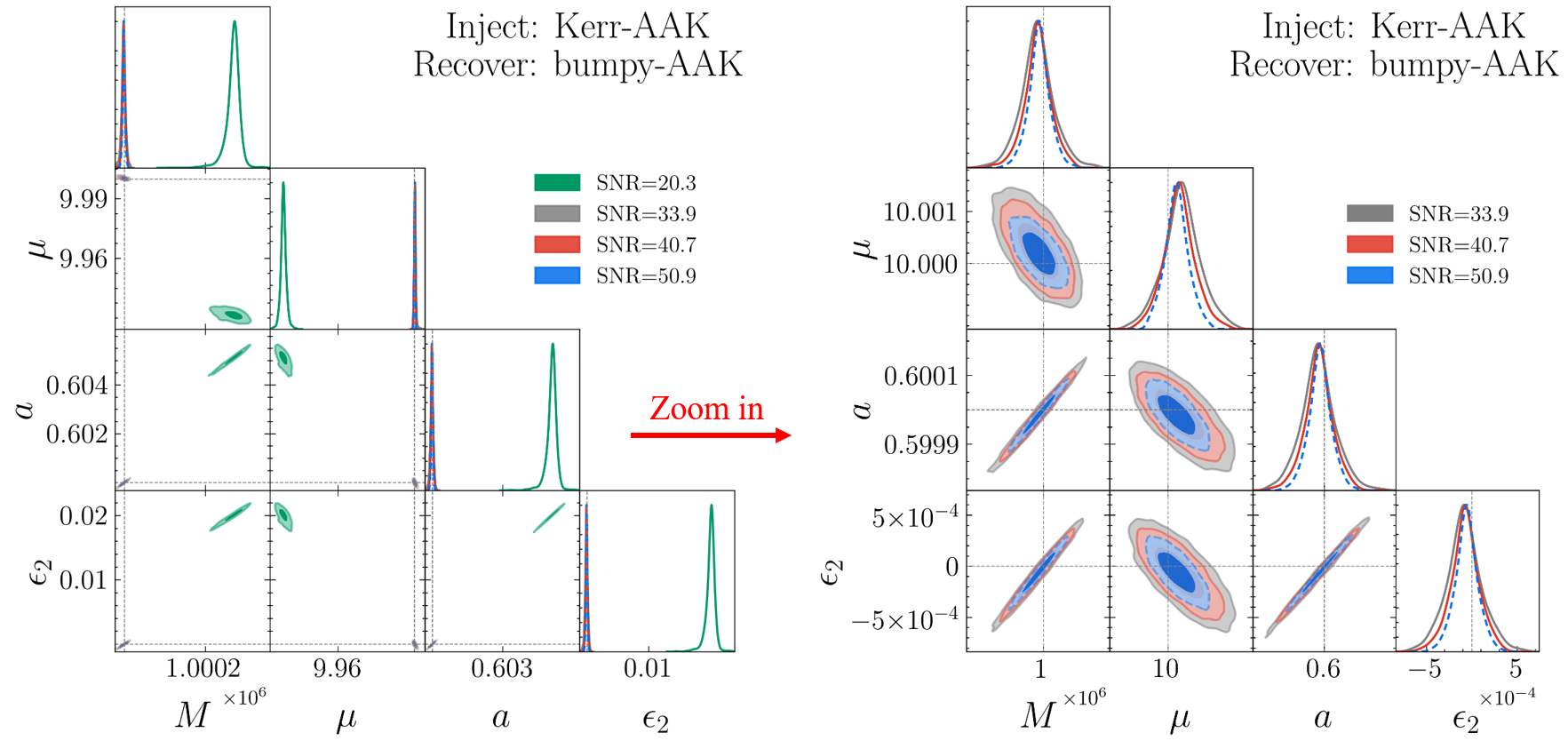}
    }
	\end{subfigure}  
    \caption{
    The posterior distribution of the parameters generated by injecting Kerr-EMRI signals with AAK models and recovering using Kerr/bumpy-AAK models. The green, grey, red, and blue parameter posteriors are generated for injected signals with SNR = 20.3, 33.9, 40.7, and 50.9, respectively. 
    The injected values are represented by the vertical black lines.
    The posterior distributions show the $1\sigma$ and $2\sigma$ contours. 
    Notice that the right part is a partially enlarged part of the left one.
    (a) inject: AAK; recover: AAK. (b) inject: AAK; recover: bumpy-AAK.
    }
    \label{fig:kk}
\end{figure*}

Given a GR signal, we recover parameters using both GR and non-GR templates. Instead of conducting the initial search for the alerted events, we assume that the source has already been identified in the data stream. 
 The injected parameters are $M=10^6 M_{\odot}$, $\mu=10 M_{\odot}$, $a=0.6, p_0=10 M$, $e_0=0.25$, $\iota_0=0.78$, $\theta_S=0.78$, $\phi_S=0.6$, $\theta_K=0.39$, $\phi_K=0$,  $\Tilde{\gamma}_0=0$, $\Phi_0=0$, and $\alpha_0=0$. The luminosity distance is set to $D=\{0.4,0.5,0.6,1.0\} \rm{Gpc}$, which is corresponding to $\rm{SNR}=\{50.9,40.7,33.9,20.3\}$. The parameters tested in this study are only $\{M, \mu, a, \epsilon_2 \}$.  The priors are uniformly distributed across all data sets as follows: the mass of SMBH $M \in U[0.99,1.01)\times 10^6 M_\odot$,  the mass of CO $\mu \in U[9.8,10.2) M_\odot$, the dimensionless spin of SMBH $a \in U[0.58,0.62)$,  and the deformetion parameter $\epsilon_2 \in U[-0.05,0.05)$ for non-GR model. 
These parameters and priors are selected to focus on studying systematic errors in EMRI waveforms when testing GR, rather than on parameter estimation techniques \cite{universe10040171,Bnbumps,PhysRevD.86.104050}. The full parameter estimation with a larger parameter space will be addressed in future work.

Fig. \ref{fig:kk} illustrates the posterior distribution of GR-EMRI signal parameters recovered using AAK templates (a) or bumpy-AAK templates (b). 
The right side of each panel shows an enlarged detail of the left side.
From the left side of Fig. \ref{fig:kk}, it is evident that as SNR increases, the posterior distributions for the parameters converge more closely to the true values, represented by the vertical black lines. This indicates that these models achieve enhanced accuracy in parameter estimation with increasing SNR.
Furthermore,  the right side of Fig. \ref{fig:kk}  illustrates that higher SNR leads to more peaked  and narrower posterior distributions, reflecting greater precision in estimating the underlying parameters. Therefore, the SNR has a significant impact on parameter estimation, with higher SNRs leading to more precise and concentrated posterior distributions.
Specifically, for golden EMRIs  ($\rm{SNR}> 50$), both models perform well, with all recovered parameters falling within the $2\sigma \    (95\%)$ credible interval, and the relative accuracy of the parameter estimation for each simulation can improve significantly, reaching up to $\sim  10^{-6}$ for $M$, $\sim 10^{-5}$ for $\mu$, and $\sim 10^{-4}$ for $a$  (see Tab. \ref{tab:kerr}).

In contrast, when the SNR is 33.9, the Kerr-AAK model (a) does not perform better than the bumpy-AAK model (b). The bumpy-AAK model benefits from a larger parameter space, which provides greater flexibility in parameter estimation. This increased flexibility allows the bumpy-AAK model to more effectively adapt to the complexities of the data, resulting in more accurate parameter recovery. However, it is important to note that the computational demands are also higher when using the bumpy-AAK model due to its larger parameter space.

\begin{table*} %[h]
\renewcommand{\arraystretch}{1.3}
\begin{ruledtabular} 
\begin{tabular}{lllllclll}
        $\rm{SNR}_{inj}$ & model& $M ( M_\odot)\   \red{(10^6)} $ & $\mu (M_\odot)\  \red{(10)}$ &  $a \  \red{(0.6)}$  & $\epsilon_2 \ \red{(0)}$ & SNR & $\log{\mathcal{Z}}$   & $\log{\mathcal{B}^{\rm bumpy}_{\rm Kerr}}$ \\
        \hline
        $20.3$ & bumpy&  $1000271^{+23.0}_{-29.8}$ & $9.9314^{+0.0025}_{-0.0026}$ & $0.6051^{+0.0003}_{-0.0004}$  & $0.0201^{+0.0010}_{-0.0012}$ & $ 20.2 $   &$-527205.755 \pm  0.133$  & $29.848 \pm 0.229$\\
           & kerr&  $999699^{+6.1}_{-4.3}$ & $9.983807^{+0.00219}_{-0.00151}$ & $0.58444^{+0.00006}_{-0.00004}$  & -- & $ 20.3 $  &$-527235.603 \pm 0.096$ & \\
         
        $33.9$ & bumpy&  $999998^{+8.2}_{-8.2}$ & $10.0003^{+0.0008}_{-0.0007}$ & $0.6000^{+0.0001}_{-0.0001}$  & $0.0001^{+0.0004}_{-0.0004}$& $ 33.9$  & $-527000.980 \pm  0.161$ &$372.128 \pm 0.268$ \\
         &  kerr& $999922^{+2.6}_{-2.6}$ & $9.98406^{+0.00089}_{-0.00090}$ & $0.59967^{+0.00002}_{-0.00002}$  & --& $ 33.9 $  &$-527373.108 \pm 0.107$  & \\
         
         $40.7$& bumpy& $999999^{+6.9}_{-6.8}$  & $ 10.0002^{+0.0006}_{-0.0006}$   & $ 0.6000^{+0.0001}_{-0.0001}$      &   $0.0001 ^{+0.0003}_{-0.0003}$ & $ 40.7$    &  $-526994.137 \pm  0.139$  & $-5.102 \pm 0.264$    \\
           & kerr& $1000000 ^{+1.2}_{-1.2}$   &  $ 10.00013^{+0.00040}_{-0.00040}$   &  $ 0.6000^{+0.0001}_{-0.0001}$     &   --  & $40.7 $   &   $-526989.035 \pm 0.125$  & \\
         
         $50.9$& bumpy& $999999^{+5.6}_{-5.6}$  & $10.0002^{+0.0005}_{-0.0005}$    & $0.6000^{+0.0001}_{-0.0001}$     & $0.0001^{+0.0002}_{-0.0002}$ & $ 50.9 $      &   $-526997.937 \pm  0.152 $  & $-8.238 \pm  0.280$     \\
           & kerr&  $1000000^{+1.0}_{-1.0}$ & $10.00010^{+0.00033}_{-0.00032}$ & $0.60000^{+0.00001}_{-0.00001}$  & -- & $ 50.9 $  & $-526989.698 \pm 0.128 $& \\
    \end{tabular}
    \caption{The $2\sigma (95\%)$  credible intervals for recovered parameters using bumpy-AAK models and Kerr-AAK models for Kerr-AAK EMRI signals. 
    The last column shows the logarithmic Bayes factor for the different signals. 
    The true values of injected signal parameters are marked in red color.}
    \label{tab:kerr}
\end{ruledtabular} 
\end{table*}

Fig. \ref{fig:kk} demonstrates that an GR-EMRI signal can be detected using both GR and non-GR templates. The question then arises: which model is statistically preferred by the data, and by how much? Tab. \ref{tab:kerr} presents the mean and $2\sigma$ standard deviation of the estimated parameters using both the bumpy-AAK and Kerr-AAK models. For model selection, we compare the evidence ($\log \mathcal{Z}$) of each case and calculate the logarithmic Bayes factor $\log_e {\mathcal{B}^{\mathrm{AG}}_{\mathrm{GR}}}$ in favor of AG model.
We find that when $\rm SNR < 40$, the value of $\log_e {\mathcal{B}^{\mathrm{AG}}_{\mathrm{GR}}}$ provides ``strong evidence" in favor of the bumpy-EMRI signal. This result could potentially lead to a misinterpretation, suggesting that a deviation from GR has been detected. However, as the SNR increases to around $ 40$ and higher, the confusion between AG and GR diminishes.

This confusion is caused by noise, which impacts the accuracy of parameter estimation, scaling roughly as  $1/\rm SNR$ \cite{PhysRevD.76.104018}.  This relationship indicates that in situations with low SNR, the higher noise levels lead to larger errors in the recovered parameters.
In addition, under low SNR conditions, the algorithm may converge to a local maximum, complicating the search for the global maximum likelihood solution. This reduces the accuracy of parameter estimation and may result in incorrect conclusions about testing GR. As SNR increases, the influence of noise diminishes, resulting in more precise and accurate parameter estimation, thereby eliminating the confusion between AG and GR.

\subsection{Given a non-GR signal}\label{sec:fundamental_nonGR}
\begin{figure*}[] 
\centering
     \begin{subfigure}[]{
     \includegraphics[scale=0.4]{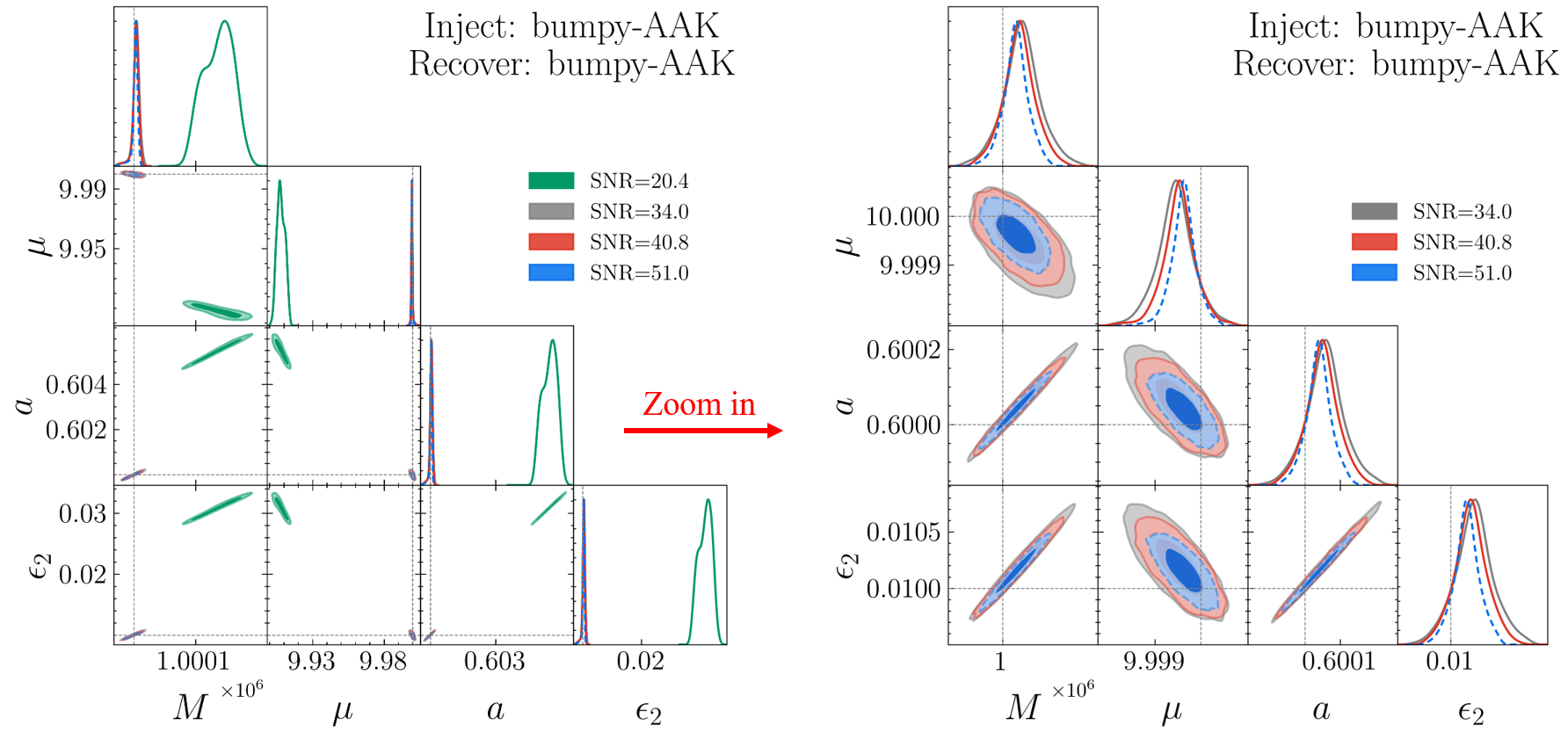}  
}
	\end{subfigure} 
    \begin{subfigure}[]{
	\includegraphics[scale=0.48]{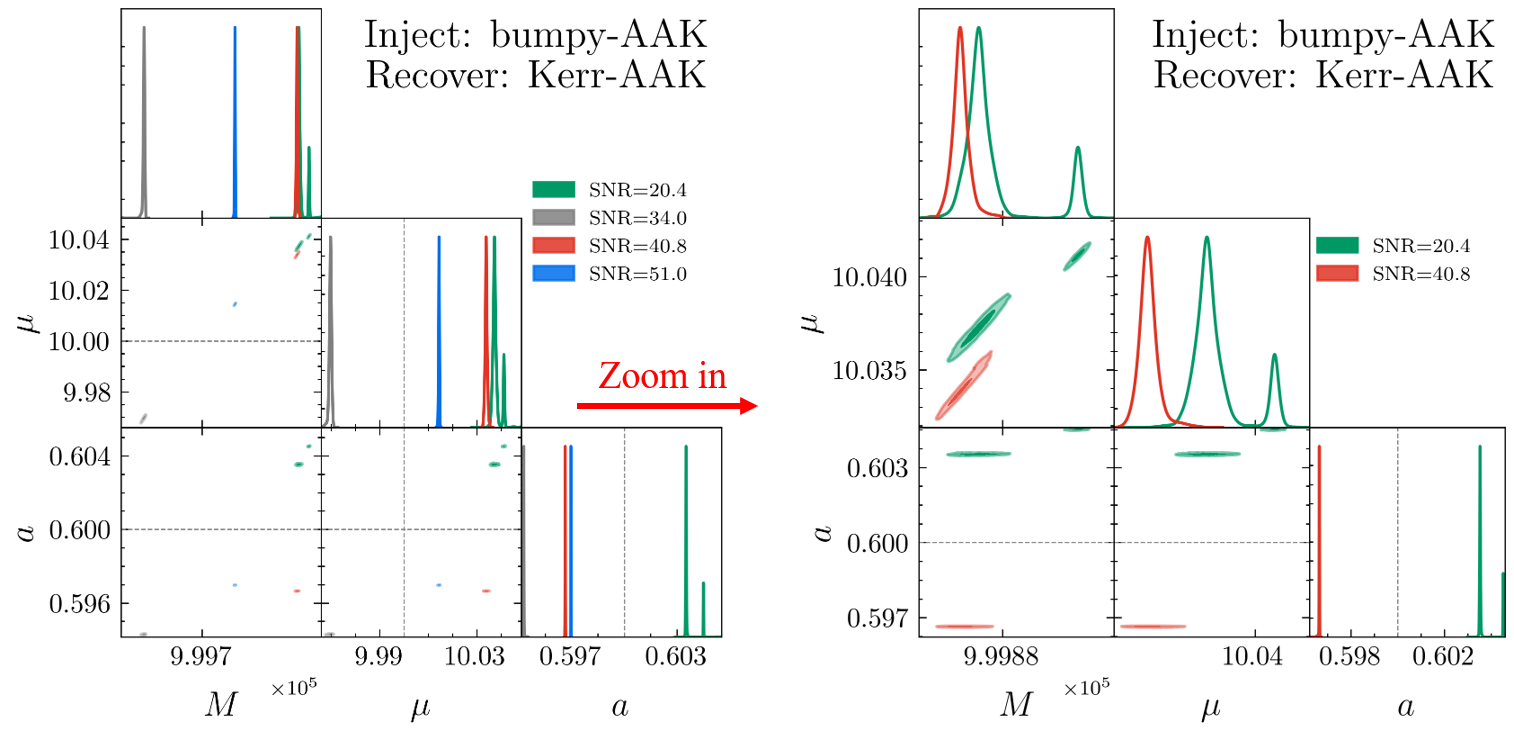}
    }
	\end{subfigure}  
    \caption{The posterior distribution on the parameters generated by injecting bumpy-EMRI signals with bumpy-AAK models and recovering using Kerr/bumpy-AAK models. The green grey, red, and blue parameter posteriors are generated for injected signals with SNR = 20.4, 34.0, 40.8, and 51.0, respectively. 
    The injected values are represented by the vertical black lines. 
    The posterior distributions show the $1\sigma$ and $2\sigma$ contours. 
    Notice that the right part is a partially enlarged part of the left one.
    (a) inject: bumpy-AAK; recover: bumpy-AAK. (b) inject: bumpy-AAK; recover: AAK.}
    \label{fig:bb}
\end{figure*}

Similar to Sec. \ref{sec:fundamental_GR}, given a non-GR signal, we also recover parameters using a GR template and AG template, respectively. 
 The injected parameters are $M=10^6 M_{\odot}$, $\mu=10 M_{\odot}$, $a=0.6, p_0=10 M$, $e_0=0.25$, $\iota_0=0.78$, $\theta_S=0.78$, $\phi_S=0.6$, $\theta_K=0.39$, $\phi_K=0$,  $\Tilde{\gamma}_0=0$, $\Phi_0=0$, $\alpha_0=0$ and $\epsilon_2=0.01$. We set the luminosity distance $D=\{0.4,0.5,0.6,1.0\} \rm{Gpc}$, which is corresponding to $\rm{SNR}=\{51.0,40.8,34.0,20.4\}$. The priors are the same as those in Sec.~\ref{sec:fundamental_GR}.

Fig. \ref{fig:bb} shows the posterior distribution on the parameters of bumpy-EMRI signals when recovering with bumpy-AAK templates (a) or AAK templates (b). 
From Fig. \ref{fig:bb} (a), we can see that for the EMRI system with $\rm{SNR}\lesssim 30$, there are significant deviations from the true parameters (vertical black lines) with respect to $2\sigma \    (95\%)$ uncertainty. However, as the SNR increases to $\gtrsim 30$, the estimates become more reliable and accurate, and the recovered parameters are all within the $2\sigma$ credible interval. 
Overall, it demonstrates that increasing the SNR improves the accuracy and precision of parameter estimation when using bumpy-AAK models to recover bumpy-EMRI signals.

We then extract parameters of bumpy-EMRI signals using Kerr-AAK models, depicted in Fig. \ref{fig:bb} (b). As the SNR increases, the posterior distributions of the parameters become more concentrated, indicating improved precision in parameter estimation. However, unlike in Fig. \ref{fig:bb} (a), even at higher SNRs, there are still noticeable deviations from the true parameters. This highlights potential limitations in using GR models for recovering non-GR signals.

While GR templates, such as Kerr-AAK models, can detect non-GR signals, the parameter estimates will exhibit biases. These biases stem from a \textit{fundamental bias} caused by the discrepancy between the GR template and the true non-GR nature of the signal. If templates are used based solely on GR models, any unexpected information the signals may contain about the nature of gravity will be filtered out, potentially leading to the misidentification of a bumpy-EMRI signal as a Kerr one. Therefore, it is crucial to develop models that effectively accommodate both GR and AG theories.

\begin{table*}[]
\renewcommand{\arraystretch}{1.3}
    \centering
    \begin{ruledtabular} 
    \begin{tabular}{lllllclll}
        $\rm{SNR}_{inj}$ & model &$M ( M_\odot)\   \red{(10^6)} $ & $\mu (M_\odot)\  \red{(10)}$ &  $a \  \red{(0.6)}$  & $\epsilon_2 \ \red{(0.01)}$  & SNR & $\log{\mathcal{Z}}$  & $\log{\mathcal{B}^{\rm bumpy}_{\rm Kerr}}$ \\
        \hline
        $20.4$ & bumpy&  $1000136^{+40.5}_{-45.2}$ & $9.9084^{+0.0051}_{-0.0044}$ & $0.6055^{+0.0005}_{-0.0006}$  & $0.0308^{+0.0018}_{-0.0021}$ &$20.2$& $-527202.650 \pm   0.103$&$-0.24 \pm 0.238$\\
        &  kerr & $999876^{+3.7}_{-3.7}$  &  $10.03725^{+0.00121}_{-0.00119}$  & $0.60352^{+0.00003}_{-0.00003}$      & --  &$ 20.4$ &  $-527202.406 \pm  0.135$&       \\
        
        $34.0$ & bumpy&  $1000004^{+8.0}_{-7.8}$ & $9.9995^{+0.0007}_{-0.0007}$ & $0.6001^{+0.0001}_{-0.0001}$  & $0.0102^{+0.0003}_{-0.0003}$& $ 34.0 $ & $-526996.411 \pm  0.145 $&$723.86\pm 0.262$\\
        &kerr & $999595^{+2.4}_{-2.4}$  & $9.96980^{+0.00081}_{-0.00083}$   & $0.59428^{+0.00002}_{-0.00002}$     & --  & $ 33.9 $  &  $ -527720.273 \pm  0.117$&      \\
        
         $40.8$& bumpy& $1000004^{+6.6}_{-6.6}$  & $9.9995^{+0.0006}_{-0.0006}$   & $0.6000^{+0.0001}_{-0.0001}$     & $0.0102^{+0.0003}_{-0.0003}$ & $ 40.8 $    &    $-526996.065 \pm  0.149$& $851.57 \pm 0.267$    \\
           & kerr& $999872^{+2.5}_{-2.5}$  & $10.03382^{+0.00082}_{-0.00082}$   & $0.59664^{+0.0001}_{-0.0001}$     &   -- & $ 40.9 $   &    $-527847.639 \pm  0.118$& \\
           
         $51.0$& bumpy&  $1000003^{+5.2}_{-5.3}$   & $9.9996^{+0.0005}_{-0.0005}$   & $0.6000^{+0.0001}_{-0.0001}$       &  $0.0102^{+0.0002}_{-0.0002}$   & $ 51.0 $  & $-526994.700 \pm  0.140$&   $465.51 \pm 0.274$    \\
         & kerr &$999760^{+1.1}_{-1.1}$  & $10.01444^{+0.00036}_{-0.00036}$   & $0.59697^{+0.00001}_{-0.00001}$     & -- & $ 51.0$   &   $-527460.212 \pm  0.134$&     \\
    \end{tabular}
    \end{ruledtabular} 
    \caption{The $2\sigma (95\%)$  credible intervals for recovered parameters using bumpy-AAK models and Kerr-AAK models.  
    The last column shows the logarithmic Bayes factor for the different signals. 
    The true values of injected signal parameters are marked in red color.
    }
    \label{tab:bumpy}
\end{table*}

Table \ref{tab:bumpy} displays the mean and standard $2\sigma$ deviation of estimated parameters utilizing both bumpy-AAK and Kerr-AAK models. We also compare the evidence ($\log \mathcal{Z}$) of each case and derive the logarithmic Bayes factor $\log_e {\mathcal{B}^{\mathrm{AG}}_{\mathrm{GR}}}$ in favor of AG for model selection. 
We observe that when $\rm{SNR}=20.3$, distinguishing between AG and GR signals becomes challenging due to the low value of $|\log_e {\mathcal{B}^{\mathrm{AG}}_{\mathrm{GR}}}|$, leading to confusion between a bumpy-EMRI signal and a Kerr one. In other words, if the signal is a bumpy-EMRI signal, using a model based on the Kerr case may result in its misinterpretation as supporting GR. 
However, as the SNR increases to $\sim 30$  and higher, the confusion between AG and GR signals dissipates.

\section{modeling error}\label{sec:modeling error}
\begin{figure*}[] 
\centering
     \begin{subfigure}[]{
     \includegraphics[scale=0.48]{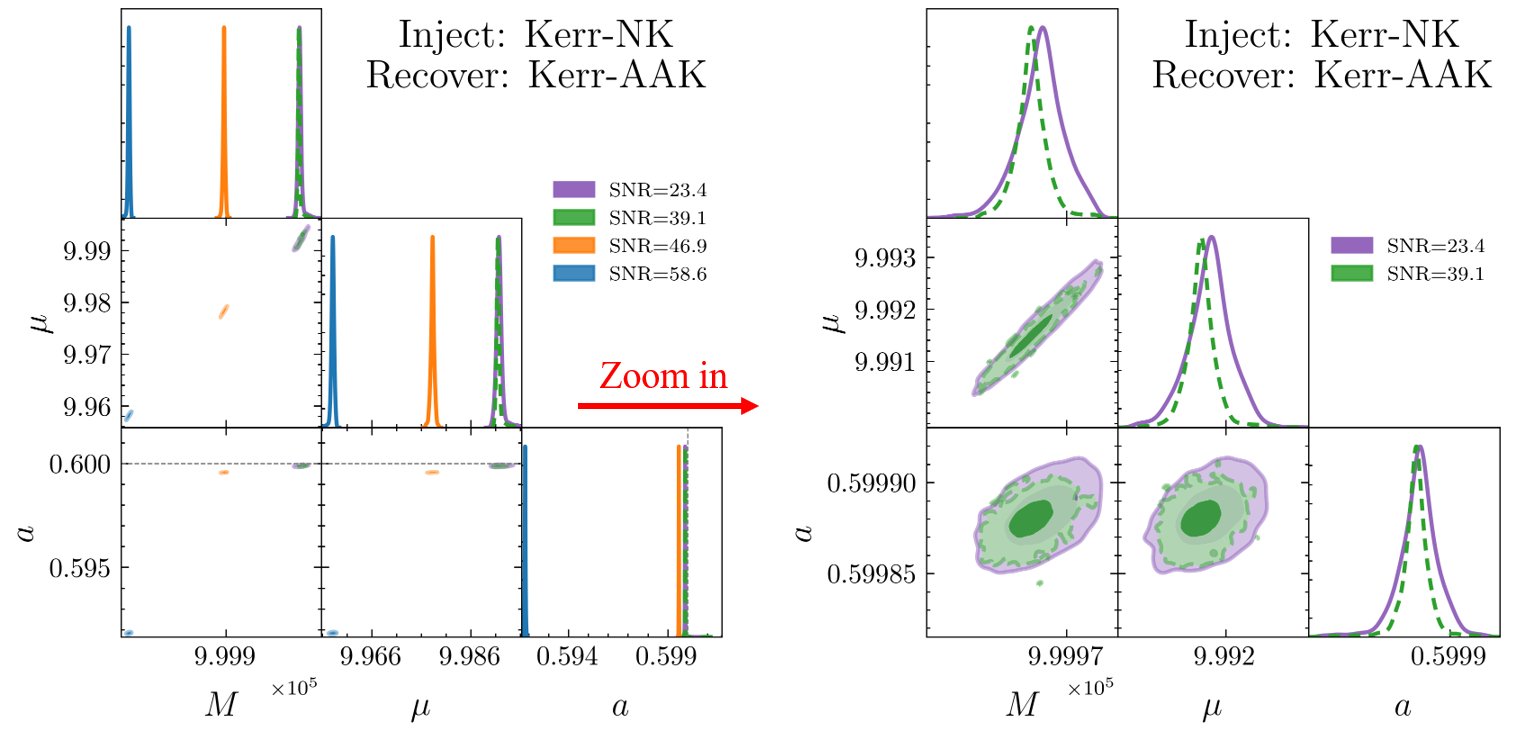}  
}
	\end{subfigure} 
    \begin{subfigure}[]{
	\includegraphics[scale=0.4]{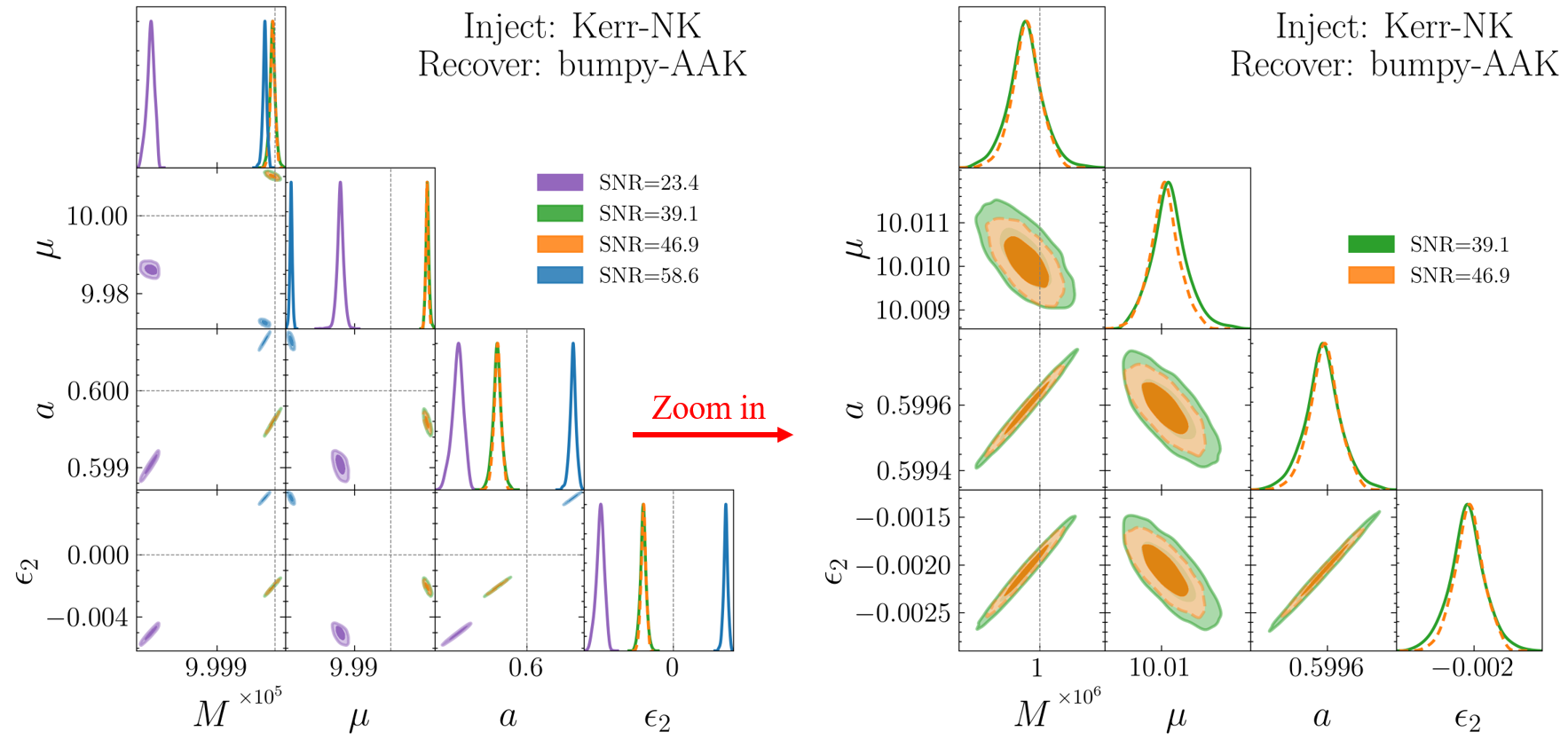}
    }
    \end{subfigure} 
\caption{The posterior distribution on the parameters generated by injecting kerr-EMRI signals with NK models and recovering using kerr-AAK models. The purple, green, orange, and blue parameter posteriors are generated for injected signals with SNR = 23.4, 39.1, 46.9, and 58.6, respectively. 
    The injected values are represented by the vertical black lines.
    The posterior distributions show the $1\sigma$ and $2\sigma$ contours. 
     Notice that the right part is a partially enlarged part of the left one.
    (a) inject: NK; recover: AAK. (b) inject: NK; recover: bumpy-AAK.}
    \label{fig:[NK-AAK]_kk}
\end{figure*}

As previously mentioned in Sec. \ref{sec:fundamental bais}, when the SNR is low, systematic errors due to fundamental biases may skew the results of the GR test. As the SNR increases, the confusion between AG and GR diminishes. However, modeling inaccuracies also become increasingly significant, especially for high-SNR sources.  For massive black hole mergers, these systematic errors from our best model waveforms could dominate over statistical errors by roughly an order of magnitude for the strongest sources \cite{template_error_2007,Hu_2023}. Testing GR using selected ``golden binaries" with high SNR is even more vulnerable to false deviations from GR.

In this section, we focus on parameter extraction errors arising from the use of less accurate template families for EMRI systems. 
 This issue can be broadly thought of as a \textit{modeling error}, where the preconception relates to physical assumptions to simplify the solutions considered or unverified assumptions about the accuracy of the solution used to model the given event. The strategies are outlined below:

(1) Given a GR-NK signal and a GR-AAK template,
how much modeling error-induced systematic error is generated in the estimation of parameters?  Can the signal even be extracted?

(2) Given a GR-NK signal and an AG-AAK template,  how much modeling error-induced systematic error is generated in the estimation of parameters? Can the signal even be extracted?

(3) Given an AG-NK signal and an AG-AAK template, how much modeling error-induced systematic error is generated in the estimation of parameters? Can the signal even be extracted?

(4) Given an AG-NK signal and a GR-AAK template, how much systematic error is generated in the estimation of parameters due to the combined effects of fundamental bias and modeling error? Can the signal even be extracted?

We inject the true reference signals using more accurate NK models and recover them with AAK models. Because the 1PA waveforms for generic orbit is still ongoing, here we only consider the 0PA waveforms, such as NK and AAK waveforms. However, it is crucial to repeat this analysis once the 1PA waveforms become available for  generic orbits.

\subsection{Given a GR signal }\label{sec:modeling error:GR}

Given a GR-EMRI signal modeled with NK model, we recover parameters using both a Kerr-AAK template and AAK template, respectively.  
The injected parameters are $M=10^6 M_{\odot}$, $\mu=10 M_{\odot}$, $a=0.6, p_0=10 M$, $e_0=0.25$, $\iota_0=0.78$, $\theta_S=0.78$, $\phi_S=0.6$, $\theta_K=0.39$, $\phi_K=0$,  $\Tilde{\gamma}_0=0$, $\Phi_0=0$, and $\alpha_0=0$. We set the luminosity distance $D=\{0.4,0.5,0.6,1.0\} \rm{Gpc}$, which refer to $\rm{SNR}= \{58.6, 46.9, 39.1,  23.4\}$, and the time of observation $T=0.5 \rm{yr}$. The priors are all  uniform for all data sets, which are: the mass of supermassive BH $M \in U[0.99,1.01)\times 10^6 M_\odot$,  the mass of CO $\mu \in U[9.8,10.2) M_\odot$, the dimensionless spin of supermassive BH $a \in U[0.58,0.62)$,  and the deformetion parameter $\epsilon_2 \in U[-0.05,0.05)$ for non-GR model.

Fig. \ref{fig:[NK-AAK]_kk} illustrates the outcomes obtained by injecting Kerr-EMRI signals using NK models and then reconstructing them using the AAK models (a) or bumpy-AAK templates (b), respectively. 
The right side of each panel shows an enlarged detail of the left side.
We find that due to modeling errors in the waveform templates, all the posterior distributions of the parameters deviate from the injected values beyond the $2\sigma (95\%)$  uncertainty. Initially, the accuracy of parameter estimation, as indicated by the peaks of the posterior distributions, improves with increasing SNR. However, when $\rm SNR > 50$, these peaks deviate even further from the injected values. This increased deviation is due to systematic errors caused by inaccuracies in our model waveforms, which are small at low SNR but become noticeable for the strongest sources, such as golden EMRIs.  Therefore, for very high SNR signals, enhancing the accuracy of the model waveforms is crucial to maintaining accuracy in parameter estimation.

Tab. \ref{tab:kerr_NK} presents the mean and $2\sigma$ standard deviation of estimated parameters using both bumpy-AAK and Kerr-AAK models. For model selection, we compare the evidence ($\log \mathcal{Z}$) of each case and derive the logarithmic Bayes factor $\log_e {\mathcal{B}^{\mathrm{AG}}_{\mathrm{GR}}}$ in favor of AG.  When recovering EMRI signals modeled with NK templates using AAK models, we observe an approximately $15\%$ reduction in the resulting event's SNR. This SNR decrease is caused by the mismatch between the NK signals and the AAK model.
Tab. \ref{tab:kerr_NK} also indicates that for systems with a high SNR, the calculated Bayesian factors tend to favor the bumpy signal hypothesis. This preference arises because the bumpy template includes an additional deformation parameter $\epsilon_2$ that is absent in the Kerr template. In such high SNR scenarios, the model bias between the NK and AAK signals may be misconstrued as the deformation parameter difference between the Kerr and bumpy signals. This misinterpretation may lead to an overestimation of the support for the bumpy signal hypothesis, potentially skewing the analysis and leading to incorrect conclusions about the underlying EMRI signal. The additional degree of freedom ($\epsilon_2$) in the bumpy template, while potentially offering a more flexible fit, also introduces another risk of confounding fundamental bias with genuine physical parameters. As a result, careful consideration is needed when interpreting these Bayesian factors, particularly in high-SNR systems where even small biases can have significant impacts on the results.

\subsection{Given a non-GR signal }\label{sec:modeling error:AG}
\begin{figure*}[] 
\centering
     \begin{subfigure}[]{
     \includegraphics[scale=0.4]{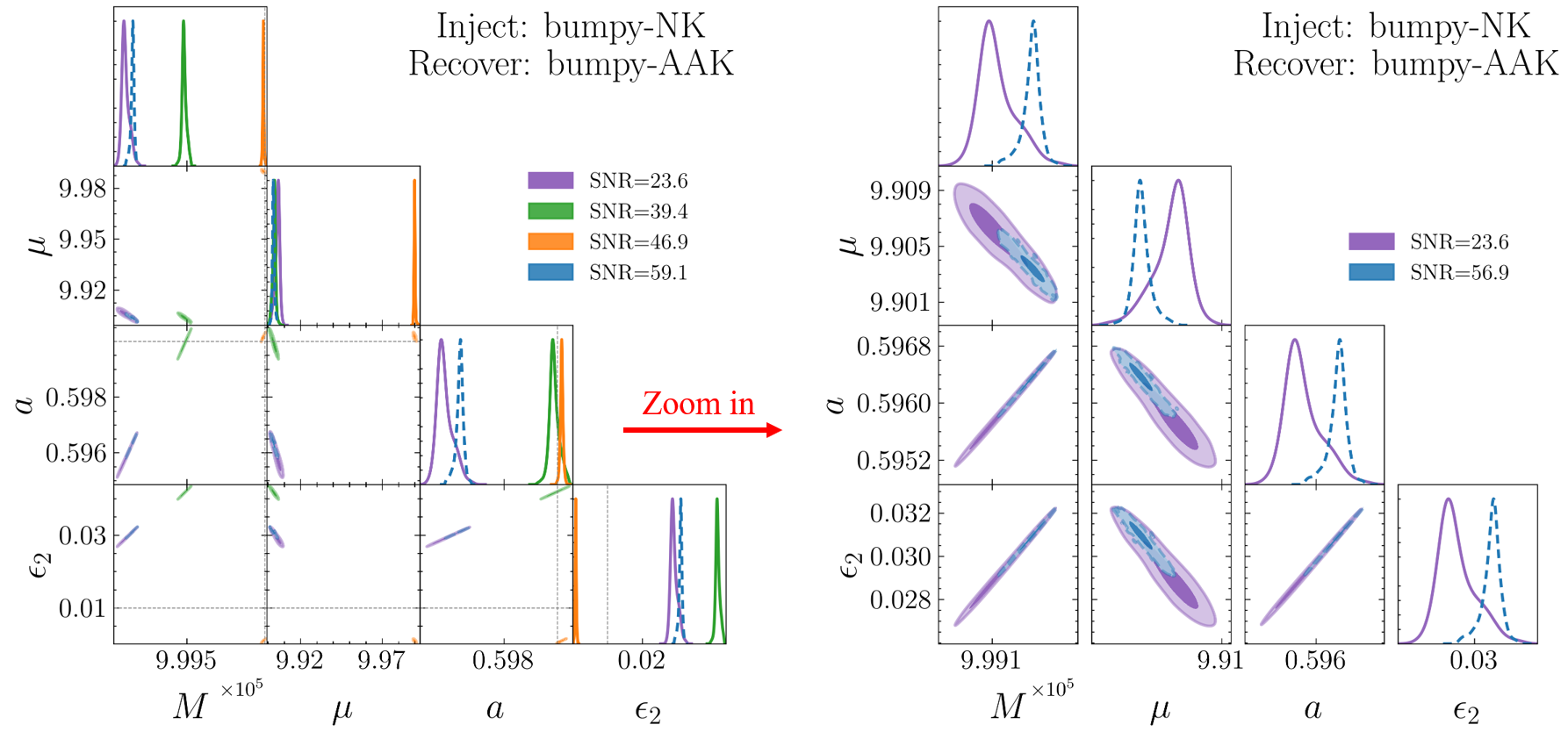}  
}
	\end{subfigure} 
    \begin{subfigure}[]{
	\includegraphics[scale=0.48]{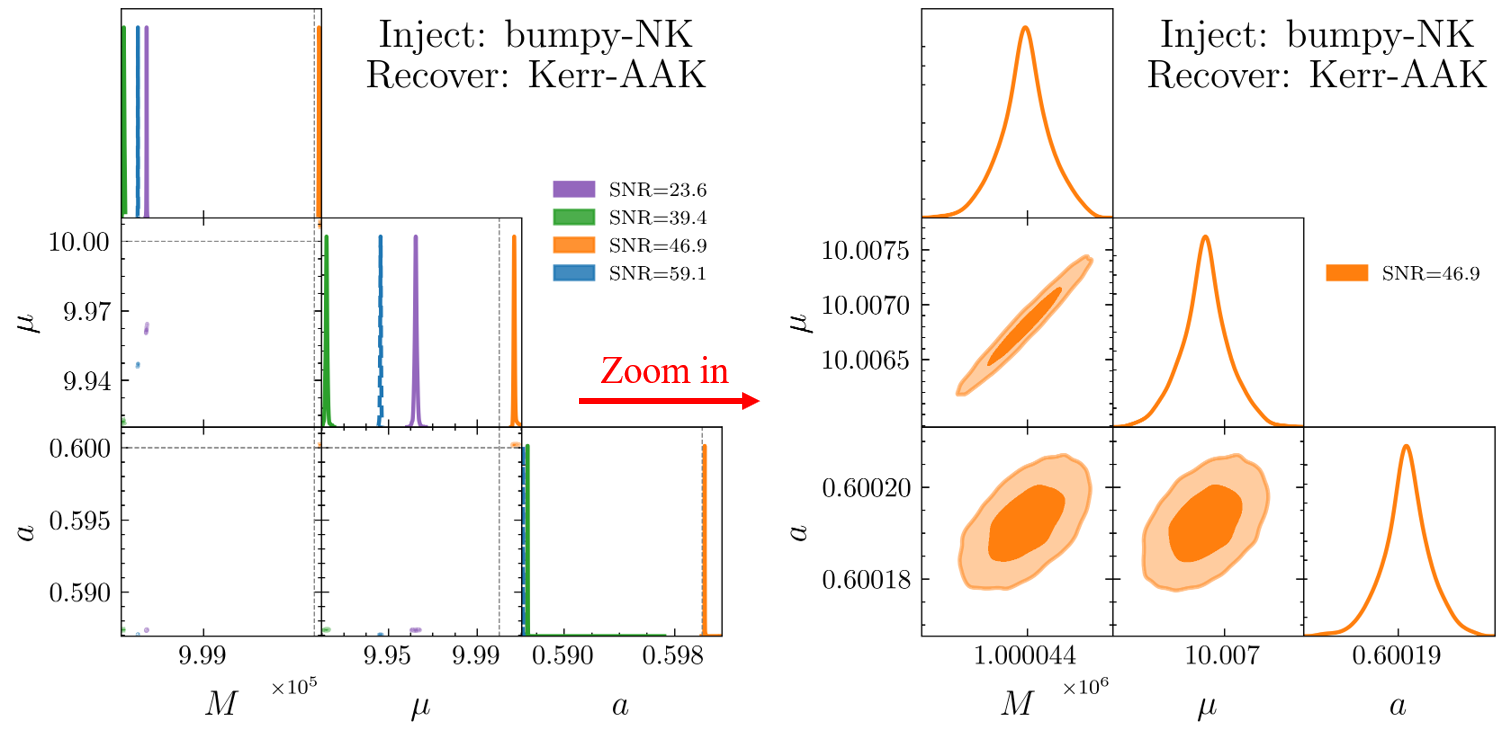}
    }
    \end{subfigure} 
\caption{The posterior distribution on the parameters generated by injecting bumpy-EMRI signals with NK models and recovering using bumpy-AAK models. The purple, green, orange, and blue parameter posteriors are generated for injected signals with SNR = 23.6, 39.4, 46.9, and 59.1, respectively. 
The injected values are represented by the vertical black lines. 
    The posterior distributions show the $1\sigma$ and $2\sigma$ contours. 
    Notice that the right part is a partially enlarged part of the left one.
    (a) inject: bumpy-NK; recover: bumpy-AAK. (b) inject: bumpy-NK; recover: AAK.}
    \label{fig:[NK-AAK]_bb}
\end{figure*}

Given an AG-EMRI signal modeled with bumpy-NK model, we recover parameters using both a bumpy-AAK template and AAK template, respectively. The injected parameters are $M=10^6 M_{\odot}$, $\mu=10 M_{\odot}$, $a=0.6, p_0=10 M$, $e_0=0.25$, $\iota_0=0.78$, $\theta_S=0.78$, $\phi_S=0.6$, $\theta_K=0.39$, $\phi_K=0$,  $\Tilde{\gamma}_0=0$, $\Phi_0=0$, $\alpha_0=0$, and $\epsilon_2=0.01$. We set the luminosity distance $D=\{0.4,0.5,0.6,1.0\} \rm{Gpc}$, which is corresponding to $\rm{SNR}=\{59.1,46.9,39.4,23.6\}$.
The priors are the same as those in Sec.~\ref{sec:modeling error:GR}.

Fig.~\ref{fig:[NK-AAK]_bb} illustrates the results obtained by injecting bumpy-EMRI signals using bumpy-NK models and then recovering them by using both the bumpy-AAK models (a) and Kerr-AAK models (b). 
Fig.\ref{fig:[NK-AAK]_bb} (a) shows discrepancies in the recovered parameters caused solely by modeling errors when using inaccurate templates. Fig.\ref{fig:[NK-AAK]_bb} (b), on the other hand, reveals the combined effects of both modeling errors and fundamental biases. 
 We find that while the signal can indeed be detected using these models, systematic errors in the waveform templates cause all the posterior distributions of the parameters to deviate from the injected values beyond the $2\sigma (95\%)$  uncertainty. Initially, the accuracy of parameter estimation improves with increasing SNR. However, in this scenario, even a slight mismatch in waveform templates becomes significant, leading to deviations in the inferred physical parameters. When $\rm SNR > 50$, these peaks deviate even further from the injected values. As mentioned earlier, this increased deviation is caused by the mismatch between the NK signal and the AAK model, which becomes noticeable in high SNR systems.

Tab. \ref{tab:bumpy_NK} displays the mean and standard $2\sigma$ deviation of estimated parameters utilizing both bumpy-AAK and Kerr-AAK models for bumpy-EMRI signals generated using bumpy-NK models. For model selection, we also compare the evidence ($\log \mathcal{Z}$) of each case and derive the logarithmic Bayes factor $\log_e {\mathcal{B}^{\mathrm{AG}}_{\mathrm{GR}}}$ in favor of AG.
We observe that when recovering bumpy-NK signals using the bumpy-AAK model, the SNR of the resulting event tends to be reduced by $\sim 15\%$, which is caused by the mismatch between the bumpy-NK signals and the bumpy-AAK model. In addition, the calculated Bayesian factors consistently favor the bumpy signal hypothesis. This preference can be attributed to the closer match between the bumpy-AAK templates and the injected bumpy-NK signals compared to the Kerr-AAK templates, leading to a more accurate recovery.

\begin{table*} %[h]
\renewcommand{\arraystretch}{1.3}
\begin{ruledtabular} 
\begin{tabular}{lllllclll}
        $\rm{SNR}_{inj}$ & model& $M ( M_\odot)\   \red{(10^6)} $ & $\mu (M_\odot)\  \red{(10)}$ &  $a \  \red{(0.6)}$  & $\epsilon_2 \ \red{(0)}$ &SNR  & $\log{\mathcal{Z}}$  & $\log{\mathcal{B}^{\rm bumpy}_{\rm Kerr}}$ \\
        \hline
        $23.4$ & bumpy&  $999787^{+14.3}_{-14.8}$ & $9.98612^{+0.00170}_{-0.00170}$ & $0.59904^{+0.00017}_{-0.00018}$  & $-0.00502^{+0.00060}_{-0.00061}$ & $ 20.3$ &$-527203.855 \pm  0.142$  & $ -168.116 \pm 0.269$\\
           & kerr&  $999969^{+2.4}_{-2.4}$ & $9.99168^{+0.00082}_{-0.00084}$ & $0.59988^{+0.00002}_{-0.00002}$  & --& $ 20.3 $ &$-527035.739 \pm  0.127 $ & \\      
        $39.1$ & bumpy&  $999996^{+9.0}_{-8.8}$ & $10.01018^{+0.00085}_{-0.00082}$ & $0.59959^{+0.00011}_{-0.00011}$  & $-0.00208^{+0.00039}_{-0.00038}$& $34.0 $ & $-527271.401 \pm  0.129$ &$ -158.305\pm 0.255$ \\
         &  kerr& $999968^{+1.4}_{-1.4}$ & $9.99148^{+0.00046}_{-0.00048}$ & $0.59988^{+0.00001}_{-0.00001}$  & --& $ 33.9 $&$ -527113.096 \pm  0.126$  & \\
         
         $46.9$& bumpy& $999996^{+7.9}_{-7.4}$  & $ 10.01008^{+0.00071}_{-0.00070}$   & $ 0.59959^{+0.00010}_{-0.00009}$      &   $-0.00206 ^{+0.00033}_{-0.00032}$ & $40.8 $  &  $ -527411.632 \pm  0.146$  & $ 420.237\pm 0.261 $    \\
           & kerr& $  999898^{+ 1.8}_{-1.9 }$   &  $ 9.97829 ^{+ 0.00062}_{-0.00063 }$   &  $ 0.59956 ^{+0.00001 }_{-0.00001 }$     &   -- & $ 40.6   $  &   $ -527831.869 \pm  0.115$  & \\
         
         $58.6$& bumpy& $999983^{+6.2}_{-6.2}$  & $9.97250^{+0.00062}_{-0.00063}$    & $0.60065^{+0.00008}_{-0.00008}$     & $0.00364^{+0.00027}_{-0.00027}$   & $50.8 $  &   $-528029.515 \pm  0.135$  & $808.883\pm 0.273 $     \\
           & kerr&  $999809^{+1.6}_{-1.6}$ & $9.95816^{+0.00050}_{-0.00050}$ & $0.59182^{+0.00004}_{-0.00003}$  & --& $50.7$ & $-528838.398 \pm  0.137$& \\
    \end{tabular}
    \caption{The $2\sigma (95\%)$  credible intervals for recovered parameters using bumpy-AAK models and Kerr-AAK models. The injected signals are generated with Kerr-NK waveforms.  
    The last column shows the logarithmic Bayes factor for the different signals. 
    The true values of injected signal parameters are marked in red color.}
    \label{tab:kerr_NK}
\end{ruledtabular} 
\end{table*}

\begin{table*} %[h]
\renewcommand{\arraystretch}{1.3}
\begin{ruledtabular} 
\begin{tabular}{lllllclll}
        $\rm{SNR}_{inj}$ & model& $M ( M_\odot)\   \red{(10^6)} $ & $\mu (M_\odot)\  \red{(10)}$ &  $a \  \red{(0.6)}$  & $\epsilon_2 \ \red{(0.01)}$ &SNR  & $\log{\mathcal{Z}}$  & $\log{\mathcal{B}^{\rm bumpy}_{\rm Kerr}}$ \\
        \hline
        $23.6$ & bumpy&  $999102^{+51.1}_{-29.7}$ & $9.90611^{+0.00189}_{-0.0026}$ & $0.59573^{+0.00064}_{-0.00037}$  & $0.02885^{+0.00213}_{-0.00123}$ & $ 20.3$ &$-527089.488 \pm  0.122$  & $ 100.897 \pm 0.247$\\
           & kerr&  $998485^{+2.8}_{-2.8}$ & $9.96244^{+0.00096}_{-0.00096}$ & $0.58735^{+0.00002}_{-0.00002}$  & --& $ 20.4 $ &$-527190.385 \pm 0.125 $ & \\          
        $39.4$ & bumpy&  $999482^{+31.2}_{-20.4}$ & $9.90394^{+0.00138}_{-0.00207}$ & $0.59987^{+0.00040}_{-0.00026}$  & $-0.04149^{+0.00133}_{-0.00086}$& $ 33.5 $ & $-527446.935 \pm 0.140$ &$ 261.960 \pm 0.259$ \\
         &  kerr& $998279^{+2.4}_{-2.4}$ & $9.92218^{+0.00077}_{-0.00078}$ & $0.58737^{+0.00001}_{-0.00001}$  & --& $ 33.8 $&$ -527708.895 \pm 0.119$  & \\
         
         $46.9$& bumpy& $999991^{+7.3}_{-7.2}$  & $ 9.98988^{+0.00061}_{-0.00063}$   & $ 0.60017^{+0.00009}_{-0.00009}$      &   $0.00098 ^{+0.00031}_{-0.00030}$ & $ 40.7 $  &  $ -527154.086 \pm 0.152$  & $ 319.089\pm 0.2286 $    \\
           & kerr& $1000043 ^{+1.3}_{-1.3}$   &  $ 10.00681^{+0.00044}_{-0.00045}$   &  $ 0.60019^{+0.0001}_{-0.0001}$     &   -- & $ 40.6 $  &   $ -527473.175 \pm 0.134$  & \\
         
         $59.1$& bumpy& $999153^{+16.1}_{-24.1}$  & $9.90337^{+0.00143}_{-0.00102}$    & $0.59637^{+0.00020}_{-0.00031}$     & $0.03096^{+0.00067}_{-0.00101}$   & $ 50.6 $  &   $-527709.242 \pm 0.152 $  & $297.688\pm 0.282 $     \\
           & kerr&  $998407^{+1.1}_{-1.1}$ & $9.94656^{+0.00038}_{-0.00039}$ & $0.58702^{+0.00001}_{-0.00001}$  & --& $  50.8$ & $-528006.930 \pm 0.130 $& \\
    \end{tabular}
    \caption{Same as Tab. \ref{tab:kerr_NK}, but for injected signals with bumpy-NK waveforms.}
    \label{tab:bumpy_NK}
\end{ruledtabular} 
\end{table*}

\section{\label{conclusion} Conclusions} 
In this work, based on bumpy-kludge waveform templates and Bayesian inference, we investigate the influence of systematic errors arising from waveform templates, which could potentially result in misleading deviations from GR. These errors can be divided into two main categories: fundamental bias and modeling error.

Firstly, we address the systematic errors arising from fundamental bias. We obtain the bumpy-AAK waveforms for testing GR with EMRI observations based on the  bumpy metric proposed by \cite{VYS11}. We compare the results simulated using AG-EMRI waveform templates (e.g., bumpy-AAK) with those using GR-EMRI waveform templates (e.g., AAK).
We find that for given AG/GR-EMRI signals, both AG and GR templates can detect these signals, but the parameter estimates may exhibit biases if inappropriate templates are used. Specifically, when templates are based solely on GR models, any unexpected information the signals may contain about the nature of gravity could be filtered out, potentially leading to the misidentification of a bumpy-EMRI signal as a Kerr one.

To accurately determine which model (AG or GR) is statistically preferred by the observational data, we employ Bayesian inference for model selection. Our results reveal that at low SNR, there is a risk of mistaking an AG-EMRI signal for a GR-EMRI one, and vice versa. This confusion arises because, at low SNR, the differences between AG and GR waveforms may be overshadowed by noise. However, as the SNR increases to around 40 and higher, the signal becomes clearer, and the confusion between AG and GR signals dissipates. At higher SNRs, the distinct characteristics of AG and GR waveforms become more pronounced, allowing for more accurate identification and reducing the risk of misinterpretation.

Secondly, we also disscuss the systematic errors stemming from modeling error.  We explore the effectiveness of using less accurate AAK models to infer the parameters of EMRI signals generated by the more accurate NK models.
We observe an approximately $15\%$ reduction in the resulting event's SNR, which is caused by the mismatch between the NK signals and the AAK model. In addition, the calculated Bayesian factors tend to favor the bumpy signal hypothesis, especially at high SNR. This preference arises because the bumpy template includes an additional deformation parameter $\epsilon_2$, which is absent in the Kerr template. In such high SNR scenarios, the detector's improved ability to identify signals makes even slight mismatches in waveform templates noticeable. The model bias between the NK and AAK signals may be misconstrued as a difference in the deformation parameter between the Kerr and bumpy signals. This misinterpretation could lead to an overestimation of the support for the bumpy signal hypothesis, potentially skewing the analysis and leading to incorrect conclusions about the underlying EMRI signal.

Therefore, it is essential to develop accurate waveforms that are suitable for reliably recovering EMRI parameters.
Models that get  $0$PA accuracy might be enough to detect most signals, but models that get  $1$PA accuracy should be enough for precise parameter extraction \cite{lisaconsortiumwaveformworkinggroup2023waveform,PhysRevD.108.064015,burke2024accuracy}.
Because the first 1PA waveforms for generic orbit is still ongoing, here we only consider the 0PA waveforms, such as NK and AAK waveforms.  However, it is important to repeat this analysis once the 1PA waveforms become available for generic orbits. 

We notice that the values of deformation parameter $\epsilon_2$ considered in our Bayesian analysis are relatively large. However, we do not know the expected values of $ \epsilon_2$, and deviations from a Kerr metric could be much smaller. In a certain sense, our test cases represent the optimistic scenario in terms of deviations from GR
in a bumpy spacetime.

In our Bayesian parameter inference, we currently model the detector noise using a Gaussian distribution as a prior. However, this simplistic approach may not accurately reflect the true nature of the noise, especially given the presence of transient noise that can affect detector sensitivity. To improve the accuracy of our analysis, it would be valuable to explore alternative priors that better capture the complex characteristics of detector noise. Furthermore, it's important to acknowledge that our current analysis overlooks confusion noise originating from the population of galactic binaries, particularly white dwarf binaries \cite{PhysRevD.107.064021,Benacquista2020}. These binaries, often unresolved individually, contribute significantly to foreground confusion noise observed by space-borne GW detectors. Integrating this consideration into our models could provide a more comprehensive understanding of the noise sources influencing our observations. In this work, we estimate only a subset of parameters within a small parameter space, which should be sufficient to verify the influence of systematic errors on the parameter estimation results. In the future, we will extend our work to estimate the full set of parameters within a larger parameter space.

To test GR or modified theories of gravity using EMRIs, it is essential to ensure that all the necessary theoretical components are in place.
Beyond GR effects may also arise from new scalar or vector fields \cite{speri2024probingfundamentalphysicsextreme,PhysRevD.110.064018,PhysRevD.107.044053,zhang2023detectingvectorchargeextreme,PhysRevLett.131.051401}. 
Thus, it is important to assess the systematic biases arising from using a GR waveform model to recover a model that includes additional field degrees of freedom.
Given that EMRI systems are inherently intertwined with their astrophysical environments, their evolution is likely to deviate from an idealized vacuum scenario \cite{Jiang2024GeneralFF}.  Environmental effects, including the interaction with possible accretion disk around the SMBH and close stellar objects near the EMRI system, may induce sizeable phase shifts to the EMRI waveform \cite{PhysRevD.89.104059,Amaro_Seoane_2018}. Therefore, beyond modeling errors and fundamental biases, it is crucial to accurately account for environmental influences to avoid misinterpreting them as signals of a GR violation.  
We will address these considerations in future work.

\section{Acknowledgements*} 

This work was supported by the National Key R\&D Program of China (Grant Nos. 2021YFC2203002), the National Natural Science Foundation of China (Grant Nos. 12173071). This work made use of the High Performance Computing Resource in the Core Facility for Advanced Research Computing at Shanghai Astronomical Observatory.

%\nocite{*} %显示未引用的文献

\bibliography{ref} 

\begin{thebibliography}{73}%
\makeatletter
\providecommand \@ifxundefined [1]{%
 \@ifx{#1\undefined}
}%
\providecommand \@ifnum [1]{%
 \ifnum #1\expandafter \@firstoftwo
 \else \expandafter \@secondoftwo
 \fi
}%
\providecommand \@ifx [1]{%
 \ifx #1\expandafter \@firstoftwo
 \else \expandafter \@secondoftwo
 \fi
}%
\providecommand \natexlab [1]{#1}%
\providecommand \enquote  [1]{``#1''}%
\providecommand \bibnamefont  [1]{#1}%
\providecommand \bibfnamefont [1]{#1}%
\providecommand \citenamefont [1]{#1}%
\providecommand \href@noop [0]{\@secondoftwo}%
\providecommand \href [0]{\begingroup \@sanitize@url \@href}%
\providecommand \@href[1]{\@@startlink{#1}\@@href}%
\providecommand \@@href[1]{\endgroup#1\@@endlink}%
\providecommand \@sanitize@url [0]{\catcode `\\12\catcode `\$12\catcode `\&12\catcode `\#12\catcode `\^12\catcode `\_12\catcode `\%12\relax}%
\providecommand \@@startlink[1]{}%
\providecommand \@@endlink[0]{}%
\providecommand \url  [0]{\begingroup\@sanitize@url \@url }%
\providecommand \@url [1]{\endgroup\@href {#1}{\urlprefix }}%
\providecommand \urlprefix  [0]{URL }%
\providecommand \Eprint [0]{\href }%
\providecommand \doibase [0]{https://doi.org/}%
\providecommand \selectlanguage [0]{\@gobble}%
\providecommand \bibinfo  [0]{\@secondoftwo}%
\providecommand \bibfield  [0]{\@secondoftwo}%
\providecommand \translation [1]{[#1]}%
\providecommand \BibitemOpen [0]{}%
\providecommand \bibitemStop [0]{}%
\providecommand \bibitemNoStop [0]{.\EOS\space}%
\providecommand \EOS [0]{\spacefactor3000\relax}%
\providecommand \BibitemShut  [1]{\csname bibitem#1\endcsname}%
\let\auto@bib@innerbib\@empty
%</preamble>
\bibitem [{\citenamefont {Abbott}\ \emph {et~al.}(2016{\natexlab{a}})\citenamefont {Abbott}, \citenamefont {Abbott}, \citenamefont {Abbott}, \citenamefont {Abernathy}, \citenamefont {Acernese}, \citenamefont {Ackley}, \citenamefont {Adams},\ and\ \citenamefont {Adams}}]{PhysRevLett.116.061102}%
  \BibitemOpen
  \bibfield  {author} {\bibinfo {author} {\bibfnamefont {B.~P.}\ \bibnamefont {Abbott}}, \bibinfo {author} {\bibfnamefont {R.}~\bibnamefont {Abbott}}, \bibinfo {author} {\bibfnamefont {T.~D.}\ \bibnamefont {Abbott}}, \bibinfo {author} {\bibfnamefont {M.~R.}\ \bibnamefont {Abernathy}}, \bibinfo {author} {\bibfnamefont {F.}~\bibnamefont {Acernese}}, \bibinfo {author} {\bibfnamefont {K.}~\bibnamefont {Ackley}}, \bibinfo {author} {\bibfnamefont {C.}~\bibnamefont {Adams}},\ and\ \bibinfo {author} {\bibfnamefont {e.~a.}\ \bibnamefont {Adams}, \bibfnamefont {T}} (\bibinfo {collaboration} {LIGO Scientific Collaboration and Virgo Collaboration}),\ }\bibfield  {title} {\bibinfo {title} {Observation of gravitational waves from a binary black hole merger},\ }\href {https://doi.org/10.1103/PhysRevLett.116.061102} {\bibfield  {journal} {\bibinfo  {journal} {Phys. Rev. Lett.}\ }\textbf {\bibinfo {volume} {116}},\ \bibinfo {pages} {061102} (\bibinfo {year} {2016}{\natexlab{a}})}\BibitemShut {NoStop}%
\bibitem [{\citenamefont {Collaboration}\ \emph {et~al.}(2015)\citenamefont {Collaboration}, \citenamefont {Aasi}, \citenamefont {Abbott}, \citenamefont {Abbott}, \citenamefont {Abbott}, \citenamefont {Abernathy}, \citenamefont {Ackley}, \citenamefont {Adams}, \citenamefont {Adams},\ and\ \citenamefont {P~Addesso}}]{Aasi_2015}%
  \BibitemOpen
  \bibfield  {author} {\bibinfo {author} {\bibfnamefont {T.~L.~S.}\ \bibnamefont {Collaboration}}, \bibinfo {author} {\bibfnamefont {J.}~\bibnamefont {Aasi}}, \bibinfo {author} {\bibfnamefont {B.~P.}\ \bibnamefont {Abbott}}, \bibinfo {author} {\bibfnamefont {R.}~\bibnamefont {Abbott}}, \bibinfo {author} {\bibfnamefont {T.}~\bibnamefont {Abbott}}, \bibinfo {author} {\bibfnamefont {M.~R.}\ \bibnamefont {Abernathy}}, \bibinfo {author} {\bibfnamefont {K.}~\bibnamefont {Ackley}}, \bibinfo {author} {\bibfnamefont {C.}~\bibnamefont {Adams}}, \bibinfo {author} {\bibfnamefont {T.}~\bibnamefont {Adams}},\ and\ \bibinfo {author} {\bibfnamefont {e.~a.}\ \bibnamefont {P~Addesso}},\ }\bibfield  {title} {\bibinfo {title} {Advanced ligo},\ }\href {https://doi.org/10.1088/0264-9381/32/7/074001} {\bibfield  {journal} {\bibinfo  {journal} {Classical and Quantum Gravity}\ }\textbf {\bibinfo {volume} {32}},\ \bibinfo {pages} {074001} (\bibinfo {year} {2015})}\BibitemShut {NoStop}%
\bibitem [{\citenamefont {Acernese}\ \emph {et~al.}(2014)\citenamefont {Acernese}, \citenamefont {Agathos},\ and\ \citenamefont {et~al.}}]{Acernese_2015}%
  \BibitemOpen
  \bibfield  {author} {\bibinfo {author} {\bibfnamefont {F.}~\bibnamefont {Acernese}}, \bibinfo {author} {\bibfnamefont {M.}~\bibnamefont {Agathos}},\ and\ \bibinfo {author} {\bibfnamefont {K.~A.}\ \bibnamefont {et~al.}},\ }\bibfield  {title} {\bibinfo {title} {Advanced virgo: a second-generation interferometric gravitational wave detector},\ }\href {https://doi.org/10.1088/0264-9381/32/2/024001} {\bibfield  {journal} {\bibinfo  {journal} {Classical and Quantum Gravity}\ }\textbf {\bibinfo {volume} {32}},\ \bibinfo {pages} {024001} (\bibinfo {year} {2014})}\BibitemShut {NoStop}%
\bibitem [{\citenamefont {Abbott}\ \emph {et~al.}(2016{\natexlab{b}})\citenamefont {Abbott}, \citenamefont {Abbott}, \citenamefont {Abbott},\ and\ \citenamefont {Abernathy}}]{PhysRevLett.116.221101}%
  \BibitemOpen
  \bibfield  {author} {\bibinfo {author} {\bibfnamefont {B.~P.}\ \bibnamefont {Abbott}}, \bibinfo {author} {\bibfnamefont {R.}~\bibnamefont {Abbott}}, \bibinfo {author} {\bibfnamefont {T.~D.}\ \bibnamefont {Abbott}},\ and\ \bibinfo {author} {\bibfnamefont {e.~a.}\ \bibnamefont {Abernathy}} (\bibinfo {collaboration} {LIGO Scientific and Virgo Collaborations}),\ }\bibfield  {title} {\bibinfo {title} {Tests of general relativity with gw150914},\ }\href {https://doi.org/10.1103/PhysRevLett.116.221101} {\bibfield  {journal} {\bibinfo  {journal} {Phys. Rev. Lett.}\ }\textbf {\bibinfo {volume} {116}},\ \bibinfo {pages} {221101} (\bibinfo {year} {2016}{\natexlab{b}})}\BibitemShut {NoStop}%
\bibitem [{\citenamefont {Yunes}\ \emph {et~al.}(2016)\citenamefont {Yunes}, \citenamefont {Yagi},\ and\ \citenamefont {Pretorius}}]{PhysRevD.94.084002}%
  \BibitemOpen
  \bibfield  {author} {\bibinfo {author} {\bibfnamefont {N.}~\bibnamefont {Yunes}}, \bibinfo {author} {\bibfnamefont {K.}~\bibnamefont {Yagi}},\ and\ \bibinfo {author} {\bibfnamefont {F.}~\bibnamefont {Pretorius}},\ }\bibfield  {title} {\bibinfo {title} {Theoretical physics implications of the binary black-hole mergers gw150914 and gw151226},\ }\href {https://doi.org/10.1103/PhysRevD.94.084002} {\bibfield  {journal} {\bibinfo  {journal} {Phys. Rev. D}\ }\textbf {\bibinfo {volume} {94}},\ \bibinfo {pages} {084002} (\bibinfo {year} {2016})}\BibitemShut {NoStop}%
\bibitem [{\citenamefont {Abbott}\ \emph {et~al.}(2018)\citenamefont {Abbott}, \citenamefont {Abbott}, \citenamefont {Abbott}, \citenamefont {Acernese}, \citenamefont {Ackley}, \citenamefont {Adams}, \citenamefont {Adams}, \citenamefont {Addesso}, \citenamefont {Adhikari},\ and\ \citenamefont {Adya}}]{PhysRevLett.120.031104}%
  \BibitemOpen
  \bibfield  {author} {\bibinfo {author} {\bibfnamefont {B.~P.}\ \bibnamefont {Abbott}}, \bibinfo {author} {\bibfnamefont {R.}~\bibnamefont {Abbott}}, \bibinfo {author} {\bibfnamefont {T.~D.}\ \bibnamefont {Abbott}}, \bibinfo {author} {\bibfnamefont {F.}~\bibnamefont {Acernese}}, \bibinfo {author} {\bibfnamefont {K.}~\bibnamefont {Ackley}}, \bibinfo {author} {\bibfnamefont {C.}~\bibnamefont {Adams}}, \bibinfo {author} {\bibfnamefont {T.}~\bibnamefont {Adams}}, \bibinfo {author} {\bibfnamefont {P.}~\bibnamefont {Addesso}}, \bibinfo {author} {\bibfnamefont {R.~X.}\ \bibnamefont {Adhikari}},\ and\ \bibinfo {author} {\bibfnamefont {e.~a.}\ \bibnamefont {Adya}, \bibfnamefont {V.~B.}} (\bibinfo {collaboration} {LIGO Scientific Collaboration and Virgo Collaboration}),\ }\bibfield  {title} {\bibinfo {title} {First search for nontensorial gravitational waves from known pulsars},\ }\href {https://doi.org/10.1103/PhysRevLett.120.031104} {\bibfield  {journal} {\bibinfo  {journal} {Phys. Rev. Lett.}\ }\textbf {\bibinfo
  {volume} {120}},\ \bibinfo {pages} {031104} (\bibinfo {year} {2018})}\BibitemShut {NoStop}%
\bibitem [{\citenamefont {Li}\ \emph {et~al.}(2024)\citenamefont {Li}, \citenamefont {Han},\ and\ \citenamefont {Yang}}]{Li_2024}%
  \BibitemOpen
  \bibfield  {author} {\bibinfo {author} {\bibfnamefont {S.}~\bibnamefont {Li}}, \bibinfo {author} {\bibfnamefont {W.-B.}\ \bibnamefont {Han}},\ and\ \bibinfo {author} {\bibfnamefont {S.-C.}\ \bibnamefont {Yang}},\ }\bibfield  {title} {\bibinfo {title} {Tests of no-hair theorem with two binary black-hole coalescences},\ }\href {https://doi.org/10.1088/1475-7516/2024/06/013} {\bibfield  {journal} {\bibinfo  {journal} {Journal of Cosmology and Astroparticle Physics}\ }\textbf {\bibinfo {volume} {2024}}\bibinfo  {number} { (06)},\ \bibinfo {pages} {013}}\BibitemShut {NoStop}%
\bibitem [{\citenamefont {Abbott}\ \emph {et~al.}(2019)\citenamefont {Abbott}, \citenamefont {Abbott},\ and\ \citenamefont {Abbott}}]{PhysRevD.100.104036}%
  \BibitemOpen
\bibfield  {number} {  }\bibfield  {author} {\bibinfo {author} {\bibfnamefont {B.~P.}\ \bibnamefont {Abbott}}, \bibinfo {author} {\bibfnamefont {R.}~\bibnamefont {Abbott}},\ and\ \bibinfo {author} {\bibfnamefont {T.~D. e.~a.}\ \bibnamefont {Abbott}} (\bibinfo {collaboration} {The LIGO Scientific Collaboration and the Virgo Collaboration}),\ }\bibfield  {title} {\bibinfo {title} {Tests of general relativity with the binary black hole signals from the ligo-virgo catalog gwtc-1},\ }\href {https://doi.org/10.1103/PhysRevD.100.104036} {\bibfield  {journal} {\bibinfo  {journal} {Phys. Rev. D}\ }\textbf {\bibinfo {volume} {100}},\ \bibinfo {pages} {104036} (\bibinfo {year} {2019})}\BibitemShut {NoStop}%
\bibitem [{\citenamefont {Amaro-Seoane}\ \emph {et~al.}(2017)\citenamefont {Amaro-Seoane}, \citenamefont {Audley}, \citenamefont {Babak}, \citenamefont {Baker}, \citenamefont {Barausse}, \citenamefont {Bender}, \citenamefont {Berti}, \citenamefont {Binetruy}, \citenamefont {Born}, \citenamefont {Bortoluzzi}, \citenamefont {Camp}, \citenamefont {Caprini}, \citenamefont {Cardoso}, \citenamefont {Colpi}, \citenamefont {Conklin}, \citenamefont {Cornish}, \citenamefont {Cutler}, \citenamefont {Danzmann}, \citenamefont {Dolesi}, \citenamefont {Ferraioli}, \citenamefont {Ferroni}, \citenamefont {Fitzsimons}, \citenamefont {Gair}, \citenamefont {Bote}, \citenamefont {Giardini}, \citenamefont {Gibert}, \citenamefont {Grimani}, \citenamefont {Halloin}, \citenamefont {Heinzel}, \citenamefont {Hertog}, \citenamefont {Hewitson}, \citenamefont {Holley-Bockelmann}, \citenamefont {Hollington}, \citenamefont {Hueller}, \citenamefont {Inchauspe}, \citenamefont {Jetzer}, \citenamefont {Karnesis}, \citenamefont {Killow},
  \citenamefont {Klein}, \citenamefont {Klipstein}, \citenamefont {Korsakova}, \citenamefont {Larson}, \citenamefont {Livas}, \citenamefont {Lloro}, \citenamefont {Man}, \citenamefont {Mance}, \citenamefont {Martino}, \citenamefont {Mateos}, \citenamefont {McKenzie}, \citenamefont {McWilliams}, \citenamefont {Miller}, \citenamefont {Mueller}, \citenamefont {Nardini}, \citenamefont {Nelemans}, \citenamefont {Nofrarias}, \citenamefont {Petiteau}, \citenamefont {Pivato}, \citenamefont {Plagnol}, \citenamefont {Porter}, \citenamefont {Reiche}, \citenamefont {Robertson}, \citenamefont {Robertson}, \citenamefont {Rossi}, \citenamefont {Russano}, \citenamefont {Schutz}, \citenamefont {Sesana}, \citenamefont {Shoemaker}, \citenamefont {Slutsky}, \citenamefont {Sopuerta}, \citenamefont {Sumner}, \citenamefont {Tamanini}, \citenamefont {Thorpe}, \citenamefont {Troebs}, \citenamefont {Vallisneri}, \citenamefont {Vecchio}, \citenamefont {Vetrugno}, \citenamefont {Vitale}, \citenamefont {Volonteri}, \citenamefont
  {Wanner}, \citenamefont {Ward}, \citenamefont {Wass}, \citenamefont {Weber}, \citenamefont {Ziemer},\ and\ \citenamefont {Zweifel}}]{amaroseoane2017laserinterferometerspaceantenna}%
  \BibitemOpen
  \bibfield  {author} {\bibinfo {author} {\bibfnamefont {P.}~\bibnamefont {Amaro-Seoane}}, \bibinfo {author} {\bibfnamefont {H.}~\bibnamefont {Audley}}, \bibinfo {author} {\bibfnamefont {S.}~\bibnamefont {Babak}}, \bibinfo {author} {\bibfnamefont {J.}~\bibnamefont {Baker}}, \bibinfo {author} {\bibfnamefont {E.}~\bibnamefont {Barausse}}, \bibinfo {author} {\bibfnamefont {P.}~\bibnamefont {Bender}}, \bibinfo {author} {\bibfnamefont {E.}~\bibnamefont {Berti}}, \bibinfo {author} {\bibfnamefont {P.}~\bibnamefont {Binetruy}}, \bibinfo {author} {\bibfnamefont {M.}~\bibnamefont {Born}}, \bibinfo {author} {\bibfnamefont {D.}~\bibnamefont {Bortoluzzi}}, \bibinfo {author} {\bibfnamefont {J.}~\bibnamefont {Camp}}, \bibinfo {author} {\bibfnamefont {C.}~\bibnamefont {Caprini}}, \bibinfo {author} {\bibfnamefont {V.}~\bibnamefont {Cardoso}}, \bibinfo {author} {\bibfnamefont {M.}~\bibnamefont {Colpi}}, \bibinfo {author} {\bibfnamefont {J.}~\bibnamefont {Conklin}}, \bibinfo {author} {\bibfnamefont {N.}~\bibnamefont {Cornish}},
  \bibinfo {author} {\bibfnamefont {C.}~\bibnamefont {Cutler}}, \bibinfo {author} {\bibfnamefont {K.}~\bibnamefont {Danzmann}}, \bibinfo {author} {\bibfnamefont {R.}~\bibnamefont {Dolesi}}, \bibinfo {author} {\bibfnamefont {L.}~\bibnamefont {Ferraioli}}, \bibinfo {author} {\bibfnamefont {V.}~\bibnamefont {Ferroni}}, \bibinfo {author} {\bibfnamefont {E.}~\bibnamefont {Fitzsimons}}, \bibinfo {author} {\bibfnamefont {J.}~\bibnamefont {Gair}}, \bibinfo {author} {\bibfnamefont {L.~G.}\ \bibnamefont {Bote}}, \bibinfo {author} {\bibfnamefont {D.}~\bibnamefont {Giardini}}, \bibinfo {author} {\bibfnamefont {F.}~\bibnamefont {Gibert}}, \bibinfo {author} {\bibfnamefont {C.}~\bibnamefont {Grimani}}, \bibinfo {author} {\bibfnamefont {H.}~\bibnamefont {Halloin}}, \bibinfo {author} {\bibfnamefont {G.}~\bibnamefont {Heinzel}}, \bibinfo {author} {\bibfnamefont {T.}~\bibnamefont {Hertog}}, \bibinfo {author} {\bibfnamefont {M.}~\bibnamefont {Hewitson}}, \bibinfo {author} {\bibfnamefont {K.}~\bibnamefont {Holley-Bockelmann}},
  \bibinfo {author} {\bibfnamefont {D.}~\bibnamefont {Hollington}}, \bibinfo {author} {\bibfnamefont {M.}~\bibnamefont {Hueller}}, \bibinfo {author} {\bibfnamefont {H.}~\bibnamefont {Inchauspe}}, \bibinfo {author} {\bibfnamefont {P.}~\bibnamefont {Jetzer}}, \bibinfo {author} {\bibfnamefont {N.}~\bibnamefont {Karnesis}}, \bibinfo {author} {\bibfnamefont {C.}~\bibnamefont {Killow}}, \bibinfo {author} {\bibfnamefont {A.}~\bibnamefont {Klein}}, \bibinfo {author} {\bibfnamefont {B.}~\bibnamefont {Klipstein}}, \bibinfo {author} {\bibfnamefont {N.}~\bibnamefont {Korsakova}}, \bibinfo {author} {\bibfnamefont {S.~L.}\ \bibnamefont {Larson}}, \bibinfo {author} {\bibfnamefont {J.}~\bibnamefont {Livas}}, \bibinfo {author} {\bibfnamefont {I.}~\bibnamefont {Lloro}}, \bibinfo {author} {\bibfnamefont {N.}~\bibnamefont {Man}}, \bibinfo {author} {\bibfnamefont {D.}~\bibnamefont {Mance}}, \bibinfo {author} {\bibfnamefont {J.}~\bibnamefont {Martino}}, \bibinfo {author} {\bibfnamefont {I.}~\bibnamefont {Mateos}}, \bibinfo
  {author} {\bibfnamefont {K.}~\bibnamefont {McKenzie}}, \bibinfo {author} {\bibfnamefont {S.~T.}\ \bibnamefont {McWilliams}}, \bibinfo {author} {\bibfnamefont {C.}~\bibnamefont {Miller}}, \bibinfo {author} {\bibfnamefont {G.}~\bibnamefont {Mueller}}, \bibinfo {author} {\bibfnamefont {G.}~\bibnamefont {Nardini}}, \bibinfo {author} {\bibfnamefont {G.}~\bibnamefont {Nelemans}}, \bibinfo {author} {\bibfnamefont {M.}~\bibnamefont {Nofrarias}}, \bibinfo {author} {\bibfnamefont {A.}~\bibnamefont {Petiteau}}, \bibinfo {author} {\bibfnamefont {P.}~\bibnamefont {Pivato}}, \bibinfo {author} {\bibfnamefont {E.}~\bibnamefont {Plagnol}}, \bibinfo {author} {\bibfnamefont {E.}~\bibnamefont {Porter}}, \bibinfo {author} {\bibfnamefont {J.}~\bibnamefont {Reiche}}, \bibinfo {author} {\bibfnamefont {D.}~\bibnamefont {Robertson}}, \bibinfo {author} {\bibfnamefont {N.}~\bibnamefont {Robertson}}, \bibinfo {author} {\bibfnamefont {E.}~\bibnamefont {Rossi}}, \bibinfo {author} {\bibfnamefont {G.}~\bibnamefont {Russano}}, \bibinfo
  {author} {\bibfnamefont {B.}~\bibnamefont {Schutz}}, \bibinfo {author} {\bibfnamefont {A.}~\bibnamefont {Sesana}}, \bibinfo {author} {\bibfnamefont {D.}~\bibnamefont {Shoemaker}}, \bibinfo {author} {\bibfnamefont {J.}~\bibnamefont {Slutsky}}, \bibinfo {author} {\bibfnamefont {C.~F.}\ \bibnamefont {Sopuerta}}, \bibinfo {author} {\bibfnamefont {T.}~\bibnamefont {Sumner}}, \bibinfo {author} {\bibfnamefont {N.}~\bibnamefont {Tamanini}}, \bibinfo {author} {\bibfnamefont {I.}~\bibnamefont {Thorpe}}, \bibinfo {author} {\bibfnamefont {M.}~\bibnamefont {Troebs}}, \bibinfo {author} {\bibfnamefont {M.}~\bibnamefont {Vallisneri}}, \bibinfo {author} {\bibfnamefont {A.}~\bibnamefont {Vecchio}}, \bibinfo {author} {\bibfnamefont {D.}~\bibnamefont {Vetrugno}}, \bibinfo {author} {\bibfnamefont {S.}~\bibnamefont {Vitale}}, \bibinfo {author} {\bibfnamefont {M.}~\bibnamefont {Volonteri}}, \bibinfo {author} {\bibfnamefont {G.}~\bibnamefont {Wanner}}, \bibinfo {author} {\bibfnamefont {H.}~\bibnamefont {Ward}}, \bibinfo {author}
  {\bibfnamefont {P.}~\bibnamefont {Wass}}, \bibinfo {author} {\bibfnamefont {W.}~\bibnamefont {Weber}}, \bibinfo {author} {\bibfnamefont {J.}~\bibnamefont {Ziemer}},\ and\ \bibinfo {author} {\bibfnamefont {P.}~\bibnamefont {Zweifel}},\ }\href {https://arxiv.org/abs/1702.00786} {\bibinfo {title} {Laser interferometer space antenna}} (\bibinfo {year} {2017}),\ \Eprint {https://arxiv.org/abs/1702.00786} {arXiv:1702.00786 [astro-ph.IM]} \BibitemShut {NoStop}%
\bibitem [{\citenamefont {rui Hu}\ and\ \citenamefont {Wu}(2017)}]{Hu2017TheTP}%
  \BibitemOpen
  \bibfield  {author} {\bibinfo {author} {\bibfnamefont {W.}~\bibnamefont {rui Hu}}\ and\ \bibinfo {author} {\bibfnamefont {Y.-L.}\ \bibnamefont {Wu}},\ }\bibfield  {title} {\bibinfo {title} {The taiji program in space for gravitational wave physics and the nature of gravity},\ }\href {https://api.semanticscholar.org/CorpusID:126068470} {\bibfield  {journal} {\bibinfo  {journal} {National Science Review}\ }\textbf {\bibinfo {volume} {4}},\ \bibinfo {pages} {685} (\bibinfo {year} {2017})}\BibitemShut {NoStop}%
\bibitem [{\citenamefont {Zhong}\ \emph {et~al.}(2023)\citenamefont {Zhong}, \citenamefont {Han}, \citenamefont {Luo},\ and\ \citenamefont {Wu}}]{zhong2023exploring}%
  \BibitemOpen
  \bibfield  {author} {\bibinfo {author} {\bibfnamefont {X.}~\bibnamefont {Zhong}}, \bibinfo {author} {\bibfnamefont {W.-B.}\ \bibnamefont {Han}}, \bibinfo {author} {\bibfnamefont {Z.}~\bibnamefont {Luo}},\ and\ \bibinfo {author} {\bibfnamefont {Y.}~\bibnamefont {Wu}},\ }\bibfield  {title} {\bibinfo {title} {Exploring the nature of black hole and gravity with an imminent merging binary of supermassive black holes},\ }\href {https://doi.org/10.1007/s11433-022-2028-7} {\bibfield  {journal} {\bibinfo  {journal} {Science China Physics, Mechanics \& Astronomy}\ }\textbf {\bibinfo {volume} {66}},\ \bibinfo {pages} {230411} (\bibinfo {year} {2023})}\BibitemShut {NoStop}%
\bibitem [{\citenamefont {Ren}\ \emph {et~al.}(2023)\citenamefont {Ren}, \citenamefont {Zhao}, \citenamefont {Cao}, \citenamefont {Guo}, \citenamefont {Han}, \citenamefont {Jin},\ and\ \citenamefont {Wu}}]{Ren_2023}%
  \BibitemOpen
  \bibfield  {author} {\bibinfo {author} {\bibfnamefont {Z.}~\bibnamefont {Ren}}, \bibinfo {author} {\bibfnamefont {T.}~\bibnamefont {Zhao}}, \bibinfo {author} {\bibfnamefont {Z.}~\bibnamefont {Cao}}, \bibinfo {author} {\bibfnamefont {Z.-K.}\ \bibnamefont {Guo}}, \bibinfo {author} {\bibfnamefont {W.-B.}\ \bibnamefont {Han}}, \bibinfo {author} {\bibfnamefont {H.-B.}\ \bibnamefont {Jin}},\ and\ \bibinfo {author} {\bibfnamefont {Y.-L.}\ \bibnamefont {Wu}},\ }\bibfield  {title} {\bibinfo {title} {Taiji data challenge for exploring gravitational wave universe},\ }\bibfield  {journal} {\bibinfo  {journal} {Frontiers of Physics}\ }\textbf {\bibinfo {volume} {18}},\ \href {https://doi.org/10.1007/s11467-023-1318-y} {10.1007/s11467-023-1318-y} (\bibinfo {year} {2023})\BibitemShut {NoStop}%
\bibitem [{\citenamefont {Luo}\ \emph {et~al.}(2016)\citenamefont {Luo}, \citenamefont {Chen}, \citenamefont {Duan}, \citenamefont {Gong}, \citenamefont {Hu}, \citenamefont {Ji}, \citenamefont {Liu}, \citenamefont {Mei}, \citenamefont {Milyukov}, \citenamefont {Sazhin}, \citenamefont {Shao}, \citenamefont {Toth}, \citenamefont {Tu}, \citenamefont {Wang}, \citenamefont {Wang}, \citenamefont {Yeh}, \citenamefont {Zhan}, \citenamefont {Zhang}, \citenamefont {Zharov},\ and\ \citenamefont {Zhou}}]{Luo_2016}%
  \BibitemOpen
  \bibfield  {author} {\bibinfo {author} {\bibfnamefont {J.}~\bibnamefont {Luo}}, \bibinfo {author} {\bibfnamefont {L.-S.}\ \bibnamefont {Chen}}, \bibinfo {author} {\bibfnamefont {H.-Z.}\ \bibnamefont {Duan}}, \bibinfo {author} {\bibfnamefont {Y.-G.}\ \bibnamefont {Gong}}, \bibinfo {author} {\bibfnamefont {S.}~\bibnamefont {Hu}}, \bibinfo {author} {\bibfnamefont {J.}~\bibnamefont {Ji}}, \bibinfo {author} {\bibfnamefont {Q.}~\bibnamefont {Liu}}, \bibinfo {author} {\bibfnamefont {J.}~\bibnamefont {Mei}}, \bibinfo {author} {\bibfnamefont {V.}~\bibnamefont {Milyukov}}, \bibinfo {author} {\bibfnamefont {M.}~\bibnamefont {Sazhin}}, \bibinfo {author} {\bibfnamefont {C.-G.}\ \bibnamefont {Shao}}, \bibinfo {author} {\bibfnamefont {V.~T.}\ \bibnamefont {Toth}}, \bibinfo {author} {\bibfnamefont {H.-B.}\ \bibnamefont {Tu}}, \bibinfo {author} {\bibfnamefont {Y.}~\bibnamefont {Wang}}, \bibinfo {author} {\bibfnamefont {Y.}~\bibnamefont {Wang}}, \bibinfo {author} {\bibfnamefont {H.-C.}\ \bibnamefont {Yeh}}, \bibinfo {author}
  {\bibfnamefont {M.-S.}\ \bibnamefont {Zhan}}, \bibinfo {author} {\bibfnamefont {Y.}~\bibnamefont {Zhang}}, \bibinfo {author} {\bibfnamefont {V.}~\bibnamefont {Zharov}},\ and\ \bibinfo {author} {\bibfnamefont {Z.-B.}\ \bibnamefont {Zhou}},\ }\bibfield  {title} {\bibinfo {title} {Tianqin: a space-borne gravitational wave detector},\ }\href {https://doi.org/10.1088/0264-9381/33/3/035010} {\bibfield  {journal} {\bibinfo  {journal} {Classical and Quantum Gravity}\ }\textbf {\bibinfo {volume} {33}},\ \bibinfo {pages} {035010} (\bibinfo {year} {2016})}\BibitemShut {NoStop}%
\bibitem [{\citenamefont {Amaro-Seoane}\ \emph {et~al.}(2007)\citenamefont {Amaro-Seoane}, \citenamefont {Gair}, \citenamefont {Freitag}, \citenamefont {Miller}, \citenamefont {Mandel}, \citenamefont {Cutler},\ and\ \citenamefont {Babak}}]{2007Intermediate}%
  \BibitemOpen
  \bibfield  {author} {\bibinfo {author} {\bibfnamefont {P.}~\bibnamefont {Amaro-Seoane}}, \bibinfo {author} {\bibfnamefont {J.~R.}\ \bibnamefont {Gair}}, \bibinfo {author} {\bibfnamefont {M.}~\bibnamefont {Freitag}}, \bibinfo {author} {\bibfnamefont {M.~C.}\ \bibnamefont {Miller}}, \bibinfo {author} {\bibfnamefont {I.}~\bibnamefont {Mandel}}, \bibinfo {author} {\bibfnamefont {C.~J.}\ \bibnamefont {Cutler}},\ and\ \bibinfo {author} {\bibfnamefont {S.}~\bibnamefont {Babak}},\ }\bibfield  {title} {\bibinfo {title} {Intermediate and extreme mass-ratio inspirals—astrophysics, science applications and detection using lisa},\ }\href {https://doi.org/10.1088/0264-9381/24/17/r01} {\bibfield  {journal} {\bibinfo  {journal} {Classical and Quantum Gravity}\ }\textbf {\bibinfo {volume} {24}},\ \bibinfo {pages} {R113–R169} (\bibinfo {year} {2007})}\BibitemShut {NoStop}%
\bibitem [{\citenamefont {Babak}\ \emph {et~al.}(2017)\citenamefont {Babak}, \citenamefont {Gair}, \citenamefont {Sesana}, \citenamefont {Barausse}, \citenamefont {Sopuerta}, \citenamefont {Berry}, \citenamefont {Berti}, \citenamefont {Amaro-Seoane}, \citenamefont {Petiteau},\ and\ \citenamefont {Klein}}]{PhysRevD.95.103012}%
  \BibitemOpen
  \bibfield  {author} {\bibinfo {author} {\bibfnamefont {S.}~\bibnamefont {Babak}}, \bibinfo {author} {\bibfnamefont {J.}~\bibnamefont {Gair}}, \bibinfo {author} {\bibfnamefont {A.}~\bibnamefont {Sesana}}, \bibinfo {author} {\bibfnamefont {E.}~\bibnamefont {Barausse}}, \bibinfo {author} {\bibfnamefont {C.~F.}\ \bibnamefont {Sopuerta}}, \bibinfo {author} {\bibfnamefont {C.~P.~L.}\ \bibnamefont {Berry}}, \bibinfo {author} {\bibfnamefont {E.}~\bibnamefont {Berti}}, \bibinfo {author} {\bibfnamefont {P.}~\bibnamefont {Amaro-Seoane}}, \bibinfo {author} {\bibfnamefont {A.}~\bibnamefont {Petiteau}},\ and\ \bibinfo {author} {\bibfnamefont {A.}~\bibnamefont {Klein}},\ }\bibfield  {title} {\bibinfo {title} {Science with the space-based interferometer lisa. v. extreme mass-ratio inspirals},\ }\href {https://doi.org/10.1103/PhysRevD.95.103012} {\bibfield  {journal} {\bibinfo  {journal} {Phys. Rev. D}\ }\textbf {\bibinfo {volume} {95}},\ \bibinfo {pages} {103012} (\bibinfo {year} {2017})}\BibitemShut {NoStop}%
\bibitem [{\citenamefont {Gair}\ \emph {et~al.}(2013)\citenamefont {Gair}, \citenamefont {Vallisneri}, \citenamefont {Larson},\ and\ \citenamefont {Baker}}]{gair2013testing}%
  \BibitemOpen
  \bibfield  {author} {\bibinfo {author} {\bibfnamefont {J.~R.}\ \bibnamefont {Gair}}, \bibinfo {author} {\bibfnamefont {M.}~\bibnamefont {Vallisneri}}, \bibinfo {author} {\bibfnamefont {S.~L.}\ \bibnamefont {Larson}},\ and\ \bibinfo {author} {\bibfnamefont {J.~G.}\ \bibnamefont {Baker}},\ }\bibfield  {title} {\bibinfo {title} {Testing general relativity with low-frequency, space-based gravitational-wave detectors},\ }\bibfield  {journal} {\bibinfo  {journal} {Living Reviews in Relativity}\ }\textbf {\bibinfo {volume} {16}},\ \href {https://doi.org/10.12942/lrr-2013-7} {10.12942/lrr-2013-7} (\bibinfo {year} {2013})\BibitemShut {NoStop}%
\bibitem [{\citenamefont {Barack}\ and\ \citenamefont {Cutler}(2007)}]{barack2007using}%
  \BibitemOpen
  \bibfield  {author} {\bibinfo {author} {\bibfnamefont {L.}~\bibnamefont {Barack}}\ and\ \bibinfo {author} {\bibfnamefont {C.}~\bibnamefont {Cutler}},\ }\bibfield  {title} {\bibinfo {title} {Using lisa extreme-mass-ratio inspiral sources to test off-kerr deviations in the geometry of massive black holes},\ }\href {https://doi.org/10.1103/PhysRevD.75.042003} {\bibfield  {journal} {\bibinfo  {journal} {Phys. Rev. D}\ }\textbf {\bibinfo {volume} {75}},\ \bibinfo {pages} {042003} (\bibinfo {year} {2007})}\BibitemShut {NoStop}%
\bibitem [{\citenamefont {Ryan}(1997)}]{PhysRevD.56.1845}%
  \BibitemOpen
  \bibfield  {author} {\bibinfo {author} {\bibfnamefont {F.~D.}\ \bibnamefont {Ryan}},\ }\bibfield  {title} {\bibinfo {title} {Accuracy of estimating the multipole moments of a massive body from the gravitational waves of a binary inspiral},\ }\href {https://doi.org/10.1103/PhysRevD.56.1845} {\bibfield  {journal} {\bibinfo  {journal} {Phys. Rev. D}\ }\textbf {\bibinfo {volume} {56}},\ \bibinfo {pages} {1845} (\bibinfo {year} {1997})}\BibitemShut {NoStop}%
\bibitem [{\citenamefont {Collins}\ and\ \citenamefont {Hughes}(2004)}]{CH04}%
  \BibitemOpen
  \bibfield  {author} {\bibinfo {author} {\bibfnamefont {N.~A.}\ \bibnamefont {Collins}}\ and\ \bibinfo {author} {\bibfnamefont {S.~A.}\ \bibnamefont {Hughes}},\ }\bibfield  {title} {\bibinfo {title} {Towards a formalism for mapping the spacetimes of massive compact objects: Bumpy black holes and their orbits},\ }\href {https://doi.org/10.1103/PhysRevD.69.124022} {\bibfield  {journal} {\bibinfo  {journal} {Phys. Rev. D}\ }\textbf {\bibinfo {volume} {69}},\ \bibinfo {pages} {124022} (\bibinfo {year} {2004})}\BibitemShut {NoStop}%
\bibitem [{\citenamefont {Fransen}\ and\ \citenamefont {Mayerson}(2022)}]{Fransen_2022}%
  \BibitemOpen
  \bibfield  {author} {\bibinfo {author} {\bibfnamefont {K.}~\bibnamefont {Fransen}}\ and\ \bibinfo {author} {\bibfnamefont {D.~R.}\ \bibnamefont {Mayerson}},\ }\bibfield  {title} {\bibinfo {title} {Detecting equatorial symmetry breaking with {LISA}},\ }\bibfield  {journal} {\bibinfo  {journal} {Physical Review D}\ }\textbf {\bibinfo {volume} {106}},\ \href {https://doi.org/10.1103/physrevd.106.064035} {10.1103/physrevd.106.064035} (\bibinfo {year} {2022})\BibitemShut {NoStop}%
\bibitem [{\citenamefont {Xin}\ \emph {et~al.}(2019)\citenamefont {Xin}, \citenamefont {Han},\ and\ \citenamefont {Yang}}]{PhysRevD.100.084055}%
  \BibitemOpen
  \bibfield  {author} {\bibinfo {author} {\bibfnamefont {S.}~\bibnamefont {Xin}}, \bibinfo {author} {\bibfnamefont {W.-B.}\ \bibnamefont {Han}},\ and\ \bibinfo {author} {\bibfnamefont {S.-C.}\ \bibnamefont {Yang}},\ }\bibfield  {title} {\bibinfo {title} {Gravitational waves from extreme-mass-ratio inspirals using general parametrized metrics},\ }\href {https://doi.org/10.1103/PhysRevD.100.084055} {\bibfield  {journal} {\bibinfo  {journal} {Phys. Rev. D}\ }\textbf {\bibinfo {volume} {100}},\ \bibinfo {pages} {084055} (\bibinfo {year} {2019})}\BibitemShut {NoStop}%
\bibitem [{\citenamefont {Zi}\ \emph {et~al.}(2021)\citenamefont {Zi}, \citenamefont {Zhang}, \citenamefont {Fan}, \citenamefont {Zhang}, \citenamefont {Hu}, \citenamefont {Shi},\ and\ \citenamefont {Mei}}]{PhysRevD.104.064008}%
  \BibitemOpen
  \bibfield  {author} {\bibinfo {author} {\bibfnamefont {T.}~\bibnamefont {Zi}}, \bibinfo {author} {\bibfnamefont {J.-d.}\ \bibnamefont {Zhang}}, \bibinfo {author} {\bibfnamefont {H.-M.}\ \bibnamefont {Fan}}, \bibinfo {author} {\bibfnamefont {X.-T.}\ \bibnamefont {Zhang}}, \bibinfo {author} {\bibfnamefont {Y.-M.}\ \bibnamefont {Hu}}, \bibinfo {author} {\bibfnamefont {C.}~\bibnamefont {Shi}},\ and\ \bibinfo {author} {\bibfnamefont {J.}~\bibnamefont {Mei}},\ }\bibfield  {title} {\bibinfo {title} {Science with the tianqin observatory: Preliminary results on testing the no-hair theorem with extreme mass ratio inspirals},\ }\href {https://doi.org/10.1103/PhysRevD.104.064008} {\bibfield  {journal} {\bibinfo  {journal} {Phys. Rev. D}\ }\textbf {\bibinfo {volume} {104}},\ \bibinfo {pages} {064008} (\bibinfo {year} {2021})}\BibitemShut {NoStop}%
\bibitem [{\citenamefont {Cárdenas-Avendaño}\ and\ \citenamefont {Sopuerta}(2024)}]{emri_GR_REVIEW}%
  \BibitemOpen
  \bibfield  {author} {\bibinfo {author} {\bibfnamefont {A.}~\bibnamefont {Cárdenas-Avendaño}}\ and\ \bibinfo {author} {\bibfnamefont {C.~F.}\ \bibnamefont {Sopuerta}},\ }\href@noop {} {\bibinfo {title} {Testing gravity with extreme-mass-ratio inspirals}} (\bibinfo {year} {2024}),\ \Eprint {https://arxiv.org/abs/2401.08085} {arXiv:2401.08085 [gr-qc]} \BibitemShut {NoStop}%
\bibitem [{\citenamefont {Hughes}(2016)}]{hughes2016adiabatic}%
  \BibitemOpen
  \bibfield  {author} {\bibinfo {author} {\bibfnamefont {S.~A.}\ \bibnamefont {Hughes}},\ }\href@noop {} {\bibinfo {title} {Adiabatic and post-adiabatic approaches to extreme mass ratio inspiral}} (\bibinfo {year} {2016}),\ \Eprint {https://arxiv.org/abs/1601.02042} {arXiv:1601.02042 [gr-qc]} \BibitemShut {NoStop}%
\bibitem [{\citenamefont {Katz}\ \emph {et~al.}(2021)\citenamefont {Katz}, \citenamefont {Chua}, \citenamefont {Speri}, \citenamefont {Warburton},\ and\ \citenamefont {Hughes}}]{PhysRevD.104.064047}%
  \BibitemOpen
  \bibfield  {author} {\bibinfo {author} {\bibfnamefont {M.~L.}\ \bibnamefont {Katz}}, \bibinfo {author} {\bibfnamefont {A.~J.~K.}\ \bibnamefont {Chua}}, \bibinfo {author} {\bibfnamefont {L.}~\bibnamefont {Speri}}, \bibinfo {author} {\bibfnamefont {N.}~\bibnamefont {Warburton}},\ and\ \bibinfo {author} {\bibfnamefont {S.~A.}\ \bibnamefont {Hughes}},\ }\bibfield  {title} {\bibinfo {title} {Fast extreme-mass-ratio-inspiral waveforms: New tools for millihertz gravitational-wave data analysis},\ }\href {https://doi.org/10.1103/PhysRevD.104.064047} {\bibfield  {journal} {\bibinfo  {journal} {Phys. Rev. D}\ }\textbf {\bibinfo {volume} {104}},\ \bibinfo {pages} {064047} (\bibinfo {year} {2021})}\BibitemShut {NoStop}%
\bibitem [{\citenamefont {Barack}\ and\ \citenamefont {Pound}(2018)}]{Barack_2018}%
  \BibitemOpen
  \bibfield  {author} {\bibinfo {author} {\bibfnamefont {L.}~\bibnamefont {Barack}}\ and\ \bibinfo {author} {\bibfnamefont {A.}~\bibnamefont {Pound}},\ }\bibfield  {title} {\bibinfo {title} {Self-force and radiation reaction in general relativity},\ }\href {https://doi.org/10.1088/1361-6633/aae552} {\bibfield  {journal} {\bibinfo  {journal} {Reports on Progress in Physics}\ }\textbf {\bibinfo {volume} {82}},\ \bibinfo {pages} {016904} (\bibinfo {year} {2018})}\BibitemShut {NoStop}%
\bibitem [{\citenamefont {Group}(2023)}]{lisaconsortiumwaveformworkinggroup2023waveform}%
  \BibitemOpen
  \bibfield  {author} {\bibinfo {author} {\bibfnamefont {L.~C. W.~W.}\ \bibnamefont {Group}},\ }\href@noop {} {\bibinfo {title} {Waveform modelling for the laser interferometer space antenna}} (\bibinfo {year} {2023}),\ \Eprint {https://arxiv.org/abs/2311.01300} {arXiv:2311.01300 [gr-qc]} \BibitemShut {NoStop}%
\bibitem [{\citenamefont {Sago}\ and\ \citenamefont {Fujita}(2015)}]{10.1093/ptep/ptv092}%
  \BibitemOpen
  \bibfield  {author} {\bibinfo {author} {\bibfnamefont {N.}~\bibnamefont {Sago}}\ and\ \bibinfo {author} {\bibfnamefont {R.}~\bibnamefont {Fujita}},\ }\bibfield  {title} {\bibinfo {title} {{Calculation of radiation reaction effect on orbital parameters in Kerr spacetime}},\ }\bibfield  {journal} {\bibinfo  {journal} {Progress of Theoretical and Experimental Physics}\ }\textbf {\bibinfo {volume} {2015}},\ \href {https://doi.org/10.1093/ptep/ptv092} {10.1093/ptep/ptv092} (\bibinfo {year} {2015}),\ \bibinfo {note} {073E03},\ \Eprint {https://arxiv.org/abs/https://academic.oup.com/ptep/article-pdf/2015/7/073E03/7698002/ptv092.pdf} {https://academic.oup.com/ptep/article-pdf/2015/7/073E03/7698002/ptv092.pdf} \BibitemShut {NoStop}%
\bibitem [{\citenamefont {Wardell}\ \emph {et~al.}(2023)\citenamefont {Wardell}, \citenamefont {Pound}, \citenamefont {Warburton}, \citenamefont {Miller}, \citenamefont {Durkan},\ and\ \citenamefont {Le~Tiec}}]{PhysRevLett.130.241402}%
  \BibitemOpen
  \bibfield  {author} {\bibinfo {author} {\bibfnamefont {B.}~\bibnamefont {Wardell}}, \bibinfo {author} {\bibfnamefont {A.}~\bibnamefont {Pound}}, \bibinfo {author} {\bibfnamefont {N.}~\bibnamefont {Warburton}}, \bibinfo {author} {\bibfnamefont {J.}~\bibnamefont {Miller}}, \bibinfo {author} {\bibfnamefont {L.}~\bibnamefont {Durkan}},\ and\ \bibinfo {author} {\bibfnamefont {A.}~\bibnamefont {Le~Tiec}},\ }\bibfield  {title} {\bibinfo {title} {Gravitational waveforms for compact binaries from second-order self-force theory},\ }\href {https://doi.org/10.1103/PhysRevLett.130.241402} {\bibfield  {journal} {\bibinfo  {journal} {Phys. Rev. Lett.}\ }\textbf {\bibinfo {volume} {130}},\ \bibinfo {pages} {241402} (\bibinfo {year} {2023})}\BibitemShut {NoStop}%
\bibitem [{\citenamefont {Pound}\ \emph {et~al.}(2020)\citenamefont {Pound}, \citenamefont {Wardell}, \citenamefont {Warburton},\ and\ \citenamefont {Miller}}]{PhysRevLett.124.021101}%
  \BibitemOpen
  \bibfield  {author} {\bibinfo {author} {\bibfnamefont {A.}~\bibnamefont {Pound}}, \bibinfo {author} {\bibfnamefont {B.}~\bibnamefont {Wardell}}, \bibinfo {author} {\bibfnamefont {N.}~\bibnamefont {Warburton}},\ and\ \bibinfo {author} {\bibfnamefont {J.}~\bibnamefont {Miller}},\ }\bibfield  {title} {\bibinfo {title} {Second-order self-force calculation of gravitational binding energy in compact binaries},\ }\href {https://doi.org/10.1103/PhysRevLett.124.021101} {\bibfield  {journal} {\bibinfo  {journal} {Phys. Rev. Lett.}\ }\textbf {\bibinfo {volume} {124}},\ \bibinfo {pages} {021101} (\bibinfo {year} {2020})}\BibitemShut {NoStop}%
\bibitem [{\citenamefont {Barack}\ and\ \citenamefont {Cutler}(2004)}]{barack2004lisa}%
  \BibitemOpen
  \bibfield  {author} {\bibinfo {author} {\bibfnamefont {L.}~\bibnamefont {Barack}}\ and\ \bibinfo {author} {\bibfnamefont {C.}~\bibnamefont {Cutler}},\ }\bibfield  {title} {\bibinfo {title} {Lisa capture sources: Approximate waveforms, signal-to-noise ratios, and parameter estimation accuracy},\ }\href {https://doi.org/10.1103/PhysRevD.69.082005} {\bibfield  {journal} {\bibinfo  {journal} {Phys. Rev. D}\ }\textbf {\bibinfo {volume} {69}},\ \bibinfo {pages} {082005} (\bibinfo {year} {2004})}\BibitemShut {NoStop}%
\bibitem [{\citenamefont {Babak}\ \emph {et~al.}(2007)\citenamefont {Babak}, \citenamefont {Fang}, \citenamefont {Gair}, \citenamefont {Glampedakis},\ and\ \citenamefont {Hughes}}]{babak2007kludge}%
  \BibitemOpen
  \bibfield  {author} {\bibinfo {author} {\bibfnamefont {S.}~\bibnamefont {Babak}}, \bibinfo {author} {\bibfnamefont {H.}~\bibnamefont {Fang}}, \bibinfo {author} {\bibfnamefont {J.~R.}\ \bibnamefont {Gair}}, \bibinfo {author} {\bibfnamefont {K.}~\bibnamefont {Glampedakis}},\ and\ \bibinfo {author} {\bibfnamefont {S.~A.}\ \bibnamefont {Hughes}},\ }\bibfield  {title} {\bibinfo {title} {``kludge'' gravitational waveforms for a test-body orbiting a kerr black hole},\ }\href {https://doi.org/10.1103/PhysRevD.75.024005} {\bibfield  {journal} {\bibinfo  {journal} {Phys. Rev. D}\ }\textbf {\bibinfo {volume} {75}},\ \bibinfo {pages} {024005} (\bibinfo {year} {2007})}\BibitemShut {NoStop}%
\bibitem [{\citenamefont {Chua}\ \emph {et~al.}(2017)\citenamefont {Chua}, \citenamefont {Moore},\ and\ \citenamefont {Gair}}]{chua2017augmented}%
  \BibitemOpen
  \bibfield  {author} {\bibinfo {author} {\bibfnamefont {A.~J.~K.}\ \bibnamefont {Chua}}, \bibinfo {author} {\bibfnamefont {C.~J.}\ \bibnamefont {Moore}},\ and\ \bibinfo {author} {\bibfnamefont {J.~R.}\ \bibnamefont {Gair}},\ }\bibfield  {title} {\bibinfo {title} {Augmented kludge waveforms for detecting extreme-mass-ratio inspirals},\ }\href {https://doi.org/10.1103/PhysRevD.96.044005} {\bibfield  {journal} {\bibinfo  {journal} {Phys. Rev. D}\ }\textbf {\bibinfo {volume} {96}},\ \bibinfo {pages} {044005} (\bibinfo {year} {2017})}\BibitemShut {NoStop}%
\bibitem [{\citenamefont {Vigeland}\ \emph {et~al.}(2011)\citenamefont {Vigeland}, \citenamefont {Yunes},\ and\ \citenamefont {Stein}}]{VYS11}%
  \BibitemOpen
  \bibfield  {author} {\bibinfo {author} {\bibfnamefont {S.}~\bibnamefont {Vigeland}}, \bibinfo {author} {\bibfnamefont {N.}~\bibnamefont {Yunes}},\ and\ \bibinfo {author} {\bibfnamefont {L.~C.}\ \bibnamefont {Stein}},\ }\bibfield  {title} {\bibinfo {title} {Bumpy black holes in alternative theories of gravity},\ }\href {https://doi.org/10.1103/PhysRevD.83.104027} {\bibfield  {journal} {\bibinfo  {journal} {Phys. Rev. D}\ }\textbf {\bibinfo {volume} {83}},\ \bibinfo {pages} {104027} (\bibinfo {year} {2011})}\BibitemShut {NoStop}%
\bibitem [{\citenamefont {Glampedakis}\ and\ \citenamefont {Babak}(2006)}]{Glampedakis_2006}%
  \BibitemOpen
  \bibfield  {author} {\bibinfo {author} {\bibfnamefont {K.}~\bibnamefont {Glampedakis}}\ and\ \bibinfo {author} {\bibfnamefont {S.}~\bibnamefont {Babak}},\ }\bibfield  {title} {\bibinfo {title} {Mapping spacetimes with lisa: inspiral of a test body in a ‘quasi-kerr’ field},\ }\href {https://doi.org/10.1088/0264-9381/23/12/013} {\bibfield  {journal} {\bibinfo  {journal} {Classical and Quantum Gravity}\ }\textbf {\bibinfo {volume} {23}},\ \bibinfo {pages} {4167} (\bibinfo {year} {2006})}\BibitemShut {NoStop}%
\bibitem [{\citenamefont {Vigeland}\ and\ \citenamefont {Hughes}(2010)}]{VH10}%
  \BibitemOpen
  \bibfield  {author} {\bibinfo {author} {\bibfnamefont {S.~J.}\ \bibnamefont {Vigeland}}\ and\ \bibinfo {author} {\bibfnamefont {S.~A.}\ \bibnamefont {Hughes}},\ }\bibfield  {title} {\bibinfo {title} {Spacetime and orbits of bumpy black holes},\ }\href {https://doi.org/10.1103/PhysRevD.81.024030} {\bibfield  {journal} {\bibinfo  {journal} {Phys. Rev. D}\ }\textbf {\bibinfo {volume} {81}},\ \bibinfo {pages} {024030} (\bibinfo {year} {2010})}\BibitemShut {NoStop}%
\bibitem [{\citenamefont {Gair}\ and\ \citenamefont {Yunes}(2011)}]{GY11}%
  \BibitemOpen
  \bibfield  {author} {\bibinfo {author} {\bibfnamefont {J.}~\bibnamefont {Gair}}\ and\ \bibinfo {author} {\bibfnamefont {N.}~\bibnamefont {Yunes}},\ }\bibfield  {title} {\bibinfo {title} {Approximate waveforms for extreme-mass-ratio inspirals in modified gravity spacetimes},\ }\href {https://doi.org/10.1103/PhysRevD.84.064016} {\bibfield  {journal} {\bibinfo  {journal} {Phys. Rev. D}\ }\textbf {\bibinfo {volume} {84}},\ \bibinfo {pages} {064016} (\bibinfo {year} {2011})}\BibitemShut {NoStop}%
\bibitem [{\citenamefont {Chua}\ \emph {et~al.}(2018)\citenamefont {Chua}, \citenamefont {Hee}, \citenamefont {Handley}, \citenamefont {Higson}, \citenamefont {Moore}, \citenamefont {Gair}, \citenamefont {Hobson},\ and\ \citenamefont {Lasenby}}]{Bnbumps}%
  \BibitemOpen
  \bibfield  {author} {\bibinfo {author} {\bibfnamefont {A.~J.~K.}\ \bibnamefont {Chua}}, \bibinfo {author} {\bibfnamefont {S.}~\bibnamefont {Hee}}, \bibinfo {author} {\bibfnamefont {W.~J.}\ \bibnamefont {Handley}}, \bibinfo {author} {\bibfnamefont {E.}~\bibnamefont {Higson}}, \bibinfo {author} {\bibfnamefont {C.~J.}\ \bibnamefont {Moore}}, \bibinfo {author} {\bibfnamefont {J.~R.}\ \bibnamefont {Gair}}, \bibinfo {author} {\bibfnamefont {M.~P.}\ \bibnamefont {Hobson}},\ and\ \bibinfo {author} {\bibfnamefont {A.~N.}\ \bibnamefont {Lasenby}},\ }\bibfield  {title} {\bibinfo {title} {{Towards a framework for testing general relativity with extreme-mass-ratio-inspiral observations}},\ }\href {https://doi.org/10.1093/mnras/sty1079} {\bibfield  {journal} {\bibinfo  {journal} {Monthly Notices of the Royal Astronomical Society}\ }\textbf {\bibinfo {volume} {478}},\ \bibinfo {pages} {28} (\bibinfo {year} {2018})},\ \Eprint {https://arxiv.org/abs/https://academic.oup.com/mnras/article-pdf/478/1/28/25133616/sty1079.pdf}
  {https://academic.oup.com/mnras/article-pdf/478/1/28/25133616/sty1079.pdf} \BibitemShut {NoStop}%
\bibitem [{\citenamefont {Vigeland}(2010)}]{PhysRevD.82.104041}%
  \BibitemOpen
  \bibfield  {author} {\bibinfo {author} {\bibfnamefont {S.~J.}\ \bibnamefont {Vigeland}},\ }\bibfield  {title} {\bibinfo {title} {Multipole moments of bumpy black holes},\ }\href {https://doi.org/10.1103/PhysRevD.82.104041} {\bibfield  {journal} {\bibinfo  {journal} {Phys. Rev. D}\ }\textbf {\bibinfo {volume} {82}},\ \bibinfo {pages} {104041} (\bibinfo {year} {2010})}\BibitemShut {NoStop}%
\bibitem [{\citenamefont {Moore}\ \emph {et~al.}(2017)\citenamefont {Moore}, \citenamefont {Chua},\ and\ \citenamefont {Gair}}]{Moore_2017}%
  \BibitemOpen
  \bibfield  {author} {\bibinfo {author} {\bibfnamefont {C.~J.}\ \bibnamefont {Moore}}, \bibinfo {author} {\bibfnamefont {A.~J.~K.}\ \bibnamefont {Chua}},\ and\ \bibinfo {author} {\bibfnamefont {J.~R.}\ \bibnamefont {Gair}},\ }\bibfield  {title} {\bibinfo {title} {Gravitational waves from extreme mass ratio inspirals around bumpy black holes},\ }\href {https://doi.org/10.1088/1361-6382/aa85fa} {\bibfield  {journal} {\bibinfo  {journal} {Classical and Quantum Gravity}\ }\textbf {\bibinfo {volume} {34}},\ \bibinfo {pages} {195009} (\bibinfo {year} {2017})}\BibitemShut {NoStop}%
\bibitem [{\citenamefont {Yunes}\ and\ \citenamefont {Pretorius}(2009{\natexlab{a}})}]{cs2009}%
  \BibitemOpen
  \bibfield  {author} {\bibinfo {author} {\bibfnamefont {N.}~\bibnamefont {Yunes}}\ and\ \bibinfo {author} {\bibfnamefont {F.}~\bibnamefont {Pretorius}},\ }\bibfield  {title} {\bibinfo {title} {Dynamical chern-simons modified gravity: Spinning black holes in the slow-rotation approximation},\ }\href {https://doi.org/10.1103/PhysRevD.79.084043} {\bibfield  {journal} {\bibinfo  {journal} {Phys. Rev. D}\ }\textbf {\bibinfo {volume} {79}},\ \bibinfo {pages} {084043} (\bibinfo {year} {2009}{\natexlab{a}})}\BibitemShut {NoStop}%
\bibitem [{\citenamefont {Yunes}\ and\ \citenamefont {Stein}(2011)}]{modifyGR_nospin}%
  \BibitemOpen
  \bibfield  {author} {\bibinfo {author} {\bibfnamefont {N.}~\bibnamefont {Yunes}}\ and\ \bibinfo {author} {\bibfnamefont {L.~C.}\ \bibnamefont {Stein}},\ }\bibfield  {title} {\bibinfo {title} {Nonspinning black holes in alternative theories of gravity},\ }\href {https://doi.org/10.1103/PhysRevD.83.104002} {\bibfield  {journal} {\bibinfo  {journal} {Phys. Rev. D}\ }\textbf {\bibinfo {volume} {83}},\ \bibinfo {pages} {104002} (\bibinfo {year} {2011})}\BibitemShut {NoStop}%
\bibitem [{\citenamefont {Cutler}\ and\ \citenamefont {Vallisneri}(2007{\natexlab{a}})}]{PhysRevD.76.104018}%
  \BibitemOpen
  \bibfield  {author} {\bibinfo {author} {\bibfnamefont {C.}~\bibnamefont {Cutler}}\ and\ \bibinfo {author} {\bibfnamefont {M.}~\bibnamefont {Vallisneri}},\ }\bibfield  {title} {\bibinfo {title} {Lisa detections of massive black hole inspirals: Parameter extraction errors due to inaccurate template waveforms},\ }\href {https://doi.org/10.1103/PhysRevD.76.104018} {\bibfield  {journal} {\bibinfo  {journal} {Phys. Rev. D}\ }\textbf {\bibinfo {volume} {76}},\ \bibinfo {pages} {104018} (\bibinfo {year} {2007}{\natexlab{a}})}\BibitemShut {NoStop}%
\bibitem [{\citenamefont {Yunes}\ and\ \citenamefont {Pretorius}(2009{\natexlab{b}})}]{PhysRevD.80.122003}%
  \BibitemOpen
  \bibfield  {author} {\bibinfo {author} {\bibfnamefont {N.}~\bibnamefont {Yunes}}\ and\ \bibinfo {author} {\bibfnamefont {F.}~\bibnamefont {Pretorius}},\ }\bibfield  {title} {\bibinfo {title} {Fundamental theoretical bias in gravitational wave astrophysics and the parametrized post-einsteinian framework},\ }\href {https://doi.org/10.1103/PhysRevD.80.122003} {\bibfield  {journal} {\bibinfo  {journal} {Phys. Rev. D}\ }\textbf {\bibinfo {volume} {80}},\ \bibinfo {pages} {122003} (\bibinfo {year} {2009}{\natexlab{b}})}\BibitemShut {NoStop}%
\bibitem [{\citenamefont {Chua}\ and\ \citenamefont {Gair}(2015)}]{aak_2015}%
  \BibitemOpen
  \bibfield  {author} {\bibinfo {author} {\bibfnamefont {A.~J.~K.}\ \bibnamefont {Chua}}\ and\ \bibinfo {author} {\bibfnamefont {J.~R.}\ \bibnamefont {Gair}},\ }\bibfield  {title} {\bibinfo {title} {Improved analytic extreme-mass-ratio inspiral model for scoping out elisa data analysis},\ }\href {https://doi.org/10.1088/0264-9381/32/23/232002} {\bibfield  {journal} {\bibinfo  {journal} {Classical and Quantum Gravity}\ }\textbf {\bibinfo {volume} {32}},\ \bibinfo {pages} {232002} (\bibinfo {year} {2015})}\BibitemShut {NoStop}%
\bibitem [{\citenamefont {Peters}\ and\ \citenamefont {Mathews}(1963)}]{PhysRev.131.435}%
  \BibitemOpen
  \bibfield  {author} {\bibinfo {author} {\bibfnamefont {P.~C.}\ \bibnamefont {Peters}}\ and\ \bibinfo {author} {\bibfnamefont {J.}~\bibnamefont {Mathews}},\ }\bibfield  {title} {\bibinfo {title} {Gravitational radiation from point masses in a keplerian orbit},\ }\href {https://doi.org/10.1103/PhysRev.131.435} {\bibfield  {journal} {\bibinfo  {journal} {Phys. Rev.}\ }\textbf {\bibinfo {volume} {131}},\ \bibinfo {pages} {435} (\bibinfo {year} {1963})}\BibitemShut {NoStop}%
\bibitem [{\citenamefont {Armstrong}\ \emph {et~al.}(1999)\citenamefont {Armstrong}, \citenamefont {Estabrook},\ and\ \citenamefont {Tinto}}]{Armstrong_1999}%
  \BibitemOpen
  \bibfield  {author} {\bibinfo {author} {\bibfnamefont {J.~W.}\ \bibnamefont {Armstrong}}, \bibinfo {author} {\bibfnamefont {F.~B.}\ \bibnamefont {Estabrook}},\ and\ \bibinfo {author} {\bibfnamefont {M.}~\bibnamefont {Tinto}},\ }\bibfield  {title} {\bibinfo {title} {Time-delay interferometry for space-based gravitational wave searches},\ }\href {https://doi.org/10.1086/308110} {\bibfield  {journal} {\bibinfo  {journal} {The Astrophysical Journal}\ }\textbf {\bibinfo {volume} {527}},\ \bibinfo {pages} {814} (\bibinfo {year} {1999})}\BibitemShut {NoStop}%
\bibitem [{\citenamefont {Wang}\ and\ \citenamefont {Hu}(2023)}]{Wang_2023}%
  \BibitemOpen
  \bibfield  {author} {\bibinfo {author} {\bibfnamefont {R.}~\bibnamefont {Wang}}\ and\ \bibinfo {author} {\bibfnamefont {B.}~\bibnamefont {Hu}},\ }\bibfield  {title} {\bibinfo {title} {Litepig: a lite parameter inference system for the gravitational wave in the millihertz band},\ }\href {https://doi.org/10.1088/1572-9494/acccdd} {\bibfield  {journal} {\bibinfo  {journal} {Communications in Theoretical Physics}\ }\textbf {\bibinfo {volume} {75}},\ \bibinfo {pages} {075402} (\bibinfo {year} {2023})}\BibitemShut {NoStop}%
\bibitem [{\citenamefont {Estabrook}\ \emph {et~al.}(2000)\citenamefont {Estabrook}, \citenamefont {Tinto},\ and\ \citenamefont {Armstrong}}]{PhysRevD.62.042002}%
  \BibitemOpen
  \bibfield  {author} {\bibinfo {author} {\bibfnamefont {F.~B.}\ \bibnamefont {Estabrook}}, \bibinfo {author} {\bibfnamefont {M.}~\bibnamefont {Tinto}},\ and\ \bibinfo {author} {\bibfnamefont {J.~W.}\ \bibnamefont {Armstrong}},\ }\bibfield  {title} {\bibinfo {title} {Time-delay analysis of lisa gravitational wave data: Elimination of spacecraft motion effects},\ }\href {https://doi.org/10.1103/PhysRevD.62.042002} {\bibfield  {journal} {\bibinfo  {journal} {Phys. Rev. D}\ }\textbf {\bibinfo {volume} {62}},\ \bibinfo {pages} {042002} (\bibinfo {year} {2000})}\BibitemShut {NoStop}%
\bibitem [{\citenamefont {Yang}\ \emph {et~al.}(2023)\citenamefont {Yang}, \citenamefont {Wang}, \citenamefont {Tan},\ and\ \citenamefont {Shao}}]{YANG2023106900}%
  \BibitemOpen
  \bibfield  {author} {\bibinfo {author} {\bibfnamefont {Z.-J.}\ \bibnamefont {Yang}}, \bibinfo {author} {\bibfnamefont {P.-P.}\ \bibnamefont {Wang}}, \bibinfo {author} {\bibfnamefont {Y.-J.}\ \bibnamefont {Tan}},\ and\ \bibinfo {author} {\bibfnamefont {C.-G.}\ \bibnamefont {Shao}},\ }\bibfield  {title} {\bibinfo {title} {Clock noise reduction in geometric time delay interferometry combinations},\ }\href {https://doi.org/https://doi.org/10.1016/j.rinp.2023.106900} {\bibfield  {journal} {\bibinfo  {journal} {Results in Physics}\ }\textbf {\bibinfo {volume} {53}},\ \bibinfo {pages} {106900} (\bibinfo {year} {2023})}\BibitemShut {NoStop}%
\bibitem [{\citenamefont {Katz}\ \emph {et~al.}(2022)\citenamefont {Katz}, \citenamefont {Bayle}, \citenamefont {Chua},\ and\ \citenamefont {Vallisneri}}]{PhysRevD.106.103001}%
  \BibitemOpen
  \bibfield  {author} {\bibinfo {author} {\bibfnamefont {M.~L.}\ \bibnamefont {Katz}}, \bibinfo {author} {\bibfnamefont {J.-B.}\ \bibnamefont {Bayle}}, \bibinfo {author} {\bibfnamefont {A.~J.~K.}\ \bibnamefont {Chua}},\ and\ \bibinfo {author} {\bibfnamefont {M.}~\bibnamefont {Vallisneri}},\ }\bibfield  {title} {\bibinfo {title} {Assessing the data-analysis impact of lisa orbit approximations using a gpu-accelerated response model},\ }\href {https://doi.org/10.1103/PhysRevD.106.103001} {\bibfield  {journal} {\bibinfo  {journal} {Phys. Rev. D}\ }\textbf {\bibinfo {volume} {106}},\ \bibinfo {pages} {103001} (\bibinfo {year} {2022})}\BibitemShut {NoStop}%
\bibitem [{\citenamefont {Tinto}\ \emph {et~al.}(2004)\citenamefont {Tinto}, \citenamefont {Estabrook},\ and\ \citenamefont {Armstrong}}]{PhysRevD.69.082001}%
  \BibitemOpen
  \bibfield  {author} {\bibinfo {author} {\bibfnamefont {M.}~\bibnamefont {Tinto}}, \bibinfo {author} {\bibfnamefont {F.~B.}\ \bibnamefont {Estabrook}},\ and\ \bibinfo {author} {\bibfnamefont {J.~W.}\ \bibnamefont {Armstrong}},\ }\bibfield  {title} {\bibinfo {title} {Time delay interferometry with moving spacecraft arrays},\ }\href {https://doi.org/10.1103/PhysRevD.69.082001} {\bibfield  {journal} {\bibinfo  {journal} {Phys. Rev. D}\ }\textbf {\bibinfo {volume} {69}},\ \bibinfo {pages} {082001} (\bibinfo {year} {2004})}\BibitemShut {NoStop}%
\bibitem [{\citenamefont {Chua}\ \emph {et~al.}(2020)\citenamefont {Chua}, \citenamefont {Katz}, \citenamefont {Warburton},\ and\ \citenamefont {Hughes}}]{chua_2020_3981654}%
  \BibitemOpen
  \bibfield  {author} {\bibinfo {author} {\bibfnamefont {A.~J.}\ \bibnamefont {Chua}}, \bibinfo {author} {\bibfnamefont {M.~L.}\ \bibnamefont {Katz}}, \bibinfo {author} {\bibfnamefont {N.}~\bibnamefont {Warburton}},\ and\ \bibinfo {author} {\bibfnamefont {S.~A.}\ \bibnamefont {Hughes}},\ }\bibfield  {title} {\bibinfo {title} {Data for fast emri waveforms},\ }\href {https://doi.org/10.5281/zenodo.3981654} {10.5281/zenodo.3981654} (\bibinfo {year} {2020})\BibitemShut {NoStop}%
\bibitem [{\citenamefont {{Speagle}}(2020)}]{2020MNRAS4933132S}%
  \BibitemOpen
  \bibfield  {author} {\bibinfo {author} {\bibfnamefont {J.~S.}\ \bibnamefont {{Speagle}}},\ }\bibfield  {title} {\bibinfo {title} {{DYNESTY: a dynamic nested sampling package for estimating Bayesian posteriors and evidences}},\ }\href {https://doi.org/10.1093/mnras/staa278} {\bibfield  {journal} {\bibinfo  {journal} {mnras}\ }\textbf {\bibinfo {volume} {493}},\ \bibinfo {pages} {3132} (\bibinfo {year} {2020})},\ \Eprint {https://arxiv.org/abs/1904.02180} {arXiv:1904.02180 [astro-ph.IM]} \BibitemShut {NoStop}%
\bibitem [{\citenamefont {Koposov}\ and\ \citenamefont {et~al}(2023)}]{sergey8408702}%
  \BibitemOpen
  \bibfield  {author} {\bibinfo {author} {\bibfnamefont {S.}~\bibnamefont {Koposov}}\ and\ \bibinfo {author} {\bibfnamefont {J.~S.}\ \bibnamefont {et~al}},\ }\href {https://doi.org/10.5281/zenodo.8408702} {\bibinfo {title} {joshspeagle/dynesty: v2.1.3}} (\bibinfo {year} {2023})\BibitemShut {NoStop}%
\bibitem [{\citenamefont {Buchner}(2023)}]{nested_sampling}%
  \BibitemOpen
  \bibfield  {author} {\bibinfo {author} {\bibfnamefont {J.}~\bibnamefont {Buchner}},\ }\bibfield  {title} {\bibinfo {title} {{Nested sampling methods}},\ }\href {https://doi.org/10.1214/23-SS144} {\bibfield  {journal} {\bibinfo  {journal} {Statistics Surveys}\ }\textbf {\bibinfo {volume} {17}},\ \bibinfo {pages} {169 } (\bibinfo {year} {2023})}\BibitemShut {NoStop}%
\bibitem [{\citenamefont {Thrane}\ and\ \citenamefont {Talbot}(2019)}]{Thrane_Talbot_2019}%
  \BibitemOpen
  \bibfield  {author} {\bibinfo {author} {\bibfnamefont {E.}~\bibnamefont {Thrane}}\ and\ \bibinfo {author} {\bibfnamefont {C.}~\bibnamefont {Talbot}},\ }\bibfield  {title} {\bibinfo {title} {An introduction to bayesian inference in gravitational-wave astronomy: Parameter estimation, model selection, and hierarchical models},\ }\href {https://doi.org/10.1017/pasa.2019.2} {\bibfield  {journal} {\bibinfo  {journal} {Publications of the Astronomical Society of Australia}\ }\textbf {\bibinfo {volume} {36}},\ \bibinfo {pages} {e010} (\bibinfo {year} {2019})}\BibitemShut {NoStop}%
\bibitem [{\citenamefont {Zou}\ \emph {et~al.}(2024)\citenamefont {Zou}, \citenamefont {Mohanty}, \citenamefont {Luo},\ and\ \citenamefont {Liu}}]{universe10040171}%
  \BibitemOpen
  \bibfield  {author} {\bibinfo {author} {\bibfnamefont {X.-B.}\ \bibnamefont {Zou}}, \bibinfo {author} {\bibfnamefont {S.~D.}\ \bibnamefont {Mohanty}}, \bibinfo {author} {\bibfnamefont {H.-G.}\ \bibnamefont {Luo}},\ and\ \bibinfo {author} {\bibfnamefont {Y.-X.}\ \bibnamefont {Liu}},\ }\bibfield  {title} {\bibinfo {title} {Search for extreme mass ratio inspirals using particle swarm optimization and reduced dimensionality likelihoods},\ }\bibfield  {journal} {\bibinfo  {journal} {Universe}\ }\textbf {\bibinfo {volume} {10}},\ \href {https://doi.org/10.3390/universe10040171} {10.3390/universe10040171} (\bibinfo {year} {2024})\BibitemShut {NoStop}%
\bibitem [{\citenamefont {Wang}\ \emph {et~al.}(2012)\citenamefont {Wang}, \citenamefont {Shang},\ and\ \citenamefont {Babak}}]{PhysRevD.86.104050}%
  \BibitemOpen
  \bibfield  {author} {\bibinfo {author} {\bibfnamefont {Y.}~\bibnamefont {Wang}}, \bibinfo {author} {\bibfnamefont {Y.}~\bibnamefont {Shang}},\ and\ \bibinfo {author} {\bibfnamefont {S.}~\bibnamefont {Babak}},\ }\bibfield  {title} {\bibinfo {title} {Extreme mass ratio inspiral data analysis with a phenomenological waveform},\ }\href {https://doi.org/10.1103/PhysRevD.86.104050} {\bibfield  {journal} {\bibinfo  {journal} {Phys. Rev. D}\ }\textbf {\bibinfo {volume} {86}},\ \bibinfo {pages} {104050} (\bibinfo {year} {2012})}\BibitemShut {NoStop}%
\bibitem [{\citenamefont {Cutler}\ and\ \citenamefont {Vallisneri}(2007{\natexlab{b}})}]{template_error_2007}%
  \BibitemOpen
  \bibfield  {author} {\bibinfo {author} {\bibfnamefont {C.}~\bibnamefont {Cutler}}\ and\ \bibinfo {author} {\bibfnamefont {M.}~\bibnamefont {Vallisneri}},\ }\bibfield  {title} {\bibinfo {title} {Lisa detections of massive black hole inspirals: Parameter extraction errors due to inaccurate template waveforms},\ }\href {https://doi.org/10.1103/PhysRevD.76.104018} {\bibfield  {journal} {\bibinfo  {journal} {Phys. Rev. D}\ }\textbf {\bibinfo {volume} {76}},\ \bibinfo {pages} {104018} (\bibinfo {year} {2007}{\natexlab{b}})}\BibitemShut {NoStop}%
\bibitem [{\citenamefont {Hu}\ and\ \citenamefont {Veitch}(2023)}]{Hu_2023}%
  \BibitemOpen
  \bibfield  {author} {\bibinfo {author} {\bibfnamefont {Q.}~\bibnamefont {Hu}}\ and\ \bibinfo {author} {\bibfnamefont {J.}~\bibnamefont {Veitch}},\ }\bibfield  {title} {\bibinfo {title} {Accumulating errors in tests of general relativity with gravitational waves: Overlapping signals and inaccurate waveforms},\ }\href {https://doi.org/10.3847/1538-4357/acbc18} {\bibfield  {journal} {\bibinfo  {journal} {The Astrophysical Journal}\ }\textbf {\bibinfo {volume} {945}},\ \bibinfo {pages} {103} (\bibinfo {year} {2023})}\BibitemShut {NoStop}%
\bibitem [{\citenamefont {Shen}\ \emph {et~al.}(2023)\citenamefont {Shen}, \citenamefont {Han}, \citenamefont {Zhang}, \citenamefont {Yang}, \citenamefont {Zhong}, \citenamefont {Jiang},\ and\ \citenamefont {Cui}}]{PhysRevD.108.064015}%
  \BibitemOpen
  \bibfield  {author} {\bibinfo {author} {\bibfnamefont {P.}~\bibnamefont {Shen}}, \bibinfo {author} {\bibfnamefont {W.-B.}\ \bibnamefont {Han}}, \bibinfo {author} {\bibfnamefont {C.}~\bibnamefont {Zhang}}, \bibinfo {author} {\bibfnamefont {S.-C.}\ \bibnamefont {Yang}}, \bibinfo {author} {\bibfnamefont {X.-Y.}\ \bibnamefont {Zhong}}, \bibinfo {author} {\bibfnamefont {Y.}~\bibnamefont {Jiang}},\ and\ \bibinfo {author} {\bibfnamefont {Q.}~\bibnamefont {Cui}},\ }\bibfield  {title} {\bibinfo {title} {Influence of mass-ratio corrections in extreme-mass-ratio inspirals for testing general relativity},\ }\href {https://doi.org/10.1103/PhysRevD.108.064015} {\bibfield  {journal} {\bibinfo  {journal} {Phys. Rev. D}\ }\textbf {\bibinfo {volume} {108}},\ \bibinfo {pages} {064015} (\bibinfo {year} {2023})}\BibitemShut {NoStop}%
\bibitem [{\citenamefont {Burke}\ \emph {et~al.}(2024)\citenamefont {Burke}, \citenamefont {Piovano}, \citenamefont {Warburton}, \citenamefont {Lynch}, \citenamefont {Speri}, \citenamefont {Kavanagh}, \citenamefont {Wardell}, \citenamefont {Pound}, \citenamefont {Durkan},\ and\ \citenamefont {Miller}}]{burke2024accuracy}%
  \BibitemOpen
  \bibfield  {author} {\bibinfo {author} {\bibfnamefont {O.}~\bibnamefont {Burke}}, \bibinfo {author} {\bibfnamefont {G.~A.}\ \bibnamefont {Piovano}}, \bibinfo {author} {\bibfnamefont {N.}~\bibnamefont {Warburton}}, \bibinfo {author} {\bibfnamefont {P.}~\bibnamefont {Lynch}}, \bibinfo {author} {\bibfnamefont {L.}~\bibnamefont {Speri}}, \bibinfo {author} {\bibfnamefont {C.}~\bibnamefont {Kavanagh}}, \bibinfo {author} {\bibfnamefont {B.}~\bibnamefont {Wardell}}, \bibinfo {author} {\bibfnamefont {A.}~\bibnamefont {Pound}}, \bibinfo {author} {\bibfnamefont {L.}~\bibnamefont {Durkan}},\ and\ \bibinfo {author} {\bibfnamefont {J.}~\bibnamefont {Miller}},\ }\href@noop {} {\bibinfo {title} {Accuracy requirements: Assessing the importance of first post-adiabatic terms for small-mass-ratio binaries}} (\bibinfo {year} {2024}),\ \Eprint {https://arxiv.org/abs/2310.08927} {arXiv:2310.08927 [gr-qc]} \BibitemShut {NoStop}%
\bibitem [{\citenamefont {Liu}\ \emph {et~al.}(2023)\citenamefont {Liu}, \citenamefont {Ruan},\ and\ \citenamefont {Guo}}]{PhysRevD.107.064021}%
  \BibitemOpen
  \bibfield  {author} {\bibinfo {author} {\bibfnamefont {C.}~\bibnamefont {Liu}}, \bibinfo {author} {\bibfnamefont {W.-H.}\ \bibnamefont {Ruan}},\ and\ \bibinfo {author} {\bibfnamefont {Z.-K.}\ \bibnamefont {Guo}},\ }\bibfield  {title} {\bibinfo {title} {Confusion noise from galactic binaries for taiji},\ }\href {https://doi.org/10.1103/PhysRevD.107.064021} {\bibfield  {journal} {\bibinfo  {journal} {Phys. Rev. D}\ }\textbf {\bibinfo {volume} {107}},\ \bibinfo {pages} {064021} (\bibinfo {year} {2023})}\BibitemShut {NoStop}%
\bibitem [{\citenamefont {Benacquista}(2020)}]{Benacquista2020}%
  \BibitemOpen
  \bibfield  {author} {\bibinfo {author} {\bibfnamefont {M.}~\bibnamefont {Benacquista}},\ }\bibinfo {title} {Lisa and the galactic population of compact binaries},\ in\ \href {https://doi.org/10.1007/978-981-15-4702-7_19-1} {\emph {\bibinfo {booktitle} {Handbook of Gravitational Wave Astronomy}}},\ \bibinfo {editor} {edited by\ \bibinfo {editor} {\bibfnamefont {C.}~\bibnamefont {Bambi}}, \bibinfo {editor} {\bibfnamefont {S.}~\bibnamefont {Katsanevas}},\ and\ \bibinfo {editor} {\bibfnamefont {K.~D.}\ \bibnamefont {Kokkotas}}}\ (\bibinfo  {publisher} {Springer Singapore},\ \bibinfo {address} {Singapore},\ \bibinfo {year} {2020})\ pp.\ \bibinfo {pages} {1--24}\BibitemShut {NoStop}%
\bibitem [{\citenamefont {Speri}\ \emph {et~al.}(2024)\citenamefont {Speri}, \citenamefont {Barsanti}, \citenamefont {Maselli}, \citenamefont {Sotiriou}, \citenamefont {Warburton}, \citenamefont {van~de Meent}, \citenamefont {Chua}, \citenamefont {Burke},\ and\ \citenamefont {Gair}}]{speri2024probingfundamentalphysicsextreme}%
  \BibitemOpen
  \bibfield  {author} {\bibinfo {author} {\bibfnamefont {L.}~\bibnamefont {Speri}}, \bibinfo {author} {\bibfnamefont {S.}~\bibnamefont {Barsanti}}, \bibinfo {author} {\bibfnamefont {A.}~\bibnamefont {Maselli}}, \bibinfo {author} {\bibfnamefont {T.~P.}\ \bibnamefont {Sotiriou}}, \bibinfo {author} {\bibfnamefont {N.}~\bibnamefont {Warburton}}, \bibinfo {author} {\bibfnamefont {M.}~\bibnamefont {van~de Meent}}, \bibinfo {author} {\bibfnamefont {A.~J.~K.}\ \bibnamefont {Chua}}, \bibinfo {author} {\bibfnamefont {O.}~\bibnamefont {Burke}},\ and\ \bibinfo {author} {\bibfnamefont {J.}~\bibnamefont {Gair}},\ }\href {https://arxiv.org/abs/2406.07607} {\bibinfo {title} {Probing fundamental physics with extreme mass ratio inspirals: a full bayesian inference for scalar charge}} (\bibinfo {year} {2024}),\ \Eprint {https://arxiv.org/abs/2406.07607} {arXiv:2406.07607 [gr-qc]} \BibitemShut {NoStop}%
\bibitem [{\citenamefont {Chen}\ \emph {et~al.}(2024)\citenamefont {Chen}, \citenamefont {Liu}, \citenamefont {Zhang},\ and\ \citenamefont {Zhang}}]{PhysRevD.110.064018}%
  \BibitemOpen
  \bibfield  {author} {\bibinfo {author} {\bibfnamefont {M.-C.}\ \bibnamefont {Chen}}, \bibinfo {author} {\bibfnamefont {H.-Y.}\ \bibnamefont {Liu}}, \bibinfo {author} {\bibfnamefont {Q.-Y.}\ \bibnamefont {Zhang}},\ and\ \bibinfo {author} {\bibfnamefont {J.}~\bibnamefont {Zhang}},\ }\bibfield  {title} {\bibinfo {title} {Probing massive fields with multiband gravitational-wave observations},\ }\href {https://doi.org/10.1103/PhysRevD.110.064018} {\bibfield  {journal} {\bibinfo  {journal} {Phys. Rev. D}\ }\textbf {\bibinfo {volume} {110}},\ \bibinfo {pages} {064018} (\bibinfo {year} {2024})}\BibitemShut {NoStop}%
\bibitem [{\citenamefont {Liang}\ \emph {et~al.}(2023)\citenamefont {Liang}, \citenamefont {Xu}, \citenamefont {Mai},\ and\ \citenamefont {Shao}}]{PhysRevD.107.044053}%
  \BibitemOpen
  \bibfield  {author} {\bibinfo {author} {\bibfnamefont {D.}~\bibnamefont {Liang}}, \bibinfo {author} {\bibfnamefont {R.}~\bibnamefont {Xu}}, \bibinfo {author} {\bibfnamefont {Z.-F.}\ \bibnamefont {Mai}},\ and\ \bibinfo {author} {\bibfnamefont {L.}~\bibnamefont {Shao}},\ }\bibfield  {title} {\bibinfo {title} {Probing vector hair of black holes with extreme-mass-ratio inspirals},\ }\href {https://doi.org/10.1103/PhysRevD.107.044053} {\bibfield  {journal} {\bibinfo  {journal} {Phys. Rev. D}\ }\textbf {\bibinfo {volume} {107}},\ \bibinfo {pages} {044053} (\bibinfo {year} {2023})}\BibitemShut {NoStop}%
\bibitem [{\citenamefont {Zhang}\ \emph {et~al.}(2023)\citenamefont {Zhang}, \citenamefont {Guo}, \citenamefont {Gong},\ and\ \citenamefont {Wang}}]{zhang2023detectingvectorchargeextreme}%
  \BibitemOpen
  \bibfield  {author} {\bibinfo {author} {\bibfnamefont {C.}~\bibnamefont {Zhang}}, \bibinfo {author} {\bibfnamefont {H.}~\bibnamefont {Guo}}, \bibinfo {author} {\bibfnamefont {Y.}~\bibnamefont {Gong}},\ and\ \bibinfo {author} {\bibfnamefont {B.}~\bibnamefont {Wang}},\ }\href {https://arxiv.org/abs/2301.05915} {\bibinfo {title} {Detecting vector charge with extreme mass ratio inspirals onto kerr black holes}} (\bibinfo {year} {2023}),\ \Eprint {https://arxiv.org/abs/2301.05915} {arXiv:2301.05915 [gr-qc]} \BibitemShut {NoStop}%
\bibitem [{\citenamefont {Barsanti}\ \emph {et~al.}(2023)\citenamefont {Barsanti}, \citenamefont {Maselli}, \citenamefont {Sotiriou},\ and\ \citenamefont {Gualtieri}}]{PhysRevLett.131.051401}%
  \BibitemOpen
  \bibfield  {author} {\bibinfo {author} {\bibfnamefont {S.}~\bibnamefont {Barsanti}}, \bibinfo {author} {\bibfnamefont {A.}~\bibnamefont {Maselli}}, \bibinfo {author} {\bibfnamefont {T.~P.}\ \bibnamefont {Sotiriou}},\ and\ \bibinfo {author} {\bibfnamefont {L.}~\bibnamefont {Gualtieri}},\ }\bibfield  {title} {\bibinfo {title} {Detecting massive scalar fields with extreme mass-ratio inspirals},\ }\href {https://doi.org/10.1103/PhysRevLett.131.051401} {\bibfield  {journal} {\bibinfo  {journal} {Phys. Rev. Lett.}\ }\textbf {\bibinfo {volume} {131}},\ \bibinfo {pages} {051401} (\bibinfo {year} {2023})}\BibitemShut {NoStop}%
\bibitem [{\citenamefont {Jiang}\ and\ \citenamefont {Han}(2024)}]{Jiang2024GeneralFF}%
  \BibitemOpen
  \bibfield  {author} {\bibinfo {author} {\bibfnamefont {Y.}~\bibnamefont {Jiang}}\ and\ \bibinfo {author} {\bibfnamefont {W.-B.}\ \bibnamefont {Han}},\ }\bibfield  {title} {\bibinfo {title} {General formalism for dirty extreme-mass-ratio inspirals},\ }\href {https://api.semanticscholar.org/CorpusID:269821783} {\bibfield  {journal} {\bibinfo  {journal} {Science China Physics, Mechanics \& Astronomy}\ } (\bibinfo {year} {2024})}\BibitemShut {NoStop}%
\bibitem [{\citenamefont {Barausse}\ \emph {et~al.}(2014)\citenamefont {Barausse}, \citenamefont {Cardoso},\ and\ \citenamefont {Pani}}]{PhysRevD.89.104059}%
  \BibitemOpen
  \bibfield  {author} {\bibinfo {author} {\bibfnamefont {E.}~\bibnamefont {Barausse}}, \bibinfo {author} {\bibfnamefont {V.}~\bibnamefont {Cardoso}},\ and\ \bibinfo {author} {\bibfnamefont {P.}~\bibnamefont {Pani}},\ }\bibfield  {title} {\bibinfo {title} {Can environmental effects spoil precision gravitational-wave astrophysics?},\ }\href {https://doi.org/10.1103/PhysRevD.89.104059} {\bibfield  {journal} {\bibinfo  {journal} {Phys. Rev. D}\ }\textbf {\bibinfo {volume} {89}},\ \bibinfo {pages} {104059} (\bibinfo {year} {2014})}\BibitemShut {NoStop}%
\bibitem [{\citenamefont {Amaro-Seoane}(2018)}]{Amaro_Seoane_2018}%
  \BibitemOpen
  \bibfield  {author} {\bibinfo {author} {\bibfnamefont {P.}~\bibnamefont {Amaro-Seoane}},\ }\bibfield  {title} {\bibinfo {title} {Relativistic dynamics and extreme mass ratio inspirals},\ }\bibfield  {journal} {\bibinfo  {journal} {Living Reviews in Relativity}\ }\textbf {\bibinfo {volume} {21}},\ \href {https://doi.org/10.1007/s41114-018-0013-8} {10.1007/s41114-018-0013-8} (\bibinfo {year} {2018})\BibitemShut {NoStop}%
\end{thebibliography}%

\end{document}